\documentclass[preprint,12pt]{elsarticle}

\usepackage{lineno}
\usepackage{float}
\usepackage{amssymb}
\usepackage{amsmath}
\usepackage{amsthm}
\usepackage{bm}
\usepackage{multirow}
\usepackage{booktabs,multirow,array}
\usepackage{siunitx}
\newcolumntype{L}[1]{>{\raggedright\arraybackslash}p{#1}} 
\newcolumntype{C}[1]{>{\centering\arraybackslash}p{#1}}   
\usepackage{placeins}
\usepackage{booktabs}
\usepackage{tabularx}
\usepackage{makecell}
\usepackage{multirow}
\usepackage{algorithm}
\usepackage{algpseudocode}
\usepackage{rotating}
\usepackage{pdflscape} 
\usepackage{booktabs}
\usepackage{tabularx}
\usepackage{makecell}
\usepackage{graphicx}  
\usepackage{xcolor}
\providecommand{\rev}[1]{#1}
\usepackage{ragged2e}  
\usepackage{amsmath}
\usepackage{bm}
\usepackage{booktabs}
\usepackage{makecell}
\usepackage{array}
\usepackage{tabularx}

\newcolumntype{Y}{>{\centering\arraybackslash}X}
\newcolumntype{Y}[1]{>{\raggedright\arraybackslash}p{#1}}
\newcolumntype{Z}[1]{>{\centering\arraybackslash}p{#1}}

\newcommand{\bmit}[1]{\bm{\mathit{#1}}} 
\newcommand{\vecv}[1]{\bmit{#1}}        
\newcommand{\matm}[1]{\bmit{#1}}        
\newcommand{\upr}[1]{\mathrm{#1}}       

\usepackage{booktabs}
\usepackage{siunitx}
\journal{}

\begin{document}

\begin{frontmatter}



\title{A finite-element-inspired bipartite graph learned simulator for manufacturability assessment in large-deformation sheet forming}
\author[1]{Yingxue Zhao}
\author[1]{Haoran Li}
\author[1]{Haosu Zhou}
\author[2]{Tobias Pfaff}
\author[1]{Nan Li\corref{cor1}}

\affiliation[1]{organization={Dyson School of Design Engineering}, 
               addressline={Imperial College London}, 
               city={London},
               country={UK}}

\affiliation[2]{organization={NVIDIA}, 
               country={UK}}
\cortext[cor1]{Corresponding author. 
E-mail address: {n.li09@imperial.ac.uk} (N. Li)}

\begin{abstract}
Explicit dynamic finite element (FE) simulations are widely used for large-deformation engineering analysis, but \rev{repeated high-fidelity} simulations remain expensive in design-space exploration and optimisation. In explicit FE analysis, nodal kinematics and element-level deformation measures evolve through a coupled node–element update process. \rev{An FE-inspired graph learned simulator can preserve this structure, enabling direct learning of nodal kinematics and elemental deformation on their native domains. Treating them as coupled physical outputs provides complementary information: nodal displacement describes geometry evolution, whereas elemental thinning identifies local deformation and critical regions. Feeding both outputs back during rollout allows each field to inform subsequent updates of the other.} This work proposes CAtt-BiGNN, a cross-attention-based bipartite graph neural network for coupled nodal–elemental learning. \rev{The bipartite graph encodes FE-guided state organisation and information flow without imposing equilibrium or constitutive constraints.} FE mesh nodes and elements are represented as distinct graph entities connected by directed node–element edges, enabling nodal displacement increments and element-level deformation states to be predicted on their native discretisation domains. \rev{An edge-aware cross-attention processor dynamically updates edge representations and weights messages using the receiving state and evolving local relations. For larger graphs, CAtt-BiUGNN uses separate nodal and elemental downsampling and upsampling to enhance long-range propagation while preserving their interaction across graph levels.} The method is evaluated on dome-shaped cold-forming and corner-shaped hot-forming benchmarks. Comparisons with node-centred baselines, an FE-inspired hypergraph model, a Physics-Attention-based neural operator, a graph Transformer, and attention ablations show improved joint accuracy in nodal displacement and elemental thinning prediction during autoregressive rollout.
\end{abstract}

\begin{graphicalabstract}
\includegraphics[width=\linewidth]{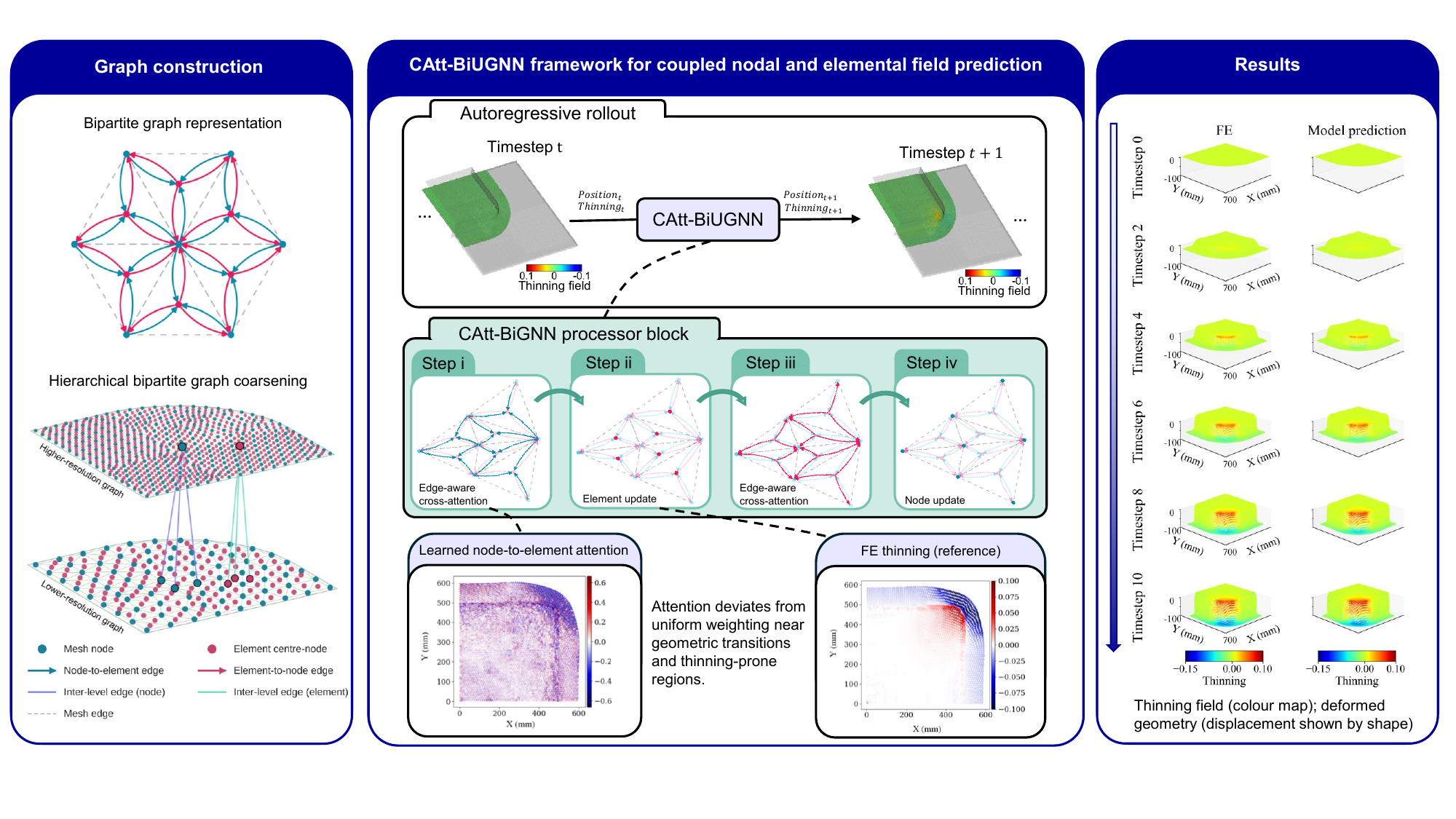}
\end{graphicalabstract}

\begin{keyword}
Sheet material forming \sep Finite-element-inspired graph learning  \sep Bipartite graph \sep Cross-attention mechanism \sep Hierarchical graph neural network \sep Manufacturability-oriented field prediction 
\end{keyword}

\end{frontmatter}

\section{Introduction}
\label{Introduction}

\rev{Explicit dynamic finite element (FE) simulations are widely used for large-deformation sheet material forming because they provide detailed predictions of geometry evolution and deformation fields~\cite{Hughes1987,bathe2006finite,belytschko2014nonlinear,Marciniak2002,Hosfordbook,zheng2018review}. However, repeated high-fidelity simulations can become computationally expensive when a large number of candidate designs or process conditions need to be evaluated~\cite{Pfrommer2018ManufacturingOptimisation,fu2023finite}. Machine-learning (ML)-based surrogate models have therefore been increasingly investigated as a complementary tool to FE analysis, particularly for rapid early-stage design feasibility assessment and design-space exploration \cite{fei2025graph,XING2025117928,ATTAR20211650,wan2024knowledge}. Such models can provide fast estimates of relevant forming responses and help identify candidate designs for subsequent detailed analysis.}

\rev{Most early ML-based forming surrogates learn a direct mapping from design or process inputs to quantities of interest. These models have been applied to rapid feasibility assessment of hot-stamped components~\cite{AttarIOP,zhao2025rapid}, deformation and thinning field prediction~\cite{AttarIOP}, face-wise drawability prediction~\cite{Stocker2025FaceWise}, and surrogate-based sensitivity analysis~\cite{Lehrer2024Drawability}. Similar approaches have also been used to support process-parameter optimisation~\cite{Pfrommer2018ManufacturingOptimisation}, component geometry optimisation under manufacturability constraints~\cite{Attar_2022,attar2023development}, blank-shape optimisation~\cite{zhou2025image}, and springback compensation~\cite{Liu2021TheoryGuidedBending}. Related developments have considered material-response prediction~\cite{ding2024constitutive,Dornheim2024Constitutive}, geometry representation~\cite{attar2023implicit,khan2025physics}, field reconstruction~\cite{Fukami2019SuperResolution,xu2022supermeshing}, and multimodal material--geometry modelling for stamping-field prediction~\cite{luo2026stampformer}. These approaches demonstrate the potential of ML surrogates for rapid forming analysis, but they mainly focus on predicting a target response directly from prescribed inputs \cite{Brzobohaty2026IncrementalForming}.}

\rev{For problems in which the physical state evolves over time, an alternative strategy is to learn the state transition itself. Temporal surrogate models have been developed to predict the evolution of physical fields in manufacturing processes~\cite{Li2026VAEALSTM}, while learned simulators predict changes in the current physical state and repeatedly use the predicted state to advance the simulation~\cite{sanchez2020learning,pfaff2021learning,Brandstetter2022MessagePassing,chen2024predicting,rubanova2024learning}. This step-wise formulation is particularly relevant to explicit dynamics, where the numerical solution also progresses through successive state updates. Rather than predicting only the final forming response, a learned simulator can therefore approximate the evolving deformation process through autoregressive rollout.}

\subsection{Existing learned simulators and remaining gap}

\rev{Graph neural networks (GNNs) are particularly suitable for this type of learned simulation because they can operate directly on irregular computational meshes and represent local interactions through graph connectivity and message passing~\cite{battaglia2018relational,gilmer2017neural,zhou2020graph,wu2020comprehensive,bronstein2021geometric,velivckovic2023everything}. Early graph-network models demonstrated that physical interactions and temporal evolution can be learned through relational message passing~\cite{battaglia2016interaction,sanchez2018graph}. This idea was subsequently extended to particle- and mesh-based learned simulators with autoregressive rollout~\cite{sanchez2020learning,pfaff2021learning}. GNN-based learned simulation has since been explored across a range of mechanics-related problems~\cite{zhao2024review,DUBEY2026118413}, including recurrent graph models for multi-step sheet-forming deformation prediction~\cite{zhao2026recurrent} and graph-based crashworthiness simulation~\cite{li2026crashworthiness,li2026maskmorph}.}

\rev{A remaining challenge for graph learned simulators is the propagation of information over large computational domains. Standard message passing is local, so long-range dependencies may require many graph-processing layers. Hierarchical and multi-scale GNNs address this issue by transferring information through coarser graph representations~\cite{Gao2019GraphUNets,deshpande2022magnet,Fortunato2022MultiScale,Cao2023BiStride}. Similar ideas have been applied to high-fidelity sheet-forming simulation through multi-level graph architectures~\cite{zhou2025multi}. More recent approaches have explored regional graph operators for arbitrary-domain PDE learning~\cite{mousavi2025rigno}. Transolver uses Physics-Attention to group mesh points into learnable slices and model interactions between them~\cite{wu2024transolver}. Transformer-based approaches have also combined mesh-based message passing with global attention for solid-mechanics simulation~\cite{iparraguirre2026meshgraphnet}, or incorporated hybrid local--global attention for mechanical-response prediction~\cite{PATRIGNANI2026118753}.}

\rev{Despite these advances, many learned simulators construct graphs primarily from mesh connectivity and represent the evolving physical state on mesh nodes~\cite{pfaff2021learning,chen2024predicting,zhao2026recurrent}. This representation is natural for nodal variables such as displacement, velocity, and acceleration \cite{vanpaemel2025physics, GAO2024112866}. However, explicit FE simulations also evolve deformation and material states at the element or integration-point level, including strain, stress, damage, and thinning~\cite{Hughes1987,bathe2006finite,belytschko2014nonlinear,Marciniak2002,mohamed2015investigation,li2014damage,politis2016prediction}. These elemental quantities can be projected onto nodes and represented in a single node-centred graph, but doing so introduces an additional mapping between the nodal and elemental domains and makes the learned state less directly aligned with the quantities' native FE domains. This suggests that the native nodal and elemental domains of the FE formulation may provide additional structural information beyond mesh connectivity alone.}

\rev{More broadly, engineering GNNs have incorporated problem-specific structural knowledge into graph construction and learning, including mechanics-based relations and domain-specific topology or validity rules~\cite{He2026PiGTN,Bleker2024LIGNN,wei2026physics}. Physical knowledge has been used to guide the representation, architecture, or information flow of the learned simulator\cite{Willard2020review,Karniadakis2021physics,dalton2023physics,vanpaemel2025physics,GAO2024112866}. For graph-based learned simulators, FE and discretisation knowledge has also been used as a physics-inspired inductive bias ~\cite{Alet2019GraphElementNetworks}. Finite-volume-informed graph networks incorporate finite-volume constraints and discretisation-inspired aggregation schemes to improve prediction on unstructured flow grids \cite{li2023finite}. Finite Element Networks (FEN) treat each mesh cell as a local computational unit and learn its contribution to the time evolution of nodal physical fields~\cite{lienen2022fen}. These cell-wise contributions are assembled to obtain nodal time derivatives, which are subsequently integrated in time. FEIH-GNN introduces explicit node--element computation through a hypergraph formulation derived from the local stiffness-matrix calculation and assembly process~\cite{GAO2024112866}. In this derivation, the element representation is associated with intermediate geometric quantities, including the element area and quadrature-point Jacobians, which contribute to the calculation and assembly of the nodal response.}

\rev{The discretisation- and FE-inspired methods discussed above demonstrate how numerical structure can guide graph learning. However, they were developed for other problem settings. The explicit dynamic forming problem considered here follows a different FE update process. Within each time increment, nodal kinematics determine the deformation of the connected elements, element-level deformation and material states are updated, and the resulting element response contributes to the subsequent nodal motion through internal-force assembly \cite{Hughes1987,bathe2006finite,belytschko2014nonlinear}. Contact conditions also evolve with the changing geometry \cite{belytschko2014nonlinear,Marciniak2002}. For the explicit dynamics considered here, the physical state is advanced without repeatedly assembling and solving a global stiffness system~\cite{bathe2006finite,belytschko2014nonlinear}. This differs from the stiffness-calculation-and-assembly process used to motivate the node--element computation in FEIH-GNN~\cite{GAO2024112866}. The relevant FE structure is therefore the update dependency between nodal kinematics, elemental states, internal-force contributions, and contact-driven nodal motion. FEN and FEIH-GNN were mainly developed and demonstrated for fluid-flow and transport problems. In these applications, element-level deformation quantities are not treated as physical state variables that are directly predicted, supervised, and updated alongside nodal kinematics during autoregressive rollout.}

\rev{This difference raises an unresolved model design challenge: translating the native nodal and elemental state domains, together with their bidirectional update dependencies, into the representation and coupling architecture of a graph learned simulator.  Preserving the two domains is important because nodal geometry evolution and element-level deformation indicators provide complementary information for forming analysis. In particular, elemental quantities such as thinning are direct manufacturability indicators and are used to identify critical deformation regions~\cite{Marciniak2002,politis2016prediction}. Representing nodal kinematics and elemental deformation fields as explicit prediction states allows both fields to be supervised against their corresponding FE outputs and fed back during rollout. However, this also creates a coupled error-propagation challenge: an error in one predicted field can affect the update of the other and accumulate over time. The required coupling mechanism should therefore exchange information between the two domains while adapting their interaction to the evolving physical state. In this work, ``finite-element-inspired'' refers to this FE-guided state organisation and information flow; equilibrium equations and constitutive relations are not imposed as explicit training constraints.}

\subsection{Proposed framework and contributions}
\rev{To address these challenges, we propose CAtt-BiGNN, a finite-element-inspired bipartite graph learned simulator for coupled nodal--elemental prediction. The methodological contribution lies in organising the bipartite state representation, directional edge updates, cross-attention, and hierarchical extension around the state domains and update dependencies of the explicit-dynamic FE formulation. Mesh nodes carry nodal kinematic and contact-related information, while element-centre nodes carry element-level deformation states. Directed node--element edges represent the local geometric relations and the two directions of physical information transfer. Because the relevance of individual node--element interactions changes with the receiving state, local geometry, and rollout time, applying the same fixed aggregation rule to all incoming interactions may be restrictive. The processor therefore updates the directed edge representations at each layer and uses edge-aware cross-attention to weight the exchanged information according to the receiving state and the evolving local relation. For larger forming meshes, CAtt-BiUGNN extends this dual-entity formulation through separate nodal and elemental downsampling and upsampling while preserving their interaction across graph levels.}

\rev{The proposed models are evaluated on two large-deformation sheet-forming cases covering cold- and hot-forming conditions. The comparisons include node-centred learned simulators, an FE-inspired hypergraph model, a Physics-Attention-based neural operator, and a graph Transformer, together with controlled bipartite and attention ablations. The evaluation considers nodal displacement, full-field and critical-region thinning, run-to-run statistical variation, computational efficiency relative to representative FE runtimes, empirical ensemble-disagreement uncertainty, and output-level area--thickness consistency during autoregressive rollout.} The main contributions of this work are summarised as follows.
\begin{itemize}

\item \rev{An explicit-dynamic FE-guided bipartite state representation is formulated for coupled nodal--elemental prediction, allowing nodal kinematics and elemental deformation states to be represented, supervised, and updated on their respective FE domains.}

\item \rev{A directional edge-aware coupling processor is developed to model the bidirectional update between the two state domains. Dynamically updated edge representations and cross-attention adapt the exchanged information to the receiving state and local node--element relation.}

\item \rev{The dual-entity formulation is extended to larger forming meshes through CAtt-BiUGNN, which processes nodal and elemental representations separately during graph downsampling and upsampling while preserving their interaction across graph levels.}

\item \rev{Controlled baseline, ablation, statistical, computational-efficiency, ensemble-disagreement, and physical-consistency analyses evaluate the proposed state representation, coupling mechanism, and hierarchical extension during autoregressive displacement and thinning prediction.}

\end{itemize}

\rev{The remainder of this paper is organised as follows. Section 2 introduces the relevant explicit FE update structure and defines the prediction problem. Sections 3--7 present the proposed framework, dataset generation, graph representation, model architecture, and training procedure. Section 8 reports the numerical comparisons and ablation studies, and Section 9 concludes the paper.}

\section{Explicit dynamic FE formulation and problem definition}
\label{Physical background of explicit FE formulation}

Before introducing the bipartite graph representation in Section~\ref{Graph representation and processing} and the model design in Section~\ref{CAtt-BiGNN model development}, this section reviews the explicit finite element (FE) update cycle that motivates the proposed graph-based design. The focus is not to reproduce the full implementation of a commercial FE solver, but to identify the mesh entities and update dependencies that are relevant to the learned simulator. In particular, nodal kinematic updates are transferred through element connectivity to update element-level deformation measures, while updated element-level quantities contribute back to subsequent nodal dynamics through FE assembly. This coupled update pattern provides the modelling motivation for representing mesh nodes and finite elements as distinct but connected graph entities.

The following notation is used throughout this section. The position of node \(v\) at time \(t^{n}\) is denoted by \(\vecv{x}_{v}^{\,n}\), and its reference position is \(\vecv{x}_{v}^{\,0}\). The accumulated nodal displacement and one-step displacement increment are defined as
\begin{equation}
\vecv{u}_{v}^{\,n}
=
\vecv{x}_{v}^{\,n}
-
\vecv{x}_{v}^{\,0},
\qquad
\Delta\vecv{u}_{v}^{\,n+1}
=
\vecv{u}_{v}^{\,n+1}
-
\vecv{u}_{v}^{\,n}
=
\vecv{x}_{v}^{\,n+1}
-
\vecv{x}_{v}^{\,n}.
\label{eq:nodal_displacement_def}
\end{equation}
Global vectors are written without the node subscript, for example \(\vecv{u}^{\,n}\) and \(\Delta\vecv{u}^{\,n+1}\). The element index is denoted by \(i\), and the integration-point index within element \(i\) is denoted by \(q\). This distinction is important because stresses, strains, and thickness variables are generally evaluated at integration points, whereas the graph model later uses an element-level representation of these quantities.

\subsection{Governing equation and explicit time integration}
\label{Governing equation and explicit time integration}

In sheet-material-forming applications, explicit dynamic integration is commonly used because it avoids the repeated global nonlinear solves required by implicit methods and can handle complex tool-blank contact efficiently~\cite{Abaqus2008,Marciniak2002,belytschko2014nonlinear}. Time is discretised as \(t^{n}=n\,\Delta t\), where \(\Delta t\) is the time increment. Quantities evaluated at \(t^{n}\) are denoted by the superscript \(n\), and half-step velocities are denoted by \(n\!\pm\!\tfrac{1}{2}\).

The semi-discrete nodal equilibrium equation at timestep \(n\) is written as
\begin{equation}
\matm{M}\,\ddot{\vecv{u}}^{\,n}
=
\vecv{P}_{\upr{ext}}^{\,n}
-
\vecv{I}_{\upr{int}}^{\,n},
\label{eq:eom_disc}
\end{equation}
where \(\matm{M}\) is the lumped mass matrix, \(\ddot{\vecv{u}}^{\,n}\) is the nodal acceleration vector, \(\vecv{P}_{\upr{ext}}^{\,n}\) is the external nodal force vector, and \(\vecv{I}_{\upr{int}}^{\,n}\) is the internal nodal force vector assembled from elemental stress contributions. In forming simulations, \(\vecv{P}_{\upr{ext}}^{\,n}\) is dominated by tool-blank contact forces, while \(\vecv{I}_{\upr{int}}^{\,n}\) depends on the current elemental stress state.

Equation~\eqref{eq:eom_disc} gives the acceleration at the current timestep after the external and internal forces have been evaluated. The acceleration is therefore not known a priori for the whole trajectory: it changes with the evolving geometry, contact state, strain, stress, and thickness-related elemental states. The displacement field must consequently be advanced step by step through the explicit time-integration scheme.

\subsection{Nodal kinematics and node-to-element strain transfer}
\label{Nodal kinematics and node-to-element strain transfer}

\subsubsection{Nodal kinematics update}
\label{Nodal kinematics update}

Given the nodal acceleration from Eq.~\eqref{eq:eom_disc}, the central-difference scheme updates the half-step velocity as
\begin{gather}
\dot{\vecv{u}}^{\,n+\tfrac{1}{2}}
=
\dot{\vecv{u}}^{\,n-\tfrac{1}{2}}
+
\ddot{\vecv{u}}^{\,n}\Delta t,
\label{eq:vel_disc}
\\
\dot{\vecv{u}}^{\,1/2}
=
\dot{\vecv{u}}^{\,0}
+
\frac{1}{2}\ddot{\vecv{u}}^{\,0}\Delta t
\qquad \text{for the first step}.
\label{eq:first_half_step_velocity}
\end{gather}
Here, \(\dot{\vecv{u}}^{\,n\pm\tfrac{1}{2}}\) denotes the nodal velocity at the half timesteps. If the blank starts from rest, \(\dot{\vecv{u}}^{\,0}=\vecv{0}\), so the first-step velocity becomes \(\dot{\vecv{u}}^{\,1/2}=\tfrac{1}{2}\ddot{\vecv{u}}^{\,0}\Delta t\).

The displacement increment over the interval \([t^{n},t^{n+1}]\) is then
\begin{equation}
\Delta\vecv{u}^{\,n+1}
=
\dot{\vecv{u}}^{\,n+\tfrac{1}{2}}\Delta t,
\label{eq:disp-increment}
\end{equation}
and the accumulated displacement and current nodal position are updated as
\begin{align}
\vecv{u}^{\,n+1}
&=
\vecv{u}^{\,n}
+
\Delta\vecv{u}^{\,n+1},
\label{eq:disp_disc}\\
\vecv{x}^{\,n+1}
&=
\vecv{x}^{\,n}
+
\Delta\vecv{u}^{\,n+1}.
\label{eq:position_update}
\end{align}
This convention distinguishes the accumulated displacement \(\vecv{u}^{\,n+1}\) from the displacement increment \(\Delta\vecv{u}^{\,n+1}\). In the surrogate formulation introduced later, if the predicted geometry is updated as \(\vecv{x}^{\,n+1}=\vecv{x}^{\,n}+\Delta\vecv{u}^{\,n+1}\) during rollout, then the network output is the displacement increment \(\Delta\vecv{u}^{\,n+1}\) rather than the accumulated displacement \(\vecv{u}^{\,n+1}\).

\subsubsection{Node-to-element strain update}
\label{Node-to-element strain update}

After the nodal kinematic update, the displacement information associated with the nodes of each element is transferred to the element integration points. Let \(\mathcal{N}(i)=\{v_{1},\ldots,v_{n_i}\}\) denote the set of nodes connected to element \(i\). If \(\vecv{u}^{\,n}\) denotes the global stacked nodal displacement vector, the element-level nodal displacement vector for element \(i\) can be extracted through the connectivity as
\begin{equation}
\vecv{u}_{\mathcal{N}(i)}^{\,n}
=
\matm{S}_{i}\vecv{u}^{\,n},
\qquad
\Delta\vecv{u}_{\mathcal{N}(i)}^{\,n+1}
=
\matm{S}_{i}\Delta\vecv{u}^{\,n+1},
\label{eq:connectivity_extraction}
\end{equation}
where \(\matm{S}_{i}\) is a Boolean extraction matrix that selects the displacement components of the nodes belonging to element \(i\). 

At integration point \(q\) of element \(i\), the strain increment is obtained from the element nodal displacement increment through the strain–displacement matrix ~\cite{belytschko2014nonlinear}:
\begin{equation}
\Delta\vecv{\varepsilon}_{iq}^{\,n+1}
\approx
\matm{B}_{iq}^{\,n}
\Delta\vecv{u}_{\mathcal{N}(i)}^{\,n+1}.
\label{eq:strain_increment}
\end{equation}

Here,
\(\Delta\vecv{\varepsilon}_{iq}^{\,n+1}\) denotes the strain-increment
vector at integration point \(q\) of element \(i\) over the interval
\([t^{n},t^{n+1}]\);
\(\Delta\vecv{u}_{\mathcal{N}(i)}^{\,n+1}\) is the stacked displacement-
increment vector of the mesh nodes connected to element \(i\); and
\(\matm{B}_{iq}^{\,n}\) is the configuration-dependent
strain--displacement operator evaluated at integration point \(q\) using
the element configuration at timestep \(n\). The approximate equality
indicates that this expression is a local linearisation of the incremental
kinematic mapping. It is introduced only to illustrate how the displacement
increments of the connected nodes determine the element-level deformation
increment. It is not intended to reproduce the complete finite-deformation
shell kinematics used by the commercial FE solver, in which the strain
measure, element configuration, local coordinate system, and finite
rotations are updated during the solution.

This transfer is one of the motivations for the bipartite graph representation introduced in Section~\ref{Graph representation and processing}. In the FE formulation, element-level quantities are not obtained from isolated element features; they are computed from the displacements of the connected nodes through element connectivity and shape-function derivatives.

\subsection{Constitutive update and thinning evaluation}
\label{Constitutive update and thinning evaluation}

With the strain increment obtained at each integration point, the stress state and material history variables are updated through the constitutive operator

\begin{equation}
\left(
\vecv{\sigma}_{iq}^{\,n+1},
\vecv{\eta}_{iq}^{\,n+1}
\right)
=
\mathcal{G}
\left(
\vecv{\sigma}_{iq}^{\,n},
\vecv{\eta}_{iq}^{\,n},
\Delta\vecv{\varepsilon}_{iq}^{\,n+1},
T_{iq}^{\,n},
\Delta t
\right).
\label{eq:constitutive_update}
\end{equation}

Here,
\(\vecv{\sigma}_{iq}^{\,n}\) and
\(\vecv{\sigma}_{iq}^{\,n+1}\) denote the stress states at integration
point \(q\) of element \(i\) before and after the constitutive update,
respectively. The vectors
\(\vecv{\eta}_{iq}^{\,n}\) and
\(\vecv{\eta}_{iq}^{\,n+1}\) collect the corresponding material history
variables, such as plastic strain and other constitutive internal variables.
The operator \(\mathcal{G}(\cdot)\) represents the finite-deformation
constitutive integration procedure adopted by the FE solver.
The input \(\Delta\vecv{\varepsilon}_{iq}^{\,n+1}\) is the strain increment
defined in Eq.~\eqref{eq:strain_increment},
\(T_{iq}^{\,n}\) is the temperature at the integration point when thermal
effects are included, and \(\Delta t=t^{n+1}-t^{n}\) is the time increment.
Depending on the adopted material model, \(\mathcal{G}(\cdot)\) may account
for material nonlinearity, strain-rate and temperature dependence, and
history-dependent constitutive behaviour\cite{shao2018strain,gao2026cohesive,li2014materials}.

In sheet forming, thickness evolution is an important elemental deformation measure. Let \(\mathrm{th}_{iq}^{\,n}\) be the current thickness associated with integration point \(q\) of element \(i\), and let \(\mathrm{th}_{iq}^{\,0}\) be its initial value. Using the through-thickness logarithmic strain convention~\cite{Marciniak2002}, the accumulated through-thickness strain is
\begin{equation}
\varepsilon_{33,iq}^{\,n}
=
\ln\left(
\frac{\mathrm{th}_{iq}^{\,n}}{\mathrm{th}_{iq}^{\,0}}
\right),
\qquad
\mathrm{th}_{iq}^{\,n+1}
=
\mathrm{th}_{iq}^{\,n}
\exp\left(\Delta\varepsilon_{33,iq}^{\,n+1}\right).
\label{eq:thickness_log_strain}
\end{equation}
Here, the subscript \(33\) denotes the through-thickness direction in the local sheet coordinate system. In a plane-stress sheet formulation, the thickness-direction strain is treated as a dependent quantity. At each explicit timestep, the thickness-direction strain increment is determined during the plane-stress constitutive update, accounting for the in-plane deformation and material response. For plastic deformation, the plastic strain increment is additionally subject to the plastic incompressibility condition ~\cite{Hosfordbook}. This connection makes thinning an elemental quantity, rather than a purely nodal quantity.

The thinning ratio is defined as
\begin{equation}
\theta_{iq}^{\,n}
=
1
-
\frac{\mathrm{th}_{iq}^{\,n}}{\mathrm{th}_{iq}^{\,0}}
=
1
-
\exp\left(\varepsilon_{33,iq}^{\,n}\right).
\label{eq:thinning_value}
\end{equation}
Here, \(\theta_{iq}^{\,n}=0\) indicates no thickness change, \(\theta_{iq}^{\,n}>0\) indicates thinning, and \(\theta_{iq}^{\,n}<0\) indicates thickening. 

\subsection{Element-to-node force assembly and contact update}
\label{Element-to-node force assembly and contact update}

After the stress field is updated at the integration points, the corresponding elemental internal force contributions are mapped back to the connected nodes. For element \(i\), the elemental internal nodal force vector is obtained from the virtual-work relation as
\begin{equation}
\vecv{f}_{i,\upr{int}}^{\,n+1}
=
\int_{\Omega_i^{\,n+1}}
\left(\matm{B}_{i}^{\,n+1}\right)^{\top}
\vecv{\sigma}_{i}^{\,n+1}
\,\mathrm{d}\Omega,
\label{eq:element-force-int}
\end{equation}
where \(\Omega_i^{\,n+1}\) is the current domain of element \(i\), \(\mathrm{d}\Omega\) is the differential measure over that domain, \(\matm{B}_{i}^{\,n+1}\) is the strain-displacement matrix over the element, and \(\vecv{\sigma}_{i}^{\,n+1}\) is the updated stress field. The transpose of \(\matm{B}_{i}^{\,n+1}\) appears because stresses are mapped to equivalent nodal forces through the virtual-work statement.

In practice, Eq.~\eqref{eq:element-force-int} is evaluated by numerical quadrature. Let \(q=1,\ldots,n_q\) index the integration points of element \(i\), where \(n_q\) is the number of integration points. The quadrature form is
\begin{equation}
\vecv{f}_{i,\upr{int}}^{\,n+1}
\approx
\sum_{q=1}^{n_q}
w_q
\left(\matm{B}_{iq}^{\,n+1}\right)^{\top}
\vecv{\sigma}_{iq}^{\,n+1}
J_{iq}^{\,n+1}.
\label{eq:element-force-quad}
\end{equation}
where \(w_q\) is the quadrature weight, \(J_{iq}^{\,n+1}\) is the corresponding Jacobian or integration-measure factor, and \(\matm{B}_{iq}^{\,n+1}\) and \(\vecv{\sigma}_{iq}^{\,n+1}\) are evaluated at integration point \(q\).

The global internal force vector is then assembled by scattering the elemental contributions into the connected nodal degrees of freedom:
\begin{equation}
\vecv{I}_{\upr{int}}^{\,n+1}
=
\sum_{i=1}^{N_e}
\matm{S}_{i}^{\top}
\vecv{f}_{i,\upr{int}}^{\,n+1}.
\label{eq:global_internal_force_assembly}
\end{equation}
where \(N_e\) is the total number of elements. Since \(\matm{S}_{i}\) extracts the degrees of freedom associated with element \(i\) from the global nodal vector, its transpose \(\matm{S}_{i}^{\top}\) scatters the element-level nodal force vector back into the corresponding entries of the global internal force vector.

In addition to the internal force vector, external nodal forces are evaluated from boundary conditions and tool-blank contact. For each slave node in contact, the contact algorithm evaluates the gap between the blank node and the tool surface, the projected contact point, and the local tool normal. These quantities determine the direction and magnitude of the contact contribution to \(\vecv{P}_{\upr{ext}}^{\,n+1}\) according to the contact formulation used by the FE solver.

The explicit update cycle is then repeated: contact and other external forces are evaluated, internal forces are assembled from the updated elemental stresses, Eq.~\eqref{eq:eom_disc} gives the next nodal acceleration, and Eqs.~\eqref{eq:vel_disc}-\eqref{eq:position_update} update the nodal kinematics. This cycle creates a bidirectional dependency between nodal and elemental quantities. Nodal displacement increments drive elemental strain, stress, and thinning updates through Eqs.~\eqref{eq:connectivity_extraction}-\eqref{eq:strain_increment}, while elemental stresses are assembled back into nodal internal forces through Eqs.~\eqref{eq:element-force-int}-\eqref{eq:global_internal_force_assembly}.

The proposed model does not explicitly solve the FE equations reviewed in this section, nor does it impose equilibrium or constitutive constraints during training. Instead, these equations are used to identify the nodal-elemental dependency pattern of the FE update. This dependency pattern is then reflected in the graph representation and message-passing architecture introduced in the following sections.

\section{Overview of the proposed CAtt-BiGNN model architecture}
\label{workflow}

\begin{figure}[htbp]
\centering
\includegraphics[width=\linewidth, trim=0mm 0mm 0mm 0mm , clip]{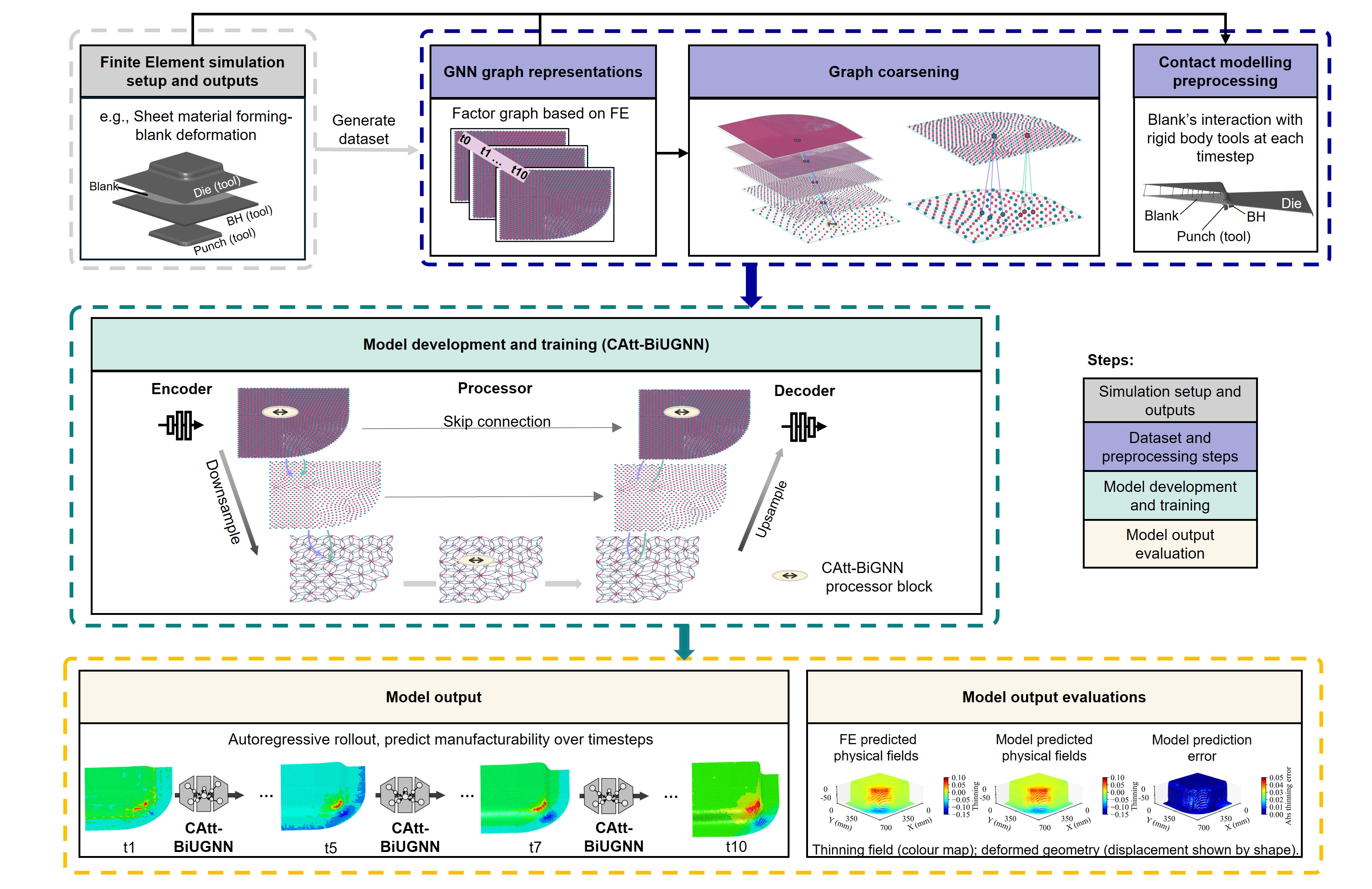}
\caption{Overview of the proposed CAtt-BiUGNN workflow for autoregressive nodal-elemental field prediction.}\label{Workflow of proposed CAtt-BiUGNN architecture}
\end{figure}

Based on the FE derivations established in Section~\ref{Physical background of explicit FE formulation}, we propose the CAtt-BiGNN model, which is inspired by explicit dynamic FE formulations. A general workflow of its hierarchical extension (CAtt-BiUGNN) is presented in Figure~\ref{Workflow of proposed CAtt-BiUGNN architecture}. CAtt-BiUGNN retains the same bipartite message-passing scheme as CAtt-BiGNN, while introducing a downsampling and upsampling mechanism to support longer-range information propagation on larger graphs. The workflow begins with the generation of simulation data (Section~\ref{Dataset generation}), from which both nodal and elemental states are extracted. These states are then preprocessed into an FE-inspired bipartite graph representation (Section~\ref{Graph representation and processing}). In this representation, nodes and elements are treated as graph entities, while graph edges encode geometric relations. The subsequent step is the model development (Section~\ref{CAtt-BiGNN model development}), which follows an encoder–processor–decoder design with skip connections. In this design, message passing is integrated with a cross-attention mechanism to adaptively couple nodal and elemental information according to the receiving state and evolving local relation. Once trained, the model performs autoregressive rollout from the undeformed configuration and dynamically recomputes contact features at each predicted timestep so that the contact descriptors remain aligned with the evolving predicted geometry. The model directly outputs nodal displacement increments and elemental thinning on their respective FE domains. This avoids projecting elemental thinning onto mesh nodes solely for prediction.  The model evaluation is presented in Section~\ref{Model performance evaluation}. It assesses field-level accuracy over time using quantitative error metrics, compares the proposed model with baseline surrogates, and examines robustness during autoregressive rollout across representative forming cases.  

\section{Dataset generation}
\label{Dataset generation}

The explicit FE solver advances through many internal integration increments. For the surrogate dataset, 11 output states were extracted over the complete forming process, following the standard output settings commonly used in commercial sheet-forming solvers. \rev{These states span the process from the undeformed configuration to the final formed state. The learned simulator therefore performs a 10-step coarse-grained autoregressive rollout, with each prediction step covering many internal FE integration increments.}

Two benchmark forming scenarios were considered in this study: a dome-shaped case representing cold forming and a corner-shaped case representing hot forming. The corresponding simulation configurations and dataset generation procedures are presented in the following subsections.

\subsection{Dome-shaped geometry}
\label{Dome shape geometry}

\begin{figure}[htbp]
\centering
\includegraphics[width=\linewidth, trim=0mm 0mm 0mm 0mm , clip]{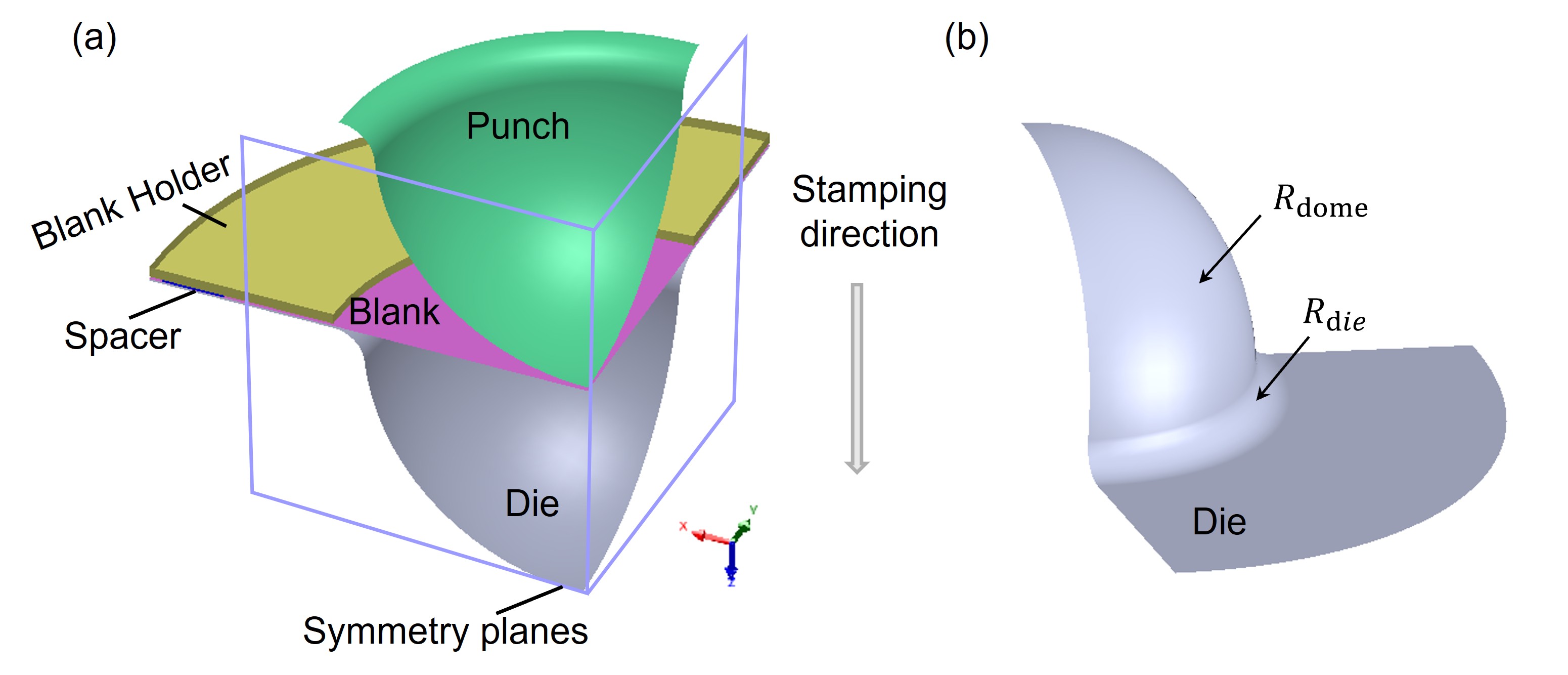}
\caption{Dome case study (a) simulation setup (b) geometry parameter variables}\label{Dome case study simulation setup}
\end{figure}

The dome-shaped geometry has a double-curvature surface, which leads to a more complex stress and deformation state than single-curvature geometries. For this reason, it is often used as a baseline to evaluate manufacturability \cite{liang2022integrated}. Figure~\ref{Dome case study simulation setup} summarises the cold-stamping configuration used in this work. The die, punch, blank holder, and spacers are modelled as rigid bodies, whereas the blank is treated as deformable. The die remains fully constrained. The punch travels vertically at a constant stamping speed of  500 mm/s to draw the blank into the die cavity. To stabilise the flange region and suppress wrinkling, a blank holding force of  200 kN is applied to the blank holder. Spacers of thickness 2.2 mm are inserted between the die and the blank holder to introduce a controlled offset, which promotes more stable material flow. The blank material follows a simplified isotropic constitutive law for cold-work Al2008, taken from the PAM-STAMP material database.  The initial blank thickness is 2 mm, and the friction coefficient is set to 0.12.

A family of dome-shaped tool geometries is generated using the geometry parameter variables in Figure~\ref{Dome case study simulation setup}(b). Two geometric parameters define the die profile:  the dome radius $R_{\mathrm{dome}}$ and the die radius $R_{\mathrm{die}}$. The DoE ranges are $60\,\text{mm} \leq R_{\mathrm{dome}} \leq 120\,\text{mm}$ and $10\,\text{mm} \leq R_{\mathrm{die}} \leq 20\,\text{mm}$. The punch geometry is constructed by applying a  2 mm offset with respect to the die.  To reduce computational cost, only one quarter of the dome is simulated. Symmetry planes are imposed as boundary conditions, and the same uniform quarter-circle blank is used for all cases. 

Dataset generation is fully automated and follows the workflow reported in \cite{ATTAR20211650}. Latin Hypercube Sampling (LHS) is used to draw tool-parameter combinations within the prescribed DoE ranges. Training, validation, and test subsets are sampled independently to improve design-space coverage and to reduce the risk of distribution leakage across splits. For each sampled design, tool geometries are created programmatically in SolidWorks, meshed using HyperMesh, and then simulated in PAM-STAMP. The resulting dataset contains 50 training cases, 25 validation cases, and 25 test cases. Each case corresponds to a unique tool configuration and provides stamping outputs over 11 \rev{saved output states}, which capture the deformation evolution under varying tool geometries. 

\rev{A dataset-size assessment using 12, 24, 36, and 50 Dome training cases showed diminishing returns from 36 to 50 cases; the complete results are provided in \ref{app:dataset_size_assessment}.}

\subsection{Corner-shaped geometry}
\label{Corner shape geometry}

\begin{figure}[htbp]
\centering
\includegraphics[width=\linewidth, trim=0mm 0mm 0mm 0mm, clip]{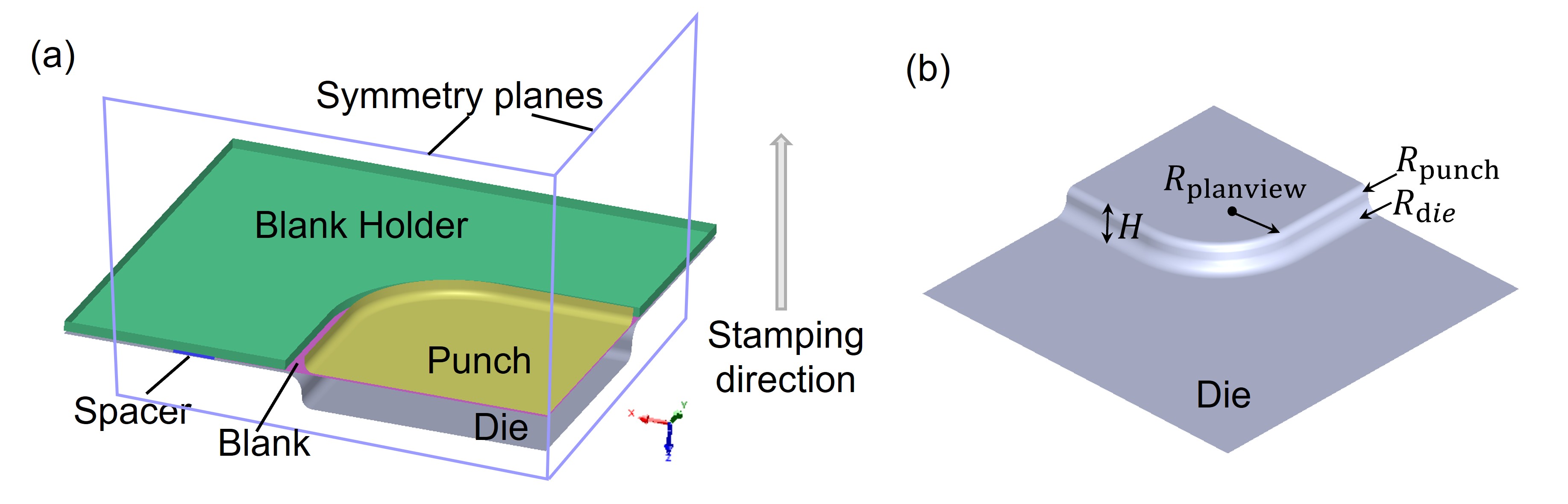}
\caption{Corner case study (a) simulation setup (b) geometry parameter variables }\label{Corner case study simulation setup}
\end{figure}

A hot-formed deep drawn corner case study was considered, as illustrated in Figure~\ref{Corner case study simulation setup}. The die geometry was defined by four design variables: the die corner radius $R_{\mathrm{die}}$, the punch radius $R_{\mathrm{punch}}$, the plan view radius $R_{\mathrm{planview}}$, and the overall height $H$. The parameter ranges were specified as $25~\text{mm} \leq R_{\mathrm{die}} \leq 40 ~\text{mm}$, $20~ \text{mm} \leq R_{\mathrm{punch}} \leq 27.3 ~\text{mm}$, $70~ \text{mm} \leq R_{\mathrm{planview}} \leq 250 ~\text{mm}$, and $60 ~\text{mm} \leq H \leq 120~ \text{mm}$. The punch geometry was defined with a constant offset of 2.3~mm with respect to the die. For all simulations, a single blank geometry was used throughout the dataset, with the blank shape kept unchanged across all cases. The sampling of tool configurations followed a Design of Experiment (DoE) strategy. Parameter combinations were generated using Latin Hypercube Sampling (LHS) within the specified bounds. For each sampled configuration, the CAD model of the tool set was generated automatically. The geometries were subsequently meshed in HyperMesh and then analysed in PAM-STAMP through an automated simulation workflow. 

An elastic–viscoplastic material model of AA6082 under hot-stamping conditions was adopted for the corner-shaped simulations \cite{ATTAR20211650, mohamed2015investigation}. The sheet was initially heated to 500 °C, whereas the forming tools were maintained at  25 °C. A stamping velocity of 500 mm/s was applied during forming. The spacer thickness was set to 2.3 mm and the blank thickness was set to 2 mm. To control material flow during forming, a blank holding force of 200 kN was applied through the blank holder in the direction opposite to the stamping motion. In the simulation setup, the punch remained fully constrained while the die translated in the negative $z$-direction, with the stroke determined by the geometric configuration (total height) of the die. 

Based on the defined DoE parameter space and process configuration, batch finite element simulations were performed in PAM-STAMP to construct the dataset. Three independent simulation batches were executed, producing 100 training samples, 25 validation samples, and 25 test samples. Each simulation was recorded at 11 discrete timesteps, yielding a temporal sequence that captures the evolution of sheet deformation under different tool geometries. 

\section{Bipartite graph representation}
\label{Graph representation and processing}

\begin{figure}[htbp]
\centering
\includegraphics[width=\linewidth, trim=0mm 0mm 0mm 0mm, clip]{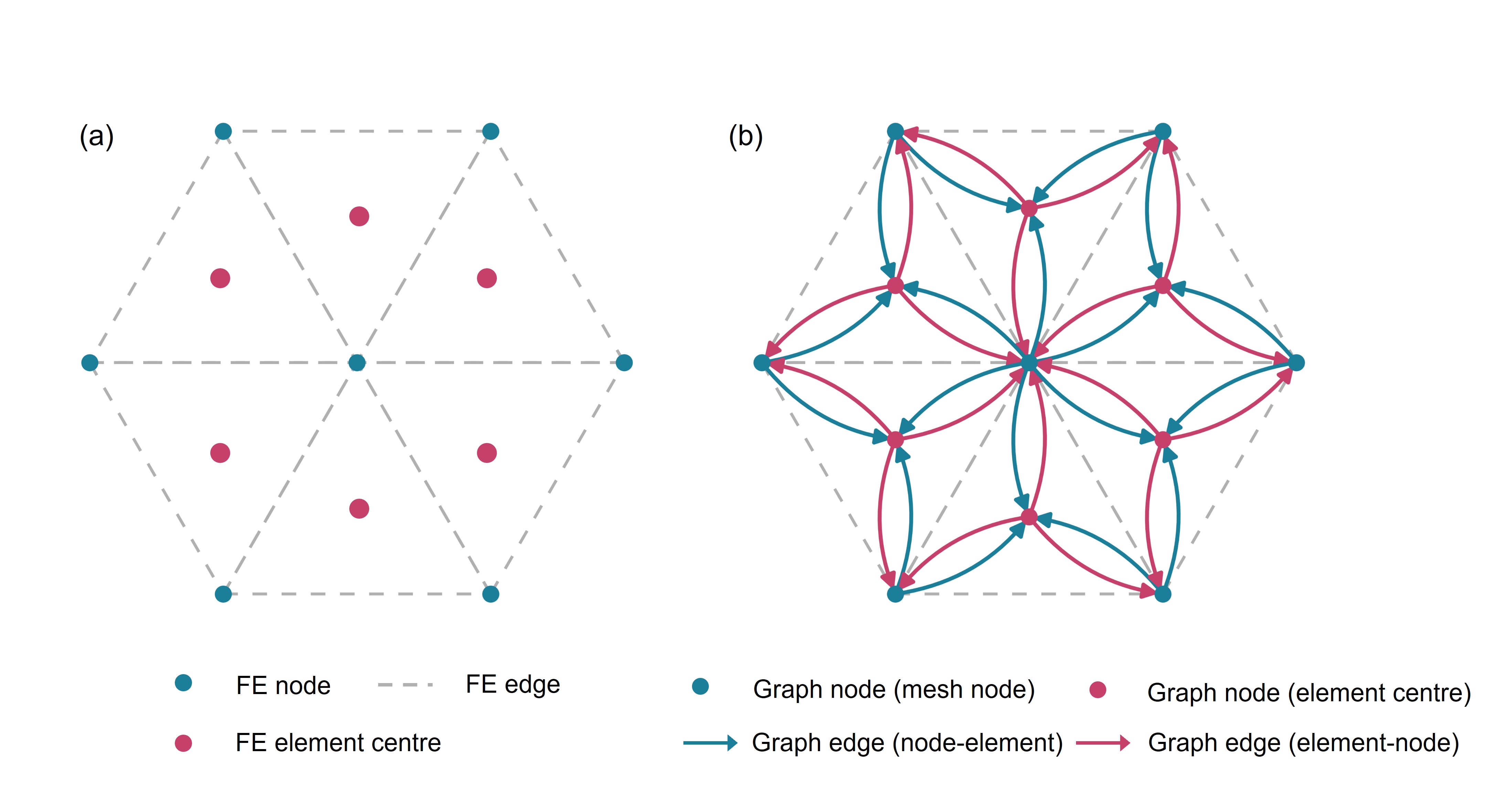}
\caption{Bipartite graph construction. (a) FE mesh with mesh nodes, FE mesh edges, and elemental quantities located at element centres or representative integration points. (b) Bipartite graph in which mesh-nodes and element-centre nodes are connected by directed node-to-element and element-to-node edges. Dashed FE mesh edges are shown only to indicate the original mesh connectivity and are not used as graph message-passing edges }\label{Graph construction method}
\end{figure}

Based on the explicit FE formulation introduced in Section~\ref{Physical background of explicit FE formulation}, we construct a bipartite graph representation that reflects the bidirectional information exchange between nodal and elemental variables. This bipartite graph representation is shown in Figure~\ref{Graph construction method}. Starting from the original FE mesh, two types of graph nodes are defined: mesh nodes, which represent nodal degrees of freedom, and element-centre nodes, which represent element-level states obtained from the corresponding elemental or integration-point quantities. Rather than adopting the original mesh edges as graph connections, the graph is built with bidirectional links between each element and its associated mesh nodes. This construction is motivated by the FE coupling described in Section~\ref{Physical background of explicit FE formulation}: nodal displacement increments are transferred to elements for strain and thickness-related updates, while elemental stress contributions are assembled back to nodes through the internal force vector. \rev{In both datasets, the deformable blank was discretised using three-node linear triangular shell elements. Consequently, each element centre-node is connected to the three mesh nodes of its parent FE element, producing three directed node-to-element edges and three directed element-to-node edges per element}.

Following the bipartite graph representation, graph input and output features are defined in Table~\ref{tab:graph-notation}. In the following, mesh node refers to a mesh node \(v\), and an element centre-node refers to a finite element \(i\). The notation is chosen to be consistent with Section~\ref{Physical background of explicit FE formulation}. The nodal position at timestep \(n\) is denoted by \(\vecv{x}_{v}^{\,n}\), the accumulated nodal displacement by \(\vecv{u}_{v}^{\,n}\), and the one-step displacement increment by \(\Delta\vecv{u}_{v}^{\,n+1}\). The element-level thinning value \(\theta_i^{\,n}\) denotes the thinning value stored for element \(i\) in the surrogate dataset. It represents the element-level FE output associated with the integration-point thinning response \(\theta_{iq}^{\,n}\) defined in Eq.~\eqref{eq:thinning_value}.

The mesh node input feature vector is constructed in a form motivated by the explicit FE formulation. The nodal displacement-increment input \(\Delta\vecv{u}_{v}^{\,n}\) represents the most recent one-step nodal motion available at timestep \(n\), consistent with the explicit kinematic update in Eq.~\eqref{eq:disp-increment}. The boundary indicator \(\vecv{b}_{v}\) represents fixed degrees of freedom in the three coordinate directions. To incorporate tool-blank contact, the contact-related feature \(\vecv{c}_{v}^{\,n}\) includes the tool-blank gap distance \(d_{v}^{\,n}\) from node \(v\) to the rigid tool surface and the local tool surface normal \(\vecv{n}_{v}^{\,n}\). These quantities correspond to the contact update discussed in Section~\ref{Element-to-node force assembly and contact update}. The mesh node output is defined as the predicted displacement increment \(\Delta\vecv{u}_{v}^{\,n+1}\), which advances the nodal position through Eq.~\eqref{eq:position_update}. During autoregressive rollout, the predicted displacement increment is used as the nodal displacement-increment input for the next step, while the updated nodal position \(\vecv{x}_{v}^{\,n+1}\) is used to recompute geometry-dependent edge features and contact-related node features. The global input \(l\) represents the stroke value that determines the travel distance of the rigid body tools.

\begin{table}[htbp]
\centering
\caption{Input and output features in the bipartite graph representation.}
\label{tab:graph-notation}
\small
\setlength{\tabcolsep}{4pt}
\renewcommand{\arraystretch}{1.12}
\begin{tabularx}{\textwidth}{l p{2.8cm} >{\RaggedRight\arraybackslash}X}
\toprule
\textbf{Symbol} & \textbf{Type} & \textbf{Features} \\
\midrule
\( \vecv{v}_{v}^{\,n},\ \vecv{c}_{v}^{\,n} \) & Mesh node input &
\( \vecv{v}_{v}^{\,n} = \big[\,\Delta\vecv{u}_{v}^{\,n},\ \vecv{b}_{v}\,\big] \) \\
& & \( \vecv{b}_{v} = (b_{v,x}, b_{v,y}, b_{v,z}) \in \{0,1\}^{3} \) \\
& & \( \vecv{c}_{v}^{\,n} = \big[\, d_{v}^{\,n},\ \vecv{n}_{v}^{\,n} \,\big] \) \\
\midrule
\( \vecv{s}_{i}^{\,n} \) & Element centre-node input &
\( \vecv{s}_{i}^{\,n} = \theta_{i}^{\,n} \) \\
\midrule
\( l \) & Global input &
\( l = \) stroke \\
\midrule
\( \vecv{e}_{v\to i}^{\,n} \) & Node-to-element edge input &
\( \vecv{e}_{v\to i}^{\,n}
=
\big[\, \|\Delta\vecv{x}_{v,i}^{\,0}\|_2,\ \Delta\vecv{x}_{v,i}^{\,0},
\|\Delta\vecv{x}_{v,i}^{\,n}\|_2,\ \Delta\vecv{x}_{v,i}^{\,n} \,\big] \) \\
& &
\( \|\Delta\vecv{x}_{v,i}^{\,n}\|_2
=
\sqrt{(\Delta x_{v,i}^{\,n})^{2}
+
(\Delta y_{v,i}^{\,n})^{2}
+
(\Delta z_{v,i}^{\,n})^{2}} \)
 \\
\midrule
\( \vecv{e}_{i\to v}^{\,n} \) & Element-to-node edge input &
\( \vecv{e}_{i\to v}^{\,n}
=
\big[\, \|\Delta\vecv{x}_{v,i}^{\,0}\|_2,\ \Delta\vecv{x}_{v,i}^{\,0},
\|\Delta\vecv{x}_{v,i}^{\,n}\|_2,\ \Delta\vecv{x}_{v,i}^{\,n} \,\big] \) \\
\midrule
\( \vecv{v}_{v}^{\,n,\mathrm{out}} \) & Mesh node output &
\( \vecv{v}_{v}^{\,n,\mathrm{out}} = \Delta\vecv{u}_{v}^{\,n+1} \) \\
\midrule
\( \vecv{s}_{i}^{\,n,\mathrm{out}} \) & Element centre-node output &
\( \vecv{s}_{i}^{\,n,\mathrm{out}} = \theta_{i}^{\,n+1} \) \\
\bottomrule
\end{tabularx}
\end{table}

Graph edge input features represent the geometric relation between mesh nodes and their associated element-centre nodes. For element centre-node \(i\), the set of connected nodes is denoted by \(\mathcal{N}(i)\), consistent with Section~\ref{Node-to-element strain update}. The element centre-node position \(\vecv{x}_{i}^{\,n}\) is taken as the element-centre position at timestep \(n\), and \(\Delta\vecv{x}_{v,i}^{\,n}=\vecv{x}_{v}^{\,n}-\vecv{x}_{i}^{\,n}\) denotes the mesh node position relative to the element centre. The Euclidean distance \(\|\Delta\vecv{x}_{v,i}^{\,n}\|_2\) is rotation-invariant, while the coordinate-difference vector \(\Delta\vecv{x}_{v,i}^{\,n}\) retains directional information. The same geometric descriptor is used for both \(\vecv{e}_{v\to i}^{\,n}\) and \(\vecv{e}_{i\to v}^{\,n}\). This is because the descriptor characterises the shared node-element pair geometry rather than a source-to-target direction vector; the two directions of information transfer are distinguished by the corresponding message-passing update. This shared descriptor also allows the two directed edge types to use the same edge-feature initialisation, reducing parameter count and computational cost. The reference-configuration edge features at \(t^{0}\) are included together with the current-configuration features at \(t^{n}\), so the model receives both a fixed geometric baseline and an updated description of the evolving geometry.

This edge design follows the FE-inspired node-element information flow described in Section~\ref{Physical background of explicit FE formulation}. In the FE formulation, nodal displacement increments are extracted through the element connectivity and transferred to element integration points, as described by Eqs.~\eqref{eq:connectivity_extraction}-\eqref{eq:strain_increment}. Conversely, elemental stress contributions are assembled back into the global internal force vector through Eq.~\eqref{eq:global_internal_force_assembly}. The bipartite graph does not explicitly evaluate these FE operators, but its two directed edge sets provide a learning structure that reflects the same node-to-element and element-to-node exchange. The use of relative geometric features is also consistent with mesh-based GNN practice~\cite{pfaff2021learning}.

In the present sheet-forming application,  the element centre-node input feature is defined as the element-level thinning indicator \(\theta_{i}^{\,n}\), corresponding to the thinning definition in Eq.~\eqref{eq:thinning_value}. The element centre-node output is the predicted thinning value \(\theta_{i}^{\,n+1}\) at the subsequent timestep. In this way, the model learns to predict nodal displacement increments and elemental thinning evolution on their native FE domains. \rev{This allows nodal displacement increments and elemental thinning to be supervised and updated on their respective FE domains, while both predicted fields are retained as inputs to subsequent autoregressive updates.}

\section{CAtt-BiGNN model development}
\label{CAtt-BiGNN model development}

This section introduces the proposed cross-attention-based bipartite graph neural network, denoted as CAtt-BiGNN, and its hierarchical extension, CAtt-BiUGNN. The graph representation introduced in Section~\ref{Graph representation and processing} contains two types of graph nodes: mesh-nodes, which carry nodal kinematic and contact information, and element-centre nodes, which carry element-level deformation information. The model exchanges information between these two domains through directed node-to-element and element-to-node edges. This design follows the node-to-element and element-to-node transfer pattern discussed in Section~\ref{Physical background of explicit FE formulation}, \rev{while remaining a data-driven surrogate that does not explicitly enforce FE equilibrium equations or constitutive relations. }

To keep the notation consistent with Sections~\ref{Physical background of explicit FE formulation} and~\ref{Graph representation and processing}, the nodal position at timestep \(n\) is denoted by \(\vecv{x}_{v}^{\,n}\), the accumulated nodal displacement by \(\vecv{u}_{v}^{\,n}\), and the one-step displacement increment by \(\Delta\vecv{u}_{v}^{\,n+1}\). Following Table~\ref{tab:graph-notation}, the model mesh node output is denoted by \(\vecv{v}_{v}^{\,n,\mathrm{out}}\) and corresponds to \(\Delta\vecv{u}_{v}^{\,n+1}\), while the model element centre-node output is denoted by \(\vecv{s}_{i}^{\,n,\mathrm{out}}\) and corresponds to \(\theta_i^{\,n+1}\). A hat symbol is reserved for latent neural-network representations, such as \(\hat{\vecv{v}}_{v,L}^{\,n}\), rather than for the physical graph input or output variables.

\subsection{General description of the CAtt-BiGNN architecture}
\label{General description to CAtt-BiGNN processor}

\begin{figure}[htbp]
\centering
\includegraphics[width=\linewidth, trim=0 0 0 0, clip]{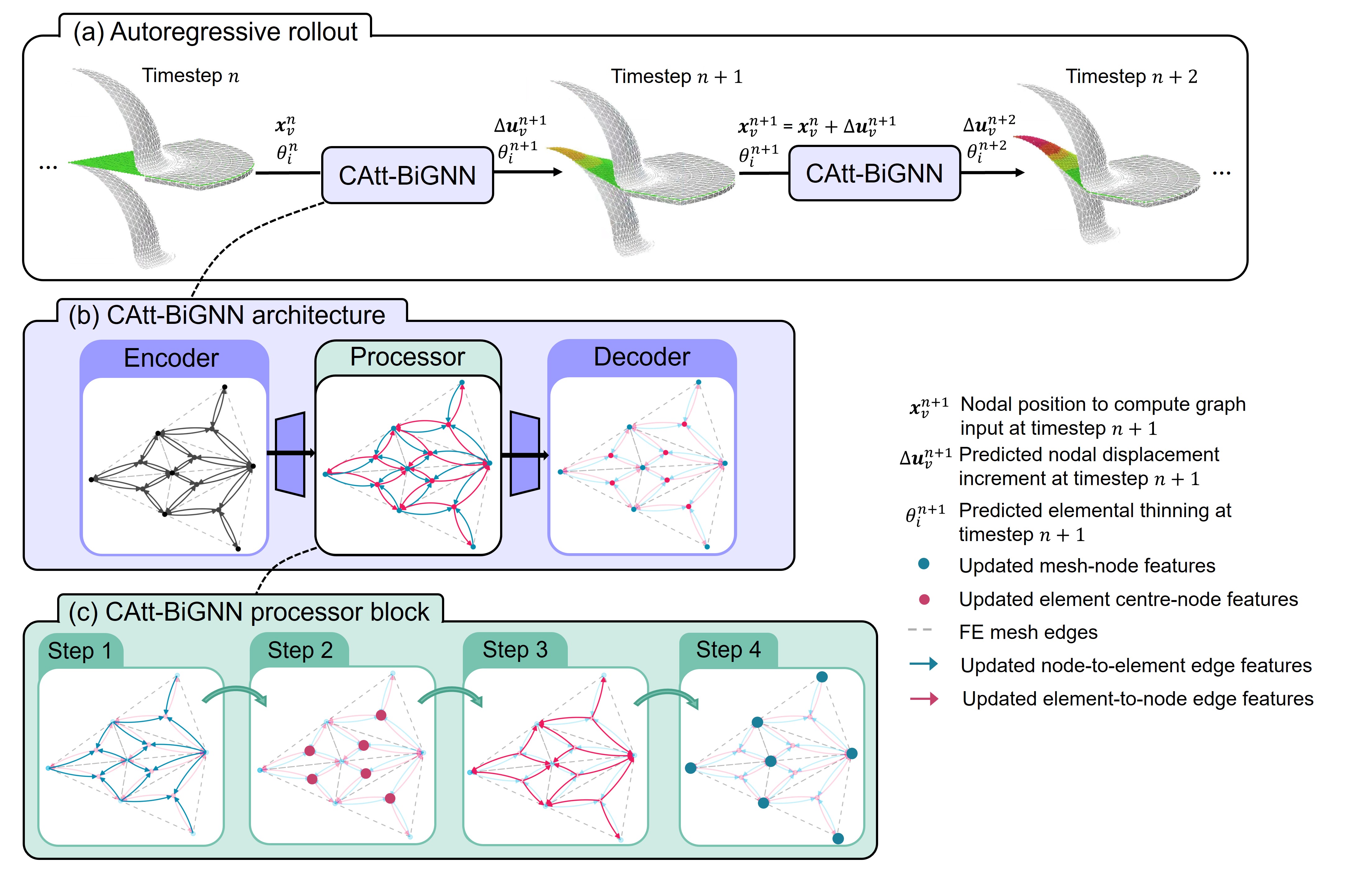}
\caption{CAtt-BiGNN architecture and autoregressive rollout. At timestep n, the encoded nodal, contact, elemental, and directed edge features are processed by node-element message-passing layers and decoded into the nodal displacement increment and elemental thinning. The nodal position is updated, and geometry-dependent features are recomputed for the next rollout step. \rev{Steps} (1)-(4) show the four processor substeps: node-to-element edge update, element centre-node update, element-to-node edge update, and mesh-node update. }\label{CAtt-BiGNN model architecture}
\end{figure}

The CAtt-BiGNN follows an encoder-processor-decoder structure, as illustrated in Figure~\ref{CAtt-BiGNN model architecture}. At timestep \(n\), the graph input contains mesh node features \(\vecv{v}_{v}^{\,n}\), contact features \(\vecv{c}_{v}^{\,n}\), element centre-node features \(\vecv{s}_{i}^{\,n}\), directed node-to-element edge features \(\vecv{e}_{v\to i}^{\,n}\), directed element-to-node edge features \(\vecv{e}_{i\to v}^{\,n}\), and the global stroke input \(l\), as defined in Table~\ref{tab:graph-notation}.

The implementation uses one encoder for mesh node features, three tool-specific encoders for contact features, one encoder for element features, one shared encoder for both directed edge types, and one encoder for the global stroke-related feature. The contact feature \(\vecv{c}_{v}^{\,n}\) in Table~\ref{tab:graph-notation} is implemented as separate die, punch, and blank-holder contact descriptors. Their encoded latent features are denoted by \(\hat{\vecv{c}}_{v,D}^{\,n}\), \(\hat{\vecv{c}}_{v,P}^{\,n}\), and \(\hat{\vecv{c}}_{v,BH}^{\,n}\), respectively. The encoded latent features before the first processor layer are

\begin{equation}
\begin{aligned}
\hat{\vecv{v}}_{v,0}^{\,n}
&=
f_{\mathrm{en}}^{v}\!\left(\vecv{v}_{v}^{\,n}\right),
&
\hat{\vecv{s}}_{i,0}^{\,n}
&=
f_{\mathrm{en}}^{s}\!\left(\vecv{s}_{i}^{\,n}\right),
\\
\hat{\vecv{c}}_{v,D}^{\,n}
&=
f_{\mathrm{en}}^{c,D}\!\left(\vecv{c}_{v,D}^{\,n}\right),
&
\hat{\vecv{c}}_{v,P}^{\,n}
&=
f_{\mathrm{en}}^{c,P}\!\left(\vecv{c}_{v,P}^{\,n}\right),
\\
\hat{\vecv{c}}_{v,BH}^{\,n}
&=
f_{\mathrm{en}}^{c,BH}\!\left(\vecv{c}_{v,BH}^{\,n}\right),
&
\hat{l}
&=
f_{\mathrm{en}}^{g}\!\left(l\right),
\\
\hat{\vecv{e}}_{v\to i,0}^{\,n}
&=
f_{\mathrm{en}}^{e}\!\left(\vecv{e}_{v\to i}^{\,n}\right),
&
\hat{\vecv{e}}_{i\to v,0}^{\,n}
&=
f_{\mathrm{en}}^{e}\!\left(\vecv{e}_{i\to v}^{\,n}\right).
\end{aligned}
\label{eq:encoder_mappings}
\end{equation}
Here, \(f_{\mathrm{en}}^{v}\), \(f_{\mathrm{en}}^{s}\), \(f_{\mathrm{en}}^{c,D}\), \(f_{\mathrm{en}}^{c,P}\), \(f_{\mathrm{en}}^{c,BH}\), \(f_{\mathrm{en}}^{g}\), and \(f_{\mathrm{en}}^{e}\) denote the encoder networks for the mesh-node feature vector, element centre-node feature vector, die contact feature vector, punch contact feature vector, blank-holder contact feature vector, global stroke input, and directed edge feature vector, respectively. The notation \(\hat{(\cdot)}\) denotes an encoded latent representation. The subscript \(0\) denotes the latent feature before the first processor layer. The same edge encoder \(f_{\mathrm{en}}^{e}\) is applied to both \(\vecv{e}_{v\to i}^{\,n}\) and \(\vecv{e}_{i\to v}^{\,n}\), because the two directed edge inputs use the same node-element pair descriptor defined in Table~\ref{tab:graph-notation}. The direction of information transfer is represented by the corresponding processor update.

After the processor and hierarchical update stages, the final latent mesh node and element centre-node states are decoded to the graph outputs defined in Table~\ref{tab:graph-notation}. In the implementation, the mesh node decoder receives the final mesh node latent feature together with the initial encoded mesh node skip feature and the three encoded contact features. The element centre-node decoder receives the final element centre-node latent feature together with the initial encoded centre-node skip feature. The decoded outputs are
\begin{equation}
\begin{aligned}
\vecv{v}_{v}^{\,n,\mathrm{out}}
&=
\Delta\vecv{u}_{v}^{\,n+1}
=
f_{\mathrm{de}}^{\mathrm{node}}
\!\left(
\left[
\hat{\vecv{v}}_{v}^{\,n,\mathrm{final}};
\hat{\vecv{v}}_{v,0}^{\,n};
\hat{\vecv{c}}_{v,D}^{\,n};
\hat{\vecv{c}}_{v,P}^{\,n};
\hat{\vecv{c}}_{v,BH}^{\,n}
\right]
\right),
\\[0.8em]
\vecv{s}_{i}^{\,n,\mathrm{out}}
&=
\theta_i^{\,n+1}
=
f_{\mathrm{de}}^{\mathrm{elem}}
\!\left(
\left[
\hat{\vecv{s}}_{i}^{\,n,\mathrm{final}};
\hat{\vecv{s}}_{i,0}^{\,n}
\right]
\right).
\end{aligned}
\label{eq:decoder_outputs}
\end{equation}
During autoregressive rollout, the decoded node output \(\vecv{v}_{v}^{\,n,\mathrm{out}}\) is accumulated to update the nodal displacement and nodal position:
\begin{equation}
\vecv{u}_{v}^{\,n+1}
=
\vecv{u}_{v}^{\,n}
+
\vecv{v}_{v}^{\,n,\mathrm{out}},
\qquad
\vecv{x}_{v}^{\,n+1}
=
\vecv{x}_{v}^{\,n}
+
\vecv{v}_{v}^{\,n,\mathrm{out}}.
\label{eq:rollout_position_update}
\end{equation}
The updated nodal positions and element states are then used to construct the graph input for the next rollout timestep. In particular, geometry-dependent edge features and contact descriptors are recomputed from the updated geometry whenever they depend on the current nodal positions.

The processor is the core of the architecture. Each processor layer performs a closed message-passing cycle between nodes and elements. First, node-to-element edge features are updated and aggregated to update the element states. Second, element-to-node edge features are updated and aggregated to update the nodal states. This sequence mirrors the FE-inspired transfer pattern in which nodal kinematic information is used to update element-level quantities, while element-level information is subsequently propagated back to nodal states. The processor is therefore FE-inspired and domain-aligned, but it does not explicitly enforce stress equilibrium or constitutive consistency. A number of processor layers are applied to extend the message-passing range. A tailored cross-attention mechanism is included to enable weighted node-element coupling, as detailed in Section~\ref{Cross-attention based edge processor block}. The model can operate on a single high-resolution graph and can also be combined with the downsampling and upsampling mechanism described in Section~\ref{Downsample/upsample mechanism}. When integrated with this multiscale structure, the model forms the hierarchical variant CAtt-BiUGNN.

\subsection{Cross-attention-based processor block}
\label{Cross-attention based edge processor block}

The cross-attention-based processor assigns adaptive weights to directed node-element messages. The purpose of the attention mechanism is to let the receiver state decide which incoming geometric relations are most relevant at the current deformation state. \rev{The edge-based key allows the attention weight to depend on the geometric relation between the node and the element. This follows the FE update, where the contribution of a node to an element depends on the position of the node within that element (Eq.~\ref{eq:strain_increment}).} In the node-to-element direction, the receiver is an element centre-node and the incoming messages originate from connected mesh nodes. In the element-to-node direction, the receiver is a mesh-node and the incoming messages originate from connected element-centre nodes.

\rev{In the controlled attention variants considered in this study, GAT-BiGNN computes attention from the states of the connected graph entities, whereas CA-BiGNN uses the receiving state as the query and the source-entity state as the key. CAtt-BiGNN instead constructs the key from the current directed-edge state, which is updated across processor layers. The attention weights therefore depend jointly on the receiving state and the evolving local relation through which information is transferred. This allows the same node–element connection to contribute differently as the coupled nodal and elemental states evolve. }
\rev{Unlike attention schemes that compute weights from the states of the connected entities, or that use edge attributes to modulate messages or attention scores, CAtt-BiGNN forms the key from the directed edge state alone, uses the updated edge state directly as the message without a separate value projection, and keeps separate edge states for the two transfer directions. Combined with the receiver-state query, this makes the normalised relevance of each interaction depend explicitly on the evolving local node--element relation.}
Consistent with Section~\ref{Graph representation and processing}, let \(\mathcal{N}(i)\) denote the set of mesh nodes connected to element centre-node \(i\). The reverse adjacency set for a mesh node \(v\) is denoted by
\(\mathcal{M}(v)=\{\,i\mid v\in\mathcal{N}(i)\,\}\). The corresponding directed edge sets can be written as
\begin{equation}
\mathcal{E}_{\mathrm{in}}^{v\to i}(i)
=
\{\,(v,i) \mid v\in\mathcal{N}(i)\,\},
\qquad
\mathcal{E}_{\mathrm{in}}^{i\to v}(v)
=
\{\,(i,v) \mid i\in\mathcal{M}(v)\,\}.
\label{eq:processor_incoming_edges}
\end{equation}

\begin{figure}[htbp]
\centering
\includegraphics[width=\linewidth, trim=0mm 0mm 0mm 0mm, clip]{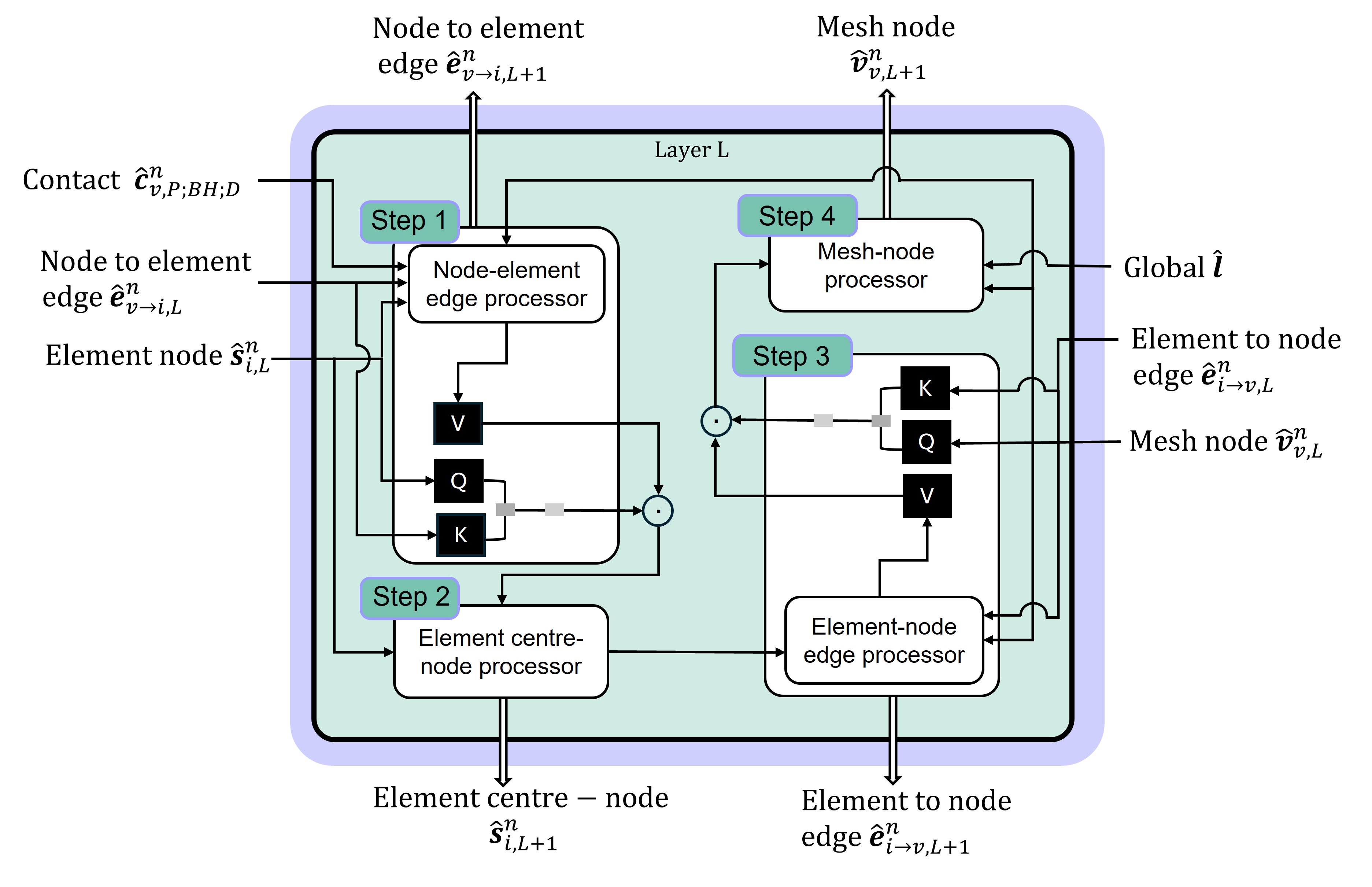}
\caption{Cross-attention-based CAtt-BiGNN processor block. Each processor layer updates node-to-element edges, element centre-node features, element-to-node edges, and mesh-node features in four residual substeps. }\label{processor}
\end{figure}

Figure~\ref{processor} can be read as four steps within one processor layer: (1) update node-to-element edge states, (2) aggregate node-to-element messages and update element centre-node states, (3) update element-to-node edge states, and (4) aggregate element-to-node messages and update mesh node states. In the attention operation, the receiver state forms the query \(\vecv{Q}\), the incoming directed edge state forms the key \(\vecv{K}\), and the updated directed edge state forms the value/message vector \(\vecv{V}\). This makes the attention weights explicitly edge-aware.

\subsubsection{Message passing with attention-modulated aggregation}
\label{Message passing with attention-modulated aggregation}

This subsection describes the data flow within one CAtt-BiGNN processor layer, following the four numbered operations in Figure~\ref{processor}. The message passing layer index is denoted by \(L\), and the timestep index is denoted by \(n\). At the beginning of layer \(L\), \(\hat{\vecv{v}}_{v,L}^{\,n}\) denotes the latent feature of mesh node \(v\), \(\hat{\vecv{s}}_{i,L}^{\,n}\) denotes the latent feature of element \(i\), \(\hat{\vecv{e}}_{v\to i,L}^{\,n}\) denotes the latent feature of the directed node-to-element edge \(v\to i\), and \(\hat{\vecv{e}}_{i\to v,L}^{\,n}\) denotes the latent feature of the directed element-to-node edge \(i\to v\). The contact latent features of node \(v\) are denoted by \(\hat{\vecv{c}}_{v,D}^{\,n}\), \(\hat{\vecv{c}}_{v,P}^{\,n}\), and \(\hat{\vecv{c}}_{v,BH}^{\,n}\), corresponding to the die, punch, and blank-holder contact descriptors.

\paragraph{Step 1: node-to-element edge update}
For each directed edge \(v\to i\), the node-to-element edge processor updates the edge latent feature:
\begin{equation}
\hat{\vecv{e}}_{v\to i,L+1}^{\,n}
=
\hat{\vecv{e}}_{v\to i,L}^{\,n}
+
f_{v\to i}^{e}
\!\left(
\hat{\vecv{v}}_{v,L}^{\,n},
\hat{\vecv{c}}_{v,D}^{\,n},
\hat{\vecv{c}}_{v,P}^{\,n},
\hat{\vecv{c}}_{v,BH}^{\,n},
\hat{\vecv{e}}_{v\to i,L}^{\,n}
\right),
\label{eq:node_to_elem_edge_update}
\end{equation}
where \(f_{v\to i}^{e}\) is a learnable node-to-element edge-update function. Step 1 produces the updated edge feature \(\hat{\vecv{e}}_{v\to i,L+1}^{\,n}\) and the attention weight \(\alpha_{v\to i,L}^{\,n}\). The updated edge feature is used as the node-to-element message,
\begin{equation}
\vecv{V}_{v\to i,L}^{\,n}
=
\hat{\vecv{e}}_{v\to i,L+1}^{\,n},
\label{eq:node_to_elem_value}
\end{equation}
The corresponding attention weight \(\alpha^{n}_{v\to i,L}\) is computed from the receiver query and directed-edge key in Section~\ref{Attention weight computation in CAtt-BiGNN}.

\paragraph{Step 2: element centre-node update}
Let \(\mathcal{N}(i)\) denote the set of mesh nodes connected to element centre-node \(i\). The messages generated in Step 1 are aggregated at element \(i\):
\begin{equation}
\vecv{m}_{i,L}^{v\to i,n}
=
\sum_{v\in\mathcal{N}(i)}
\alpha_{v\to i,L}^{\,n}
\vecv{V}_{v\to i,L}^{\,n}.
\label{eq:elem_agg}
\end{equation}
Here, \(\vecv{m}_{i,L}^{v\to i,n}\) is the aggregated node-to-element message. This message is used to update the element centre-node latent feature:
\begin{equation}
\hat{\vecv{s}}_{i,L+1}^{\,n}
=
\hat{\vecv{s}}_{i,L}^{\,n}
+
f_{v\to i}^{s}
\!\left(
\hat{\vecv{s}}_{i,L}^{\,n},
\vecv{m}_{i,L}^{v\to i,n}
\right),
\label{eq:elem_update}
\end{equation}
where \(f_{v\to i}^{s}\) is a learnable element-update function. Step 2 produces the updated element centre-node feature \(\hat{\vecv{s}}_{i,L+1}^{\,n}\), which is then used in the element-to-node edge update.

\paragraph{Step 3: element-to-node edge update}
For each directed edge \(i\to v\), the element-to-node edge processor updates the edge latent feature using the updated element centre-node feature from Step 2:
\begin{equation}
\hat{\vecv{e}}_{i\to v,L+1}^{\,n}
=
\hat{\vecv{e}}_{i\to v,L}^{\,n}
+
f_{i\to v}^{e}
\!\left(
\hat{\vecv{s}}_{i,L+1}^{\,n},
\hat{\vecv{v}}_{v,L}^{\,n},
\hat{\vecv{e}}_{i\to v,L}^{\,n}
\right),
\label{eq:elem_to_node_edge_update}
\end{equation}
where \(f_{i\to v}^{e}\) is a learnable element-to-node edge-update function. Step 3 produces the updated edge feature \(\hat{\vecv{e}}_{i\to v,L+1}^{\,n}\) and the attention weight \(\alpha_{i\to v,L}^{\,n}\). The updated edge feature is used as the element-to-node message,
\begin{equation}
\vecv{V}_{i\to v,L}^{\,n}
=
\hat{\vecv{e}}_{i\to v,L+1}^{\,n},
\label{eq:elem_to_node_value}
\end{equation}
while \(\alpha_{i\to v,L}^{\,n}\) controls its contribution to the receiver mesh node. The computation of \(\alpha_{i\to v,L}^{\,n}\) is given in Section~\ref{Attention weight computation in CAtt-BiGNN}.

\paragraph{Step 4: mesh-node update}
Let \(\mathcal{M}(v)=\{i\mid v\in\mathcal{N}(i)\}\) denote the set of elements connected to mesh node \(v\). The messages generated in Step 3 are aggregated at mesh node \(v\):
\begin{equation}
\vecv{m}_{v,L}^{i\to v,n}
=
\sum_{i\in\mathcal{M}(v)}
\alpha_{i\to v,L}^{\,n}
\vecv{V}_{i\to v,L}^{\,n}.
\label{eq:node_agg}
\end{equation}
Here, \(\vecv{m}_{v,L}^{i\to v,n}\) is the aggregated element-to-node message. This message, together with the encoded global stroke input \(\hat{l}\), is used to update the mesh-node latent feature:
\begin{equation}
\hat{\vecv{v}}_{v,L+1}^{\,n}
=
\hat{\vecv{v}}_{v,L}^{\,n}
+
f_{i\to v}^{v}
\!\left(
\hat{\vecv{v}}_{v,L}^{\,n},
\vecv{m}_{v,L}^{i\to v,n},
\hat{l}
\right),
\label{eq:node_update}
\end{equation}
where \(f_{i\to v}^{v}\) is a learnable mesh-node update function. Step 4 produces the updated mesh-node feature \(\hat{\vecv{v}}_{v,L+1}^{\,n}\), completing one bidirectional node-element message-passing layer.

\subsubsection{Attention weight computation in CAtt-BiGNN}
\label{Attention weight computation in CAtt-BiGNN}

Section~\ref{Message passing with attention-modulated aggregation} introduced the attention weights \(\alpha_{v\to i,L}^{\,n}\) and \(\alpha_{i\to v,L}^{\,n}\), which are generated by the directed edge processors and then used in the element and mesh-node updates. This subsection describes how these weights are computed. The main design choice in CAtt-BiGNN is that the query is constructed from the receiver state, while the key is constructed from the incoming directed-edge state before the edge update. This is because the receiver node or element determines what information is required, whereas the edge feature carries the local node-element geometric relation. The updated directed edge feature is then used as the value/message vector in the aggregation step.

For clarity, the node-to-element direction is described first. The element-to-node direction follows the same principle with the receiver changed from an element to a mesh node.

\paragraph{Query and key}
For a directed edge from mesh node \(v\) to element \(i\), the receiver is element \(i\). The query vector is therefore computed from the element latent feature \(\hat{\vecv{s}}_{i,L}^{\,n}\):
\begin{equation}
\vecv{Q}_{i,L}^{\,n}
=
f_q^{s}
\!\left(
\hat{\vecv{s}}_{i,L}^{\,n}
\right),
\qquad
\vecv{K}_{v\to i,L}^{\,n}
=
f_k^{s}
\!\left(
\hat{\vecv{e}}_{v\to i,L}^{\,n}
\right).
\label{eq:QK_node_to_elem}
\end{equation}
Here, \(\vecv{Q}_{i,L}^{\,n}\) is the query vector of element \(i\), and \(\vecv{K}_{v\to i,L}^{\,n}\) is the key vector of the incoming directed edge \(v\to i\). The functions \(f_q^{s}\) and \(f_k^{s}\) are learnable projection functions that map the element latent feature and the directed edge latent feature into the query and key spaces, respectively.

The value/message vector is the updated node-to-element edge feature produced by Step 1 in Section~\ref{Message passing with attention-modulated aggregation}:
\begin{equation}
\vecv{V}_{v\to i,L}^{\,n}
=
\hat{\vecv{e}}_{v\to i,L+1}^{\,n}.
\label{eq:V_node_to_elem_attention}
\end{equation}
No additional value-projection layer is used; the learned edge update itself provides the message to be weighted and aggregated.

\paragraph{Scaled dot-product score}
The unnormalised compatibility score for edge \(v\to i\) is computed by a scaled dot product \cite{vaswani2017attention}:
\begin{equation}
\rho_{v\to i,L}^{\,n}
=
\frac{
\left(
\vecv{Q}_{i,L}^{\,n}
\right)^{\top}
\vecv{K}_{v\to i,L}^{\,n}
}{
\sqrt{d_h}
},
\label{eq:att_score}
\end{equation}
where \(\rho_{v\to i,L}^{\,n}\) is the unnormalised attention score, and \(d_h\) is the hidden dimension of the query and key vectors.

\paragraph{Softmax over incoming edges}
For each receiver element centre-node \(i\), the softmax is evaluated over all incoming node-to-element edges from the connected mesh node set \(\mathcal{N}(i)\):
\begin{equation}
\alpha_{v\to i,L}^{\,n}
=
\frac{
\exp
\!\left(
\rho_{v\to i,L}^{\,n}
\right)
}{
\sum\limits_{v'\in\mathcal{N}(i)}
\exp
\!\left(
\rho_{v'\to i,L}^{\,n}
\right)
}.
\label{eq:att_weight}
\end{equation}
Here, \(\alpha_{v\to i,L}^{\,n}\) is the normalised attention weight assigned to the directed edge \(v\to i\). Because the weights are softmax-normalised over the incoming edges of the same receiver element centre-node, they satisfy
\begin{equation}
\sum_{v\in\mathcal{N}(i)}
\alpha_{v\to i,L}^{\,n}
=
1.
\label{eq:att_normalisation}
\end{equation}
These weights are then used in Eq.~\eqref{eq:elem_agg} to aggregate the updated node-to-element edge messages.

\paragraph{Element-to-node attention}
The element-to-node direction is computed analogously. For a directed edge \(i\to v\), the receiver is mesh node \(v\), so the query is computed from \(\hat{\vecv{v}}_{v,L}^{\,n}\), while the key is computed from the directed edge feature \(\hat{\vecv{e}}_{i\to v,L}^{\,n}\):
\begin{equation}
\vecv{Q}_{v,L}^{\,n}
=
f_q^{v}
\!\left(
\hat{\vecv{v}}_{v,L}^{\,n}
\right),
\qquad
\vecv{K}_{i\to v,L}^{\,n}
=
f_k^{v}
\!\left(
\hat{\vecv{e}}_{i\to v,L}^{\,n}
\right),
\qquad
\vecv{V}_{i\to v,L}^{\,n}
=
\hat{\vecv{e}}_{i\to v,L+1}^{\,n}.
\label{eq:QKV_elem_to_node}
\end{equation}
Here, \(\vecv{Q}_{v,L}^{\,n}\) is the query vector of mesh node \(v\), and \(\vecv{K}_{i\to v,L}^{\,n}\) is the key vector of the incoming directed edge \(i\to v\). The functions \(f_q^{v}\) and \(f_k^{v}\) are learnable projection functions that map the mesh node latent feature and the directed edge latent feature into the query and key spaces, respectively. The corresponding unnormalised score and normalised attention weight are
\begin{align}
\rho_{i\to v,L}^{\,n}
&=
\frac{
\left(
\vecv{Q}_{v,L}^{\,n}
\right)^{\top}
\vecv{K}_{i\to v,L}^{\,n}
}{
\sqrt{d_h}
},
\\
\alpha_{i\to v,L}^{\,n}
&=
\frac{
\exp
\!\left(
\rho_{i\to v,L}^{\,n}
\right)
}{
\sum\limits_{i'\in\mathcal{M}(v)}
\exp
\!\left(
\rho_{i'\to v,L}^{\,n}
\right)
}.
\label{eq:att_weight_elem_to_node}
\end{align}
Here, \(\rho_{i\to v,L}^{\,n}\) is the unnormalised attention score, \(\mathcal{M}(v)=\{i\mid v\in\mathcal{N}(i)\}\) is the set of element-centre nodes connected to mesh node \(v\). The resulting weights \(\alpha_{i\to v,L}^{\,n}\) are used in Eq.~\eqref{eq:node_agg} to aggregate the updated element-to-node edge messages.

Overall, this attention design makes the node-element coupling edge-aware. The receiver state determines the query, the incoming directed edge state determines the key, and the updated directed edge feature provides the message being weighted. \rev{This allows the processor to adapt the strength of node–element information exchange jointly according to the receiving state and the evolving local relation encoded by each directed edge. }

\subsection{Downsampling and upsampling mechanism}
\label{Downsample/upsample mechanism}

\begin{figure}[htbp]
\centering
\includegraphics[width=\linewidth, trim=0mm  0mm 0mm 0mm, clip]{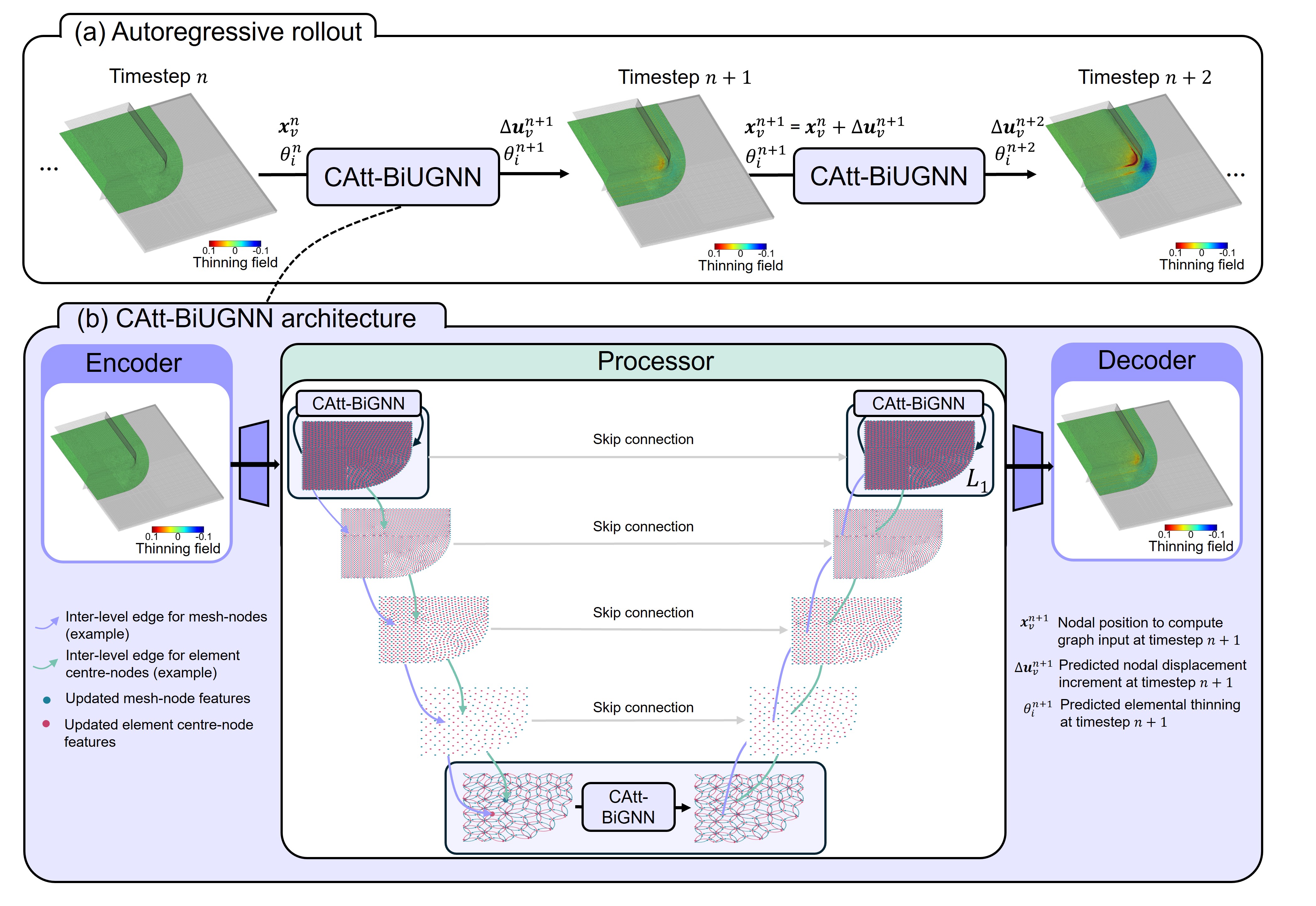}
\caption{Hierarchical CAtt-BiUGNN architecture. The original fine bipartite graph is encoded and progressively downsampled to lower-resolution graph levels. CAtt-BiGNN processor blocks operate within the hierarchy, while skip connections preserve intermediate latent features. Upsampling transfers the updated lower-resolution-level information back to the fine graph before decoding the nodal displacement increments and elemental thinning values.}
\label{Architecture of CAtt-BiUGNN}
\end{figure}

For large sheet-forming graphs, applying many CAtt-BiGNN processor layers directly on the finest graph can be computationally expensive and may limit the effective receptive field. To address this, the proposed model is extended to a hierarchical architecture, denoted as CAtt-BiUGNN. As shown in Figure~\ref{Architecture of CAtt-BiUGNN}, the model first applies CAtt-BiGNN processor layers on the original fine graph, then transfers the mesh node and element centre-node latent features to coarser graph levels, applies additional CAtt-BiGNN processor layers at the coarsest level, and finally upsamples the features back to the fine graph before decoding.

The graph hierarchy is indexed by \(\ell=0,1,\ldots,L_g\), where \(\ell=0\) denotes the original high-resolution graph and \(\ell=L_g\) denotes the lowest-resolution graph. The timestep index is denoted by \(n\). Superscript \(V\) denotes quantities associated with mesh nodes, and superscript \(S\) denotes quantities associated with element-centre nodes. The symbols \(\downarrow\) and \(\uparrow\) denote downsampling and upsampling operations, respectively. Mesh nodes are indexed by \(v\) on level \(\ell\) and by \(u\) on level \(\ell+1\). Element-centre nodes are indexed by \(i\) on level \(\ell\) and by \(j\) on level \(\ell+1\).

\begin{figure}[htbp]
\centering
\includegraphics[width=\linewidth, trim=0mm 0mm 0mm 0mm, clip]{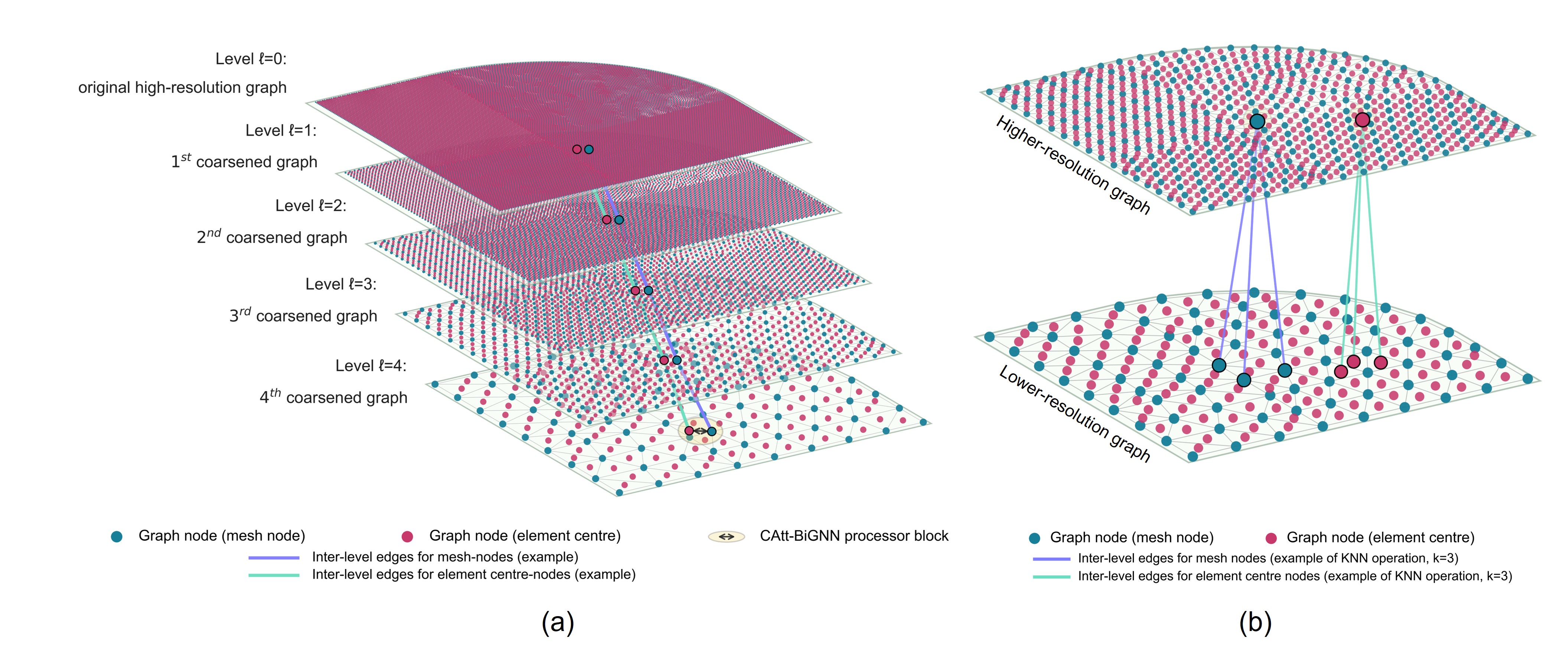}
\caption{Bipartite graph coarsening and inter-level mapping used in the downsampling and upsampling mechanism. (a) A hierarchy of bipartite graphs is constructed from the original high-resolution graph at level ({\(\ell=0\)}) to the fourth coarsened graph at level ({\(\ell=4\)}). Mesh nodes and element centres are represented as two types of graph entities at each level. Example inter-level edges show how mesh node and element-centre features are transferred between adjacent graph levels. (b) Illustration of the $k$-nearest-neighbour inter-level mapping between a higher-resolution graph and a lower-resolution graph, with $k=3$ shown as an example. Separate mappings are constructed for mesh nodes and element centres. These mappings are used for downsampling and upsampling in the hierarchical CAtt-BiUGNN architecture.}
\label{Graph coarsening method used in downsampling/upsampling operations}
\end{figure}

Coarsening is applied separately to mesh nodes and element-centre nodes. At each coarsened graph level, the corresponding mesh node and element centre-node sets form a bipartite graph, so the CAtt-BiGNN processor can be applied. In the corner-shaped case study, four coarsening operations are used, producing five graph levels including the original high-resolution graph, as illustrated in Figure~\ref{Graph coarsening method used in downsampling/upsampling operations}. \rev{Specifically, the same blank geometry was remeshed in commercial meshing software using progressively larger element sizes to generate graph levels with decreasing spatial resolution. Each remeshed blank defines one bipartite graph level, in which the mesh nodes and element-centre nodes are obtained from the corresponding FE mesh, and each element centre-node is connected to the mesh nodes belonging to that element. Because the blank geometry is identical across all corner samples, the same preconstructed five-level graph hierarchy is reused for all training, validation, and test cases.}

Inter-level edges define how information is transferred between adjacent graph levels. A k-nearest-neighbour search is used to construct sparse inter-level mappings. The number of nearest neighbours used in the hierarchy is denoted by \(K_{\mathrm{nn}}\). In the implementation, each higher-resolution-level node is connected to its \(K_{\mathrm{nn}}\) nearest lower-resolution-level nodes. \rev{For each higher-resolution entity, the 3 nearest lower-resolution entities of the same type are identified using a KD-Tree. Separate mappings are constructed for mesh nodes and element-centre nodes, such that each higher-resolution entity is connected to its three nearest entities of the same type at the adjacent lower-resolution level. These mappings are precomputed from the undeformed graph geometries and reused throughout training and autoregressive rollout. During downsampling, the higher-resolution latent features are transformed by non-shared learnable inter-level edge weights, summed at the connected lower-resolution entity, and passed through a LeakyReLU activation function. During upsampling, the same inter-level edges are used in the reverse direction, and the propagated feature is concatenated with the stored skip feature from the corresponding downsampling stage before transformation and aggregation.}

For downsampling from level \(\ell\) to level \(\ell+1\), let \(\mathcal{E}_{\downarrow,\ell}^{V}\) denote the set of inter-level mesh node edges from higher-resolution mesh nodes to lower-resolution mesh nodes, and let \(\mathcal{E}_{\downarrow,\ell}^{S}\) denote the set of inter-level edges from higher-resolution element-centre nodes to lower-resolution element-centre nodes. For a lower-resolution mesh node \(u\) and a lower-resolution element centre-node \(j\), the incoming higher-resolution-level neighbour sets are
\begin{align}
\mathcal{N}_{\downarrow}^{V}(u)
&=
\{\,v \mid (v,u)\in\mathcal{E}_{\downarrow,\ell}^{V}\,\},
\label{eq:down_neighbour_mesh}\\
\mathcal{N}_{\downarrow}^{S}(j)
&=
\{\,i \mid (i,j)\in\mathcal{E}_{\downarrow,\ell}^{S}\,\}.
\label{eq:down_neighbour_element}
\end{align}
Here, \(\mathcal{N}_{\downarrow}^{V}(u)\) is the set of higher-resolution-level mesh nodes connected to lower-resolution mesh node \(u\), and \(\mathcal{N}_{\downarrow}^{S}(j)\) is the set of higher-resolution-level element-centre nodes connected to lower-resolution element centre-node \(j\).

The downsampling operation is applied separately to mesh node and element centre-node latent features. Let \(\hat{\vecv{v}}_{v,\ell}^{\,n}\) denote the latent feature of higher-resolution-level mesh node \(v\), and let \(\hat{\vecv{v}}_{u,\ell+1}^{\,n}\) denote the resulting latent feature of lower-resolution mesh node \(u\). Similarly, let \(\hat{\vecv{s}}_{i,\ell}^{\,n}\) denote the latent feature of higher-resolution-level element centre-node \(i\), and let \(\hat{\vecv{s}}_{j,\ell+1}^{\,n}\) denote the resulting latent feature of lower-resolution element centre-node \(j\). The two downsampling operations are
\begin{align}
\hat{\vecv{v}}_{u,\ell+1}^{\,n}
&=
\phi
\!\left(
\vecv{a}_{\downarrow,\ell}^{V}
+
\sum_{v\in\mathcal{N}_{\downarrow}^{V}(u)}
\matm{W}_{v\to u,\ell}^{\downarrow,V}
\hat{\vecv{v}}_{v,\ell}^{\,n}
\right),
\label{eq:ds-mesh}\\
\hat{\vecv{s}}_{j,\ell+1}^{\,n}
&=
\phi
\!\left(
\vecv{a}_{\downarrow,\ell}^{S}
+
\sum_{i\in\mathcal{N}_{\downarrow}^{S}(j)}
\matm{W}_{i\to j,\ell}^{\downarrow,S}
\hat{\vecv{s}}_{i,\ell}^{\,n}
\right).
\label{eq:ds-element}
\end{align}
Here, \(\matm{W}_{v\to u,\ell}^{\downarrow,V}\) is the learnable edge-specific weight matrix for the mesh node inter-level edge \((v,u)\), and \(\matm{W}_{i\to j,\ell}^{\downarrow,S}\) is the corresponding edge-specific weight matrix for the element inter-level edge \((i,j)\). The vectors \(\vecv{a}_{\downarrow,\ell}^{V}\) and \(\vecv{a}_{\downarrow,\ell}^{S}\) are learnable bias vectors for mesh node and element centre-node downsampling, respectively. The activation function is denoted by \(\phi(\cdot)\).

Upsampling transfers the updated lower-resolution-level features back to higher-resolution graph levels. The upsampling edge sets are the reverse of the corresponding downsampling mappings. For upsampling from level \(\ell+1\) back to level \(\ell\), let \(\mathcal{E}_{\uparrow,\ell}^{V}\) denote the set of inter-level edges from lower-resolution mesh nodes to higher-resolution mesh nodes, and let \(\mathcal{E}_{\uparrow,\ell}^{S}\) denote the set of inter-level edges from lower-resolution element centre-node to higher-resolution element centre-node. For a higher-resolution mesh node \(v\) and a higher-resolution element centre-node \(i\), the connected lower-resolution-level neighbour sets are
\begin{align}
\mathcal{N}_{\uparrow}^{V}(v)
&=
\{\,u \mid (u,v)\in\mathcal{E}_{\uparrow,\ell}^{V}\,\},
\label{eq:up_neighbour_mesh}\\
\mathcal{N}_{\uparrow}^{S}(i)
&=
\{\,j \mid (j,i)\in\mathcal{E}_{\uparrow,\ell}^{S}\,\}.
\label{eq:up_neighbour_element}
\end{align}
Here, \(\mathcal{N}_{\uparrow}^{V}(v)\) is the set of lower-resolution mesh nodes connected to higher-resolution mesh node \(v\), and \(\mathcal{N}_{\uparrow}^{S}(i)\) is the set of lower-resolution element centre-node connected to higher-resolution element centre-node \(i\).

Skip connections are used during upsampling to preserve features from the corresponding downsampling stage. Let \(\hat{\vecv{v}}_{u,\ell+1}^{\,n,\mathrm{up}}\) denote the updated lower-resolution mesh node feature propagated from the coarser part of the network, and let \(\hat{\vecv{v}}_{u,\ell+1}^{\,n,\mathrm{skip}}\) denote the stored lower-resolution mesh node feature from the corresponding downsampling stage. Similarly, let \(\hat{\vecv{s}}_{j,\ell+1}^{\,n,\mathrm{up}}\) and \(\hat{\vecv{s}}_{j,\ell+1}^{\,n,\mathrm{skip}}\) denote the updated and stored lower-resolution element centre-node features. The two upsampling operations are
\begin{align}
\hat{\vecv{v}}_{v,\ell}^{\,n,\mathrm{up}}
&=
\phi
\!\left(
\vecv{a}_{\uparrow,\ell}^{V}
+
\sum_{u\in\mathcal{N}_{\uparrow}^{V}(v)}
\matm{W}_{u\to v,\ell}^{\uparrow,V}
\left[
\hat{\vecv{v}}_{u,\ell+1}^{\,n,\mathrm{up}};
\hat{\vecv{v}}_{u,\ell+1}^{\,n,\mathrm{skip}}
\right]
\right),
\label{eq:us-mesh}\\
\hat{\vecv{s}}_{i,\ell}^{\,n,\mathrm{up}}
&=
\phi
\!\left(
\vecv{a}_{\uparrow,\ell}^{S}
+
\sum_{j\in\mathcal{N}_{\uparrow}^{S}(i)}
\matm{W}_{j\to i,\ell}^{\uparrow,S}
\left[
\hat{\vecv{s}}_{j,\ell+1}^{\,n,\mathrm{up}};
\hat{\vecv{s}}_{j,\ell+1}^{\,n,\mathrm{skip}}
\right]
\right).
\label{eq:us-element}
\end{align}
Here, \([\cdot;\cdot]\) denotes feature concatenation. The matrices \(\matm{W}_{u\to v,\ell}^{\uparrow,V}\) and \(\matm{W}_{j\to i,\ell}^{\uparrow,S}\) are learnable edge-specific weight matrices for mesh node and element centre-node upsampling, respectively. The vectors \(\vecv{a}_{\uparrow,\ell}^{V}\) and \(\vecv{a}_{\uparrow,\ell}^{S}\) are the corresponding bias vectors.

After the final upsampling operation returns information to the original higher-resolution graph, the resulting higher-resolution-level mesh node and element latent features are passed to the decoder described in Section~\ref{General description to CAtt-BiGNN processor}. The hierarchical architecture therefore combines local node-element coupling on the higher-resolution graph with longer-range information propagation through lower-resolution graph levels.

\section{Model training}
\label{Model training}

\subsection{Training setup}
\label{Training setup}

The models were trained on different computing platforms depending on the case study. For the dome-shaped case study, which contains fewer mesh nodes and elements, training was performed on a workstation equipped with an NVIDIA Quadro 5000 GPU with 16 GB of GPU memory. For the corner-shaped case study, training was conducted on Google Colab using an L4 GPU with 22.5 GB of GPU memory. All Dome models were trained for 1000 epochs. In the Corner case, UMGN, RUGNN and the four BiUGNN variants were trained for 500 epochs. \rev{ Transolver and GNN-Transformer had not converged after 500 epochs and were therefore trained for 1500 epochs.}

All models were trained using the Adam optimiser. The learning rate followed an exponential decay schedule from \(10^{-4}\) to \(10^{-6}\), consistent with the baseline architectures~\cite{pfaff2021learning,zhao2026recurrent}. Standardisation was applied to the input and output fields before training to improve numerical conditioning and stabilise convergence. \rev{A batch size of 1 was used for all training runs.}

For the dome-shaped case study, no downsampling or upsampling was used, allowing a direct comparison between CAtt-BiGNN and the conventional mesh-based node-centred GNN baselines. The dome models used 20 message-passing layers with a latent feature dimension of 128. For the larger corner-shaped case study, the proposed CAtt-BiUGNN incorporates hierarchical graph downsampling and upsampling, as described in Section~\ref{Downsample/upsample mechanism}, to enable information propagation over longer spatial distances. \rev{The Corner CAtt-BiUGNN used latent feature dimensions of 32, 32, 64, 128, and 128 from graph levels \(\ell=0\) to \(\ell=4\), respectively. Two CAtt-BiGNN processor layers were applied on the finest graph before downsampling, 10 processor layers were applied on the coarsest graph, and two additional processor layers were applied on the finest graph after upsampling. In all models, the encoder and latent-processing MLPs contained three linear layers, with LeakyReLU activation after the first two layers and Layer Normalisation after the final latent layer. The output decoders map the latent features directly to the physical output dimensions. }

\rev{To assess the sensitivity of the results to training randomness, multiple trained runs were evaluated. For the dome-shaped case study, each model was trained and evaluated independently using random seeds 0, 1, and 2. The main corner-shaped baselines were trained and evaluated independently using random seeds 0, 1, and 2. The results are reported in Section~\ref{Model performance evaluation}.}

The training strategy used teacher forcing. At each prediction transition from saved state \(n\) to saved state \(n+1\), the ground-truth graph state at \(t^n\) was used as the model input, and the model was trained to predict the next-step nodal displacement increments and elemental thinning value, consistent with the graph-output convention defined in Table~\ref{tab:graph-notation} and the decoder formulation in Section~\ref{General description to CAtt-BiGNN processor}. During autoregressive validation and testing, the predicted displacement increment was accumulated to update the nodal position, and the updated geometry was then used to recompute geometry-dependent edge features and contact-related node features for the next timestep. 

The total training loss combines the nodal displacement increment loss and the elemental thinning loss. Let \(M\) denote the number of trajectories, \(N_p\) the number of prediction transitions per trajectory, \(V\) the number of mesh nodes, and \(N_e\) the number of element centre-nodes. If 11 states are recorded for one rollout, then \(N_p=10\) prediction transitions. A superscript star denotes the FE reference value, while quantities without a star denote model predictions. In the implementation, the losses are evaluated in the standardised output space; the symbol \(\widetilde{(\cdot)}\) denotes the corresponding standardised quantity.

The displacement increment loss is
\begin{equation}
\mathcal{L}_{\upr{disp}}
=
\frac{1}{3 M N_p V}
\sum_{m=1}^{M}
\sum_{n=0}^{N_p-1}
\sum_{v=1}^{V}
\sum_{d\in\{x,y,z\}}
\left(
\widetilde{\Delta u}_{v,d}^{\,m,n+1}
-
\widetilde{\Delta u}_{v,d}^{*\,m,n+1}
\right)^2 .
\label{eq:disp_loss}
\end{equation}
Here, \(\Delta u_{v,d}^{\,m,n+1}\) is the predicted displacement increment of mesh node \(v\) in Cartesian direction \(d\) for trajectory \(m\), and \(\Delta u_{v,d}^{*\,m,n+1}\) is the corresponding FE reference increment. This definition follows the rollout update in which the predicted increment, rather than the accumulated displacement, advances the nodal position.

The thinning loss is
\begin{equation}
\mathcal{L}_{\upr{thin}}
=
\frac{1}{M N_p N_e}
\sum_{m=1}^{M}
\sum_{n=0}^{N_p-1}
\sum_{i=1}^{N_e}
\left(
\widetilde{\theta}_{i}^{\,m,n+1}
-
\widetilde{\theta}_{i}^{*\,m,n+1}
\right)^2 ,
\label{eq:thin_loss}
\end{equation}
where \(\theta_i^{\,m,n+1}\) and \(\theta_i^{*\,m,n+1}\) are the predicted and FE reference elemental thinning values for element centre-node \(i\), respectively. The total loss is the unweighted sum of the two terms:
\begin{equation}
\mathcal{L}_{\upr{total}}
=
\mathcal{L}_{\upr{disp}}
+
\mathcal{L}_{\upr{thin}} .
\label{eq:total_loss}
\end{equation}
For evaluation and visualisation, the predicted outputs were transformed back from the standardised space to their original physical units before computing the reported physical error metrics.

\section{Model performance evaluation}
\label{Model performance evaluation}

The evaluation is organised to test the modelling choices introduced in the previous sections. The dome-shaped case study uses a smaller graph and is used to isolate the effects of the bipartite node--element representation and the edge-aware attention mechanism without hierarchical downsampling and upsampling. The corner-shaped case study evaluates whether the hierarchical CAtt-BiUGNN architecture preserves the benefits of edge-aware nodal--elemental coupling on a larger graph under more complex hot-forming conditions. \rev{In addition to the baseline and ablation comparisons, the evaluation quantifies run-to-run variability, examines the sensitivity of the thinning metrics to their prescribed parameters, and assesses the output-level area--thickness consistency and empirical error evolution during autoregressive rollout.}

\subsection{Dome-shaped case study}
\label{Dome shape case study}
\paragraph{\rev{Compared models}}
The dome-shaped case uses relatively small graphs and therefore does not require hierarchical downsampling or upsampling. This provides a controlled setting in which the effects of the nodal--elemental representation and attention mechanism can be examined directly. \rev{Seven configurations are compared. The external baselines are MeshGraphNet (MGN), which performs message passing on the node-centred mesh-connectivity \cite{pfaff2021learning}; Recurrent GNN (RGNN), which introduces recurrent graph updates to MGN \cite{chen2024predicting}; and FEIH-GNN, which introduces explicit node--element computation through a hypergraph formulation derived from the local stiffness-matrix calculation and assembly process \cite{GAO2024112866}. The architectural ablations comprise Vanilla-BiGNN with uniform aggregation, GAT-BiGNN with GAT-style attention \cite{velivckovic2017graph}, and CA-BiGNN with cross-attention Keys constructed from source-node features. In contrast, the proposed CAtt-BiGNN constructs the Keys from directed edge features, as described in Section~\ref{Cross-attention based edge processor block}. Vanilla-, GAT- and CA-BiGNN correspond to the relation-specific message-passing and attention schemes used in bipartite GNNs, applied to the same node–element graph. They therefore serve as bipartite-graph ablations within a controlled architecture. All configurations use the same training/test split and ten-step autoregressive rollout protocol.}

\paragraph{Evaluation metrics}

The evaluation metrics are designed to assess the autoregressive prediction accuracy of both nodal geometry evolution and element level thinning response. The main quantitative metrics are evaluated at the final rollout timestep because the final formed geometry and thinning distribution are the primary quantities used for manufacturability assessment. 
Two displacement metrics are used to evaluate final geometry prediction.
The mean positional Euclidean error measures the average Euclidean distance between the predicted and FE reference nodal positions at the final rollout. The position mean absolute error (MAE) measures the average absolute coordinate-wise difference between the predicted and FE reference final nodal positions over all mesh nodes and three Cartesian directions.

Two thinning-related metrics are used to evaluate element-level thinning prediction at the final rollout timestep. The threshold-based thinning error is the mean absolute error over elements whose FE reference thinning magnitude exceeds $\tau$. This setting focuses the evaluation on regions with a meaningful thinning response and reduces the influence of near-zero thinning regions. \rev{ Results are reported for $\tau\in\{0.0025,0.005,0.01\}$ to assess sensitivity of the non-negligible thinning region.} The top-$p$\% critical relative thinning error further evaluates the prediction accuracy in the most thinning-critical region. For each test case, elements are ranked according to their FE reference thinning values at the final rollout timestep, and the top $p$\% elements are selected as the critical set. The relative error is calculated between the mean predicted and FE-reference thinning values over the same selected elements. \rev{Results are reported for $p\in\{2.5,5,10\}$ to assess sensitivity in the critical thinning region.}

\paragraph{Results}
Table~\ref{tab:dome-model-comparison} reports the final-timestep mean and sample standard deviation across three independently trained runs. For each run, the metrics are first averaged over the same 25 test trajectories. \rev{To assess variability across independent training runs, each compared model was trained independently using three random seeds. Table~\ref{tab:dome-model-comparison} reports the autoregressive-rollout results from these three runs. The reported mean and sample standard deviation are then calculated across the three run-level values.}
\begin{table*}[t]
\caption{\rev{Baseline and architectural-ablation comparison for the
dome-shaped case study at the final rollout timestep. Values are the mean $\pm$ sample standard
 deviation across three independent training runs; lower values are better. 
Bold values indicate the best result among all evaluated models.}}
\label{tab:dome-model-comparison}

\centering
\begingroup
\footnotesize
\setlength{\tabcolsep}{3.2pt}
\renewcommand{\arraystretch}{1.20}

\newcommand{\numcell}[2]{%
  \begin{tabular}[c]{@{}c@{}}
    \rev{$#1$}\\[-1pt]
    \rev{$\pm #2$}
  \end{tabular}}

\newcommand{\bestnumcell}[2]{%
  \begin{tabular}[c]{@{}c@{}}
    \rev{$\mathbf{#1}$}\\[-1pt]
    \rev{$\mathbf{\pm #2}$}
  \end{tabular}}

\newcommand{\pctcell}[2]{%
  \begin{tabular}[c]{@{}c@{}}
    \rev{$#1\%$}\\[-1pt]
    \rev{$\pm #2\%$}
  \end{tabular}}

\newcommand{\bestpctcell}[2]{%
  \begin{tabular}[c]{@{}c@{}}
    \rev{$\mathbf{#1\%}$}\\[-1pt]
    \rev{$\mathbf{\pm #2\%}$}
  \end{tabular}}

\resizebox{\textwidth}{!}{%
\begin{tabular}{@{}
l
cc
@{\hspace{5pt}}
ccc
@{\hspace{5pt}}
ccc
@{}}

\toprule

&
\multicolumn{2}{c}{
  \rev{\makecell[c]{\textbf{Displacement}\\\textbf{evaluation}}}
}
&
\multicolumn{3}{c}{
  \rev{\makecell[c]{\textbf{Threshold-based}\\\textbf{thinning error}}}
}
&
\multicolumn{3}{c}{
  \rev{\makecell[c]{\textbf{Top-$p\%$ critical}\\
  \textbf{relative thinning error}}}
}
\\

\cmidrule(lr){2-3}
\cmidrule(lr){4-6}
\cmidrule(l){7-9}

\rev{\textbf{Model}}
&
\rev{\makecell[c]{Mean positional\\Euclidean error\\(mm)}}
&
\rev{\makecell[c]{Positional\\MAE (mm)}}
&
\rev{\makecell[c]{$\tau=$\\0.0025}}
&
\rev{\makecell[c]{$\tau=$\\0.005}}
&
\rev{\makecell[c]{$\tau=$\\0.01}}
&
\rev{\makecell[c]{$p=$\\2.5\%}}
&
\rev{\makecell[c]{$p=$\\5\%}}
&
\rev{\makecell[c]{$p=$\\10\%}}
\\

\midrule


\rev{MGN}
&
\numcell{0.489910}{0.020781}
&
\numcell{0.230821}{0.006642}
&
\numcell{0.005305}{0.001241}
&
\numcell{0.006684}{0.001554}
&
\numcell{0.008469}{0.002333}
&
\pctcell{5.313}{3.499}
&
\pctcell{4.982}{3.644}
&
\pctcell{4.171}{3.235}
\\

\rev{RGNN}
&
\numcell{0.434748}{0.165268}&
\numcell{0.205426}{0.078363}&
\numcell{0.004687}{0.000640}&
\numcell{0.005805}{0.000759}
&
\numcell{0.007111}{0.000896}
&
\pctcell{3.376}{0.871}
&
\pctcell{3.084}{0.720}
&
\pctcell{2.586}{0.653}
\\

\rev{FEIH-GNN}
&
\numcell{0.543251}{0.056437}
&
\numcell{0.254525}{0.023953}
&
\numcell{0.004483}{0.000243}
&
\numcell{0.005328}{0.000437}&
\numcell{0.006283}{0.000609}&
\pctcell{3.830}{0.959}
&
\pctcell{3.262}{0.564}
&
\pctcell{2.559}{0.398}\\

\midrule


\rev{\makecell[l]{Vanilla-\\BiGNN}}
&
\numcell{0.459880}{0.052682}
&
\numcell{0.214552}{0.020985}
&
\numcell{0.003019}{0.000237}
&
\numcell{0.003630}{0.000315}
&
\numcell{0.004336}{0.000373}
&
\pctcell{1.678}{0.101}
&
\bestpctcell{1.456}{0.081}
&
\bestpctcell{1.199}{0.032}
\\

\rev{GAT-BiGNN}
&
\numcell{0.485619}{0.030659}
&
\numcell{0.228792}{0.012507}
&
\numcell{0.004169}{0.000521}
&
\numcell{0.004916}{0.000643}
&
\numcell{0.005896}{0.000776}
&
\pctcell{2.534}{0.823}
&
\pctcell{2.340}{0.597}
&
\pctcell{1.973}{0.393}
\\

\rev{CA-BiGNN}
&
\numcell{0.496587}{0.026658}
&
\numcell{0.231842}{0.013053}
&
\numcell{0.003600}{0.000018}
&
\numcell{0.004248}{0.000138}
&
\numcell{0.005098}{0.000267}
&
\pctcell{2.115}{0.177}
&
\pctcell{1.902}{0.184}
&
\pctcell{1.633}{0.144}
\\

\rev{\makecell[l]{CAtt-BiGNN\\(proposed)}}
&
\bestnumcell{0.382526}{0.034364}
&
\bestnumcell{0.180889}{0.015170}
&
\bestnumcell{0.002761}{0.000256}
&
\bestnumcell{0.003402}{0.000321}
&
\bestnumcell{0.004261}{0.000439}
&
\bestpctcell{1.631}{0.135}
&
\pctcell{1.562}{0.207}
&
\pctcell{1.387}{0.295}\\

\bottomrule
\end{tabular}%
}

\endgroup
\end{table*}

Among the node-centred baselines, RGNN improves on MGN for all reported final-timestep metrics. \rev{FEIH-GNN gives lower threshold-based thinning errors than both node-centred baselines. This indicates the benefit of element-level computation for thinning prediction. Its displacement errors are higher than those of the node-centred baselines. These results suggest that, in this forming problem, node–element connectivity alone may not improve both fields together.}

\rev{Vanilla-BiGNN introduces separate nodal and elemental states with bidirectional node-to-element and element-to-node message passing, but uses uniform aggregation. It improves on FEIH-GNN for both displacement metrics, all three threshold-based thinning metrics, and all three critical relative thinning metrics. Relative to the node-centred models, it also gives substantially lower thinning errors and a lower displacement error than MGN, although RGNN retains a lower mean displacement error.} This may result from the added coupling between the elemental thinning field and the nodal displacement field. The coupling was expected to facilitate joint learning between the nodal and elemental states; however, the model does not fully capture this dependency effectively. This suggests that errors in one predicted field may propagate through the coupling and affect the other field during rollout. 

\rev{The attention ablations demonstrate that the message-weighting mechanism materially affects this balance. GAT-BiGNN and CA-BiGNN do not consistently improve on Vanilla-BiGNN across all measures. Vanilla-BiGNN gives the lowest critical relative thinning errors at $p=5\%$ and $p=10\%$, but its displacement prediction accuracy shows little improvement compared to node-centred models. By constructing attention Keys from the directed-edge features, CAtt-BiGNN obtains the lowest mean positional Euclidean error, positional MAE, and threshold-based thinning errors at all three values of $\tau$. It also gives the lowest critical relative thinning error at $p=2.5\%$ and the second-lowest values at $p=5\%$ and $p=10\%$. Overall, CAtt-BiGNN ranks first in six of the eight reported measures and provides the strongest joint balance between geometry evolution and elemental thinning prediction.}

\rev{Table~\ref{tab:dome-computational-efficiency} compares the computational requirements of all Dome models. Relative to the representative FE wall-clock time of approximately 180~s per trajectory, the learned models reduce runtime by 99.43--99.89\%. CAtt-BiGNN requires 0.965~s for a complete ten-step rollout, including graph-feature computation and updates at each step. Although its deeper bipartite message-passing architecture increases training time and peak memory relative to the node-centred baselines, it remains substantially faster than FE simulation and provides the strongest overall prediction accuracy in Table~\ref{tab:dome-model-comparison}.}

\begin{table*}[t]
\caption{\rev{Computational efficiency of all evaluated models for the dome-shaped case study. Inference time is reported per trajectory for the complete 10-step autoregressive rollout. }}
\label{tab:dome-computational-efficiency}
\centering
\begingroup
\footnotesize
\setlength{\tabcolsep}{4.5pt}
\renewcommand{\arraystretch}{1.18}
\resizebox{\textwidth}{!}{%
\begin{tabular}{@{}lrrrrr@{}}
\toprule
\rev{\textbf{Model}} &
\rev{\makecell[c]{\textbf{Training time}\\\textbf{(s/epoch)}}} &
\rev{\makecell[c]{\textbf{Peak training}\\\textbf{memory (GiB)}}} &
\rev{\makecell[c]{\textbf{Inference time}\\\textbf{(s/trajectory)}}} &
\rev{\makecell[c]{\textbf{FE time}\\\textbf{(s/trajectory)}}} &
\rev{\makecell[c]{\textbf{FE runtime}\\\textbf{reduction (\%)}}} \\
\midrule
\rev{MGN}                       & \rev{23.85}  & \rev{0.352} & \rev{0.205} & \rev{180} & \rev{99.89\%} \\
\rev{RGNN}                      & \rev{29.00}  & \rev{0.858} & \rev{0.220} & \rev{180} & \rev{99.88\%} \\
\rev{FEIH-GNN}                  & \rev{33.48}  & \rev{0.578} & \rev{0.449} & \rev{180} & \rev{99.75\%} \\
\rev{Vanilla-BiGNN}             & \rev{55.32}  & \rev{0.966} & \rev{0.828} & \rev{180} & \rev{99.54\%} \\
\rev{GAT-BiGNN}                 & \rev{74.86} & \rev{0.811} & \rev{1.029} & \rev{180} & \rev{99.43\%} \\
\rev{CA-BiGNN}                  & \rev{69.21} & \rev{1.014} & \rev{1.024} & \rev{180} & \rev{99.43\%} \\
\rev{CAtt-BiGNN (proposed)}     & \rev{71.36} & \rev{0.952} & \rev{0.965} & \rev{180} & \rev{99.46\%} \\
\bottomrule
\end{tabular}%
}
\parbox{\textwidth}{\raggedright\scriptsize \rev{FE runtime reduction is calculated as $(1-t_{\mathrm{inference}}/t_{\mathrm{FE}})\times100\%$. }}
\endgroup
\end{table*}

\begin{figure}[htbp]
\centering
\includegraphics[width=\linewidth, trim=0mm 0mm 0mm 0mm, clip]{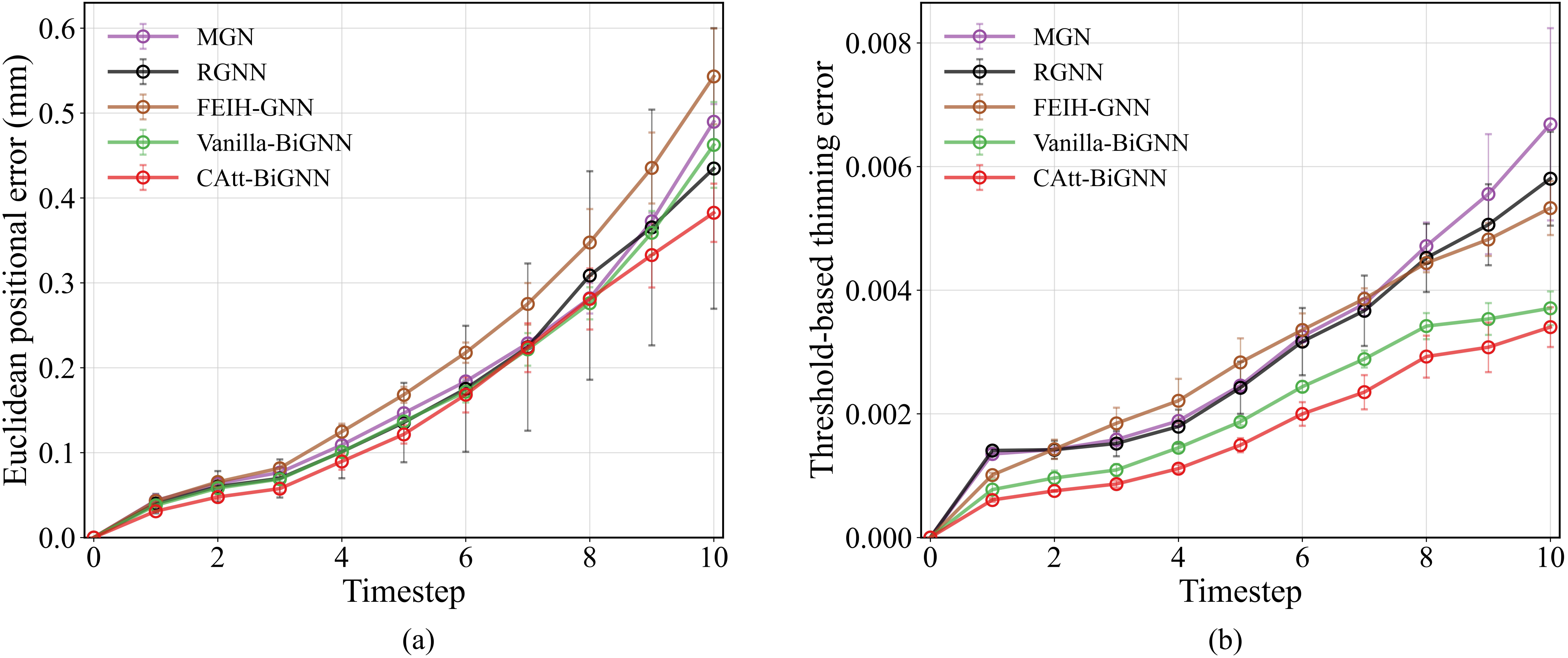}
\caption{Dome case: evolution of test-set errors over the 10 autoregressive rollout transitions. Lines show the mean and error bars show the sample standard deviation across three independently trained runs with different random seeds. (a) Euclidean positional error in mm. (b) Threshold-based thinning error at $\tau=0.005$.}
\label{dome_mean_over_time}
\end{figure}

\rev{Figure~\ref{dome_mean_over_time} shows how the Euclidean positional error and the threshold-based thinning error at $\tau=0.005$ evolve during rollout. CAtt-BiGNN remains among the lowest-error models during the early and middle rollout stages and gives the lowest positional error over timesteps 8--10. For threshold-based thinning, it gives the lowest mean error at every rollout timestep. The time-resolved results are therefore consistent with the final-timestep comparison: the edge-aware bipartite processor maintains the strongest thinning accuracy throughout rollout while becoming increasingly advantageous for geometry prediction at later deformation stages.}

Figure~\ref{dome_results_vis_all} provides a qualitative comparison of the final rollout predictions for the dome-shaped case. \rev{The displayed fields correspond to the maximum-radius test geometry for each learned model.} The node-centred baselines reproduce the overall shape but show visible error concentrations in regions with high curvature and geometric transition. \rev{FEIH-GNN predicts thinning directly on the elemental domain. Compared with the node-centred baselines, it gives lower thinning errors near the top of the dome. This improvement is less evident in its displacement field, and the main displacement- and thinning-error regions appear at different locations. Vanilla-BiGNN also improves the thinning field relative to the node-centred baselines, but its displacement prediction remains comparable to theirs. Vanilla-BiGNN explicitly couples the nodal and elemental states through bidirectional message passing inspired by the explicit-dynamics FE formulation. The similar spatial patterns in its displacement and thinning errors may suggest that the model is aware of the coupling of node--element relations. However, its uniform aggregation assigns equal importance to all incoming messages and therefore cannot adapt the coupling to different local node--element relations.} CAtt-BiGNN gives the most balanced visual agreement with the FE reference across both displacement and thinning fields. \rev{Its edge-aware cross-attention provides adaptive weighting of the coupled node--element messages according to the evolving directed-edge relations. For this case, CAtt-BiGNN also gives the lowest case-averaged positional error and absolute thinning error. Because the visualisation represents only one test geometry, the test-set metrics aggregated across three independently trained seeds in Table~\ref{tab:dome-model-comparison} provide the basis for the overall model comparison.}

\begin{figure}[htbp]
\centering
\includegraphics[width=\linewidth, trim=0mm 0mm 0mm 0mm, clip]{Fig.10.jpg}
\caption{Final-timestep visual comparison for the dome test case with the largest dome radius in the test set using the rollout predictions. Panel (a) compares deformed nodal positions and Euclidean position-error maps. Panel (b) compares elemental thinning fields and absolute thinning-error maps. Colour bars indicate position error in mm and dimensionless thinning error.}
\label{dome_results_vis_all}
\end{figure}

\subsection{Corner-shaped case study}
\label{Corner shape case study}

\rev{The corner-shaped case study compares hierarchical message-passing models, a Physics-Attention-based neural operator, and a graph Transformer on the larger forming problem. The architectural ablation separately examines the bipartite representation and attention mechanism within the common BiUGNN hierarchy. }

\paragraph{\rev{Compared models}}
\rev{Eight} model configurations are evaluated. \rev{The main-model comparison comprises UMGN, which extends MGN with graph downsampling and upsampling \cite{zhao2026recurrent}; RUGNN, which adds recurrent updates to UMGN \cite{zhao2026recurrent}; Transolver \cite{wu2024transolver}, a Physics-Attention-based neural operator; GNN-Transformer, adapted from the node-centred MGN-Transformer to use a node--element bipartite graph while retaining its global Transformer processor; and the proposed CAtt-BiUGNN. These models have three independently trained runs with different random seeds and are compared in Table~\ref{tab:corner-baselines}. The architectural ablation in Table~\ref{tab:corner-ablations} compares Vanilla-BiUGNN, GAT-BiUGNN, CA-BiUGNN, and CAtt-BiUGNN. These models use the same multilevel node--element architecture but different bipartite message-passing schemes. Vanilla-BiUGNN uses MLP-based edge updates followed by uniform aggregation. GAT-BiUGNN uses additive attention computed from the source and target entity features. CA-BiUGNN uses scaled dot-product cross-attention, with queries obtained from the target features and keys from the source mesh-node or element-centre features. CAtt-BiUGNN instead obtains the keys from the directed edge features. All models use the same test set and autoregressive evaluation protocol.}

\paragraph{Evaluation metrics}

\rev{The evaluation metrics follow Section~\ref{Dome shape case study}. In the Corner case, both thinning and thickening are pronounced. For each test case, elements are ranked according to their FE-reference thinning values, and the top and bottom $p\%$ elements are selected separately as the critical thinning and thickening regions. Following the Dome case, the relative error is calculated between the mean predicted and FE-reference thinning values in each region, and the two errors are averaged. Results are reported for $p\in\{0.5,1,2\}$ and $\tau\in\{0.0025,0.005,0.01\}$. Smaller values of $p$ are used than in the Dome case because the Corner mesh contains about 6 times as many elements and its critical thinning region is more localised, so the Dome percentages would select a region much larger than the critical region.} Each simulation consists of ten rollout timesteps. \rev{The main-model comparison results are reported as the mean and sample standard deviation across three independently trained runs with different random seeds. The architectural ablation is reported separately using the common seed-0 autoregressive rollouts for all four BiUGNN variants.}

\paragraph{Results}
\rev{Table~\ref{tab:corner-baselines} shows that CAtt-BiUGNN achieves the lowest mean errors across all eight reported measures in the main-model comparison. Among the external baselines, UMGN gives the lowest displacement errors, while GNN-Transformer achieves the lowest errors across all six thinning-related measures, with displacement errors comparable to UMGN. Transolver gives lower critical-region relative errors than RUGNN, although its displacement and threshold-based thinning errors are higher.}

\rev{The evaluated baselines employ different representations and information-processing mechanisms. UMGN and RUGNN use node-centred hierarchical graph architectures, while Transolver employs Physics-Attention to model interactions through learned slices. The adapted GNN-Transformer also uses a node–element bipartite graph, but combines local graph processing with a global Transformer processor. In contrast, CAtt-BiUGNN uses dynamically updated directed edges and edge-aware cross-attention to model nodal–elemental interactions within its hierarchical architecture. The improved joint prediction accuracy suggests that the design of the coupling mechanism is important for the present forming problem. The contribution of the proposed edge-aware coupling is further examined through the controlled ablation in Table 5.}

\begin{table*}[t]
\caption{\rev{Baseline-model comparison for the corner-shaped case study using autoregressive rollouts at the final timestep. The mean $\pm$ sample standard deviation is calculated across independent seed runs. Lower values are better, and bold values indicate the best result.}}
\label{tab:corner-baselines}
\centering
\begingroup
\footnotesize
\setlength{\tabcolsep}{2.5pt}
\renewcommand{\arraystretch}{1.20}

\newcommand{\numcell}[2]{%
  \begin{tabular}[c]{@{}c@{}}
    \rev{$#1$}\\[-1pt]
    \rev{$\pm #2$}
  \end{tabular}%
}
\newcommand{\bestnumcell}[2]{%
  \begin{tabular}[c]{@{}c@{}}
    \rev{$\mathbf{#1}$}\\[-1pt]
    \rev{$\mathbf{\pm #2}$}
  \end{tabular}%
}
\newcommand{\pctcell}[2]{%
  \begin{tabular}[c]{@{}c@{}}
    \rev{$#1\%$}\\[-1pt]
    \rev{$\pm #2\%$}
  \end{tabular}%
}
\newcommand{\bestpctcell}[2]{%
  \begin{tabular}[c]{@{}c@{}}
    \rev{$\mathbf{#1\%}$}\\[-1pt]
    \rev{$\mathbf{\pm #2\%}$}
  \end{tabular}%
}

\resizebox{\textwidth}{!}{%
\begin{tabular}{@{}l cc @{\hspace{5pt}} ccc @{\hspace{5pt}} ccc@{}}
\toprule
&
\multicolumn{2}{c}{%
  \rev{\makecell[c]{\textbf{Displacement}\\\textbf{evaluation}}}
} &
\multicolumn{3}{c}{%
  \rev{\makecell[c]{\textbf{Threshold-based}\\\textbf{thinning error}}}
} &
\multicolumn{3}{c}{%
  \rev{\makecell[c]{\textbf{Top-$p\%$ mean critical relative}\\
  \textbf{thinning error}}}
} \\
\cmidrule(lr){2-3}
\cmidrule(lr){4-6}
\cmidrule(l){7-9}

\rev{\textbf{Model}} &
\rev{\makecell[c]{Mean positional\\Euclidean error\\(mm)}} &
\rev{\makecell[c]{Positional\\MAE (mm)}} &
\rev{\makecell[c]{$\tau=$\\0.0025}} &
\rev{\makecell[c]{$\tau=$\\0.005}} &
\rev{\makecell[c]{$\tau=$\\0.01}} &
\rev{\makecell[c]{$p=$\\0.5\%}} &
\rev{\makecell[c]{$p=$\\1\%}} &
\rev{\makecell[c]{$p=$\\2\%}} \\
\midrule

\rev{UMGN} &
\numcell{0.403945}{0.080826} &
\numcell{0.188071}{0.036315} &
\numcell{0.003252}{0.000277} &
\numcell{0.003887}{0.000345} &
\numcell{0.004640}{0.000377} &
\pctcell{16.125}{1.200}
&
\pctcell{13.271}{1.302}
&
\pctcell{10.101}{1.324} \\

\rev{RUGNN} &
\numcell{0.575215}{0.151339} &
\numcell{0.268517}{0.071055} &
\numcell{0.003704}{0.000343} &
\numcell{0.004436}{0.000411} &
\numcell{0.005262}{0.000471} &
\pctcell{16.926}{0.705}
&
\pctcell{13.946}{0.647}
&
\pctcell{10.818}{0.538} \\

\rev{Transolver} &
\numcell{1.044050}{0.034837} &
\numcell{0.491874}{0.019369} &
\numcell{0.004076}{0.000048} &
\numcell{0.004736}{0.000075} &
\numcell{0.005723}{0.000118} &
\pctcell{13.744}{1.109}
&
\pctcell{11.654}{1.157}
&
\pctcell{9.227}{1.691} \\

\rev{GNN-Transformer} &
\numcell{0.433753}{0.012053} &
\numcell{0.204154}{0.004905} &
\numcell{0.002988}{0.000050} &
\numcell{0.003523}{0.000058} &
\numcell{0.004367}{0.000076} &
\pctcell{11.274}{0.094}
&
\pctcell{9.021}{0.180}
&
\pctcell{6.662}{0.434} \\

\rev{CAtt-BiUGNN (proposed)} &
\bestnumcell{0.352305}{0.028390} &
\bestnumcell{0.164084}{0.013519} &
\bestnumcell{0.001712}{0.000084} &
\bestnumcell{0.001963}{0.000104} &
\bestnumcell{0.002218}{0.000134} &
\bestpctcell{4.023}{0.547}
&
\bestpctcell{3.520}{0.516}
&
\bestpctcell{3.246}{0.380} \\

\bottomrule
\end{tabular}%
}
\endgroup
\end{table*}

\rev{Within the controlled BiUGNN ablation in Table~\ref{tab:corner-ablations}, CAtt-BiUGNN gives the lowest errors for all eight reported measures. GAT-BiUGNN improves both displacement measures, all three threshold-based thinning measures, and the critical-region errors at $p=0.5\%$ and $p=1\%$ relative to Vanilla-BiUGNN, while its $p=2\%$ error is slightly higher. CA-BiUGNN reduces all thinning-related errors relative to Vanilla-BiUGNN, but gives higher displacement errors.}

\begin{table*}[t]
\caption{\rev{Architectural ablation study for the corner-shaped case study using autoregressive
rollouts at the final timestep; lower values are better. Bold values indicate the best result.}}
\label{tab:corner-ablations}
\centering
\begingroup
\footnotesize
\setlength{\tabcolsep}{3.2pt}
\renewcommand{\arraystretch}{1.20}
\resizebox{\textwidth}{!}{%
\begin{tabular}{@{}l cc @{\hspace{5pt}} ccc @{\hspace{5pt}} ccc@{}}
\toprule
&
\multicolumn{2}{c}{\rev{\makecell[c]{\textbf{Displacement}\\\textbf{evaluation}}}} &
\multicolumn{3}{c}{\rev{\makecell[c]{\textbf{Threshold-based}\\\textbf{thinning error}}}} &
\multicolumn{3}{c}{\rev{\makecell[c]{\textbf{Top-$p\%$ mean critical relative}\\\textbf{thinning error}}}} \\
\cmidrule(lr){2-3}
\cmidrule(lr){4-6}
\cmidrule(l){7-9}
\rev{\textbf{Model}} &
\rev{\makecell[c]{Mean positional\\Euclidean error\\(mm)}} &
\rev{\makecell[c]{Positional\\MAE (mm)}} &
\rev{\makecell[c]{$\tau=$\\0.0025}} &
\rev{\makecell[c]{$\tau=$\\0.005}} &
\rev{\makecell[c]{$\tau=$\\0.01}} &
\rev{\makecell[c]{$p=$\\0.5\%}} &
\rev{\makecell[c]{$p=$\\1\%}} &
\rev{\makecell[c]{$p=$\\2\%}} \\
\midrule
\rev{Vanilla-BiUGNN} &
\rev{0.550565} & \rev{0.250835} &
\rev{0.002731} & \rev{0.003194} & \rev{0.003469} &
\rev{4.642\%} & \rev{3.956\%} & \rev{3.565\%} \\
\rev{GAT-BiUGNN} &
\rev{0.430733} & \rev{0.199060} &
\rev{0.001826} & \rev{0.002084} & \rev{0.002318} &
\rev{3.998\%} & \rev{3.808\%} & \rev{3.585\%} \\
\rev{CA-BiUGNN} &
\rev{0.576968} & \rev{0.263821} &
\rev{0.001941} & \rev{0.002226} & \rev{0.002500} &
\rev{3.531\%} & \rev{3.371\%} & \rev{3.386\%} \\
\rev{CAtt-BiUGNN (proposed)} &
\rev{\textbf{0.320854}} & \rev{\textbf{0.148962}} &
\rev{\textbf{0.001618}} & \rev{\textbf{0.001846}} & \rev{\textbf{0.002077}} &
\rev{\textbf{3.448\%}} &
\rev{\textbf{3.019\%}} &
\rev{\textbf{2.955\%}} \\
\bottomrule
\end{tabular}%
}
\endgroup
\end{table*}

\rev{The advantage of CAtt-BiUGNN is more evident in the corner-shaped case.
The thinning response is highly localised around the corner and contact-driven
regions, while its spatial pattern is not directly reflected by the overall
geometry deformation. Accurate prediction therefore requires the model to
preserve both the nodal geometry evolution and the local element-level response.
In CAtt-BiUGNN, the attention weights are conditioned on the directed
node--element edge features. This allows the contribution of individual
node--element interactions to vary with the local geometric relation.
This may help the model identify locally important interactions for thinning
prediction without degrading the nodal displacement prediction.
The results therefore suggest that the proposed edge-aware cross-attention
provides a more balanced coupling of geometry and thinning fields in this case.}

To examine whether the improved accuracy is mainly attributable to the number of trainable parameters, an additional parameter-budget-controlled comparison is provided in \ref{appF}. In this comparison, the hidden dimensions of UMGN and RUGNN are adjusted from 128 to 212, yielding trainable-parameter counts comparable to those of the BiUGNN variants. Increasing the parameter budget does not consistently improve the node-centred baselines. CAtt-BiUGNN gives the lowest errors for all four reported measures in the parameter-budget-controlled comparison.

\rev{Table~\ref{tab:corner_computational_efficiency} reports the computational requirements of the Corner models. Transolver and GNN-Transformer require less peak GPU memory and shorter inference times than CAtt-BiUGNN. CAtt-BiUGNN requires 1.285 s per rollout but achieves the lowest prediction errors. Relative to the representative FE runtime of 480 s, it still provides a speed-up of approximately 373.5×. These results highlight the accuracy–cost trade-off of the proposed hierarchical bipartite architecture.}

\begin{table*}[t]
\caption{\rev{Computational efficiency of all evaluated models for the corner-shaped case study. Inference time is reported per trajectory for the complete 10-step autoregressive rollout.}}
\label{tab:corner_computational_efficiency}
\centering
\begingroup
\footnotesize
\setlength{\tabcolsep}{4.5pt}
\renewcommand{\arraystretch}{1.18}
\resizebox{\textwidth}{!}{%
\begin{tabular}{@{}lrrrrr@{}}
\toprule
\rev{\textbf{Model}} &
\rev{\makecell[c]{\textbf{Training time}\\\textbf{(s/epoch)}}} &
\rev{\makecell[c]{\textbf{Peak training}\\\textbf{memory (GiB)}}} &
\rev{\makecell[c]{\textbf{Inference time}\\\textbf{(s/trajectory)}}} &
\rev{\makecell[c]{\textbf{FE time}\\\textbf{(s/trajectory)}}} &
\rev{\makecell[c]{\textbf{FE runtime}\\\textbf{reduction (\%)}}} \\
\midrule
\rev{UMGN}                       & \rev{49.25}  & \rev{1.590} & \rev{1.025} & \rev{480} & \rev{99.79\%} \\
\rev{RUGNN}                      & \rev{65.99}  & \rev{1.753} & \rev{0.986} & \rev{480} & \rev{99.79\%} \\
\rev{Transolver}                     & \rev{24.89}    & \rev{0.447}   & \rev{0.427}   & \rev{480} & \rev{99.91\%}   \\
\rev{GNN-Transformer}                      & \rev{37.62}  & \rev{0.815} & \rev{0.482} & \rev{480} & \rev{99.90\%} \\
\rev{Vanilla-BiUGNN}             & \rev{91.55} & \rev{4.477} & \rev{1.193} & \rev{480} & \rev{99.75\%} \\
\rev{GAT-BiUGNN}                 & \rev{132.92} & \rev{4.485} & \rev{1.312} & \rev{480} & \rev{99.73\%} \\
\rev{CA-BiUGNN}                  & \rev{125.56} & \rev{4.494} & \rev{1.307} & \rev{480} & \rev{99.73\%} \\
\rev{CAtt-BiUGNN (proposed)}     & \rev{129.11} & \rev{4.497} & \rev{1.285} & \rev{480} & \rev{99.73\%} \\
\bottomrule
\end{tabular}%
}
\parbox{\textwidth}{\raggedright\scriptsize \rev{FE runtime reduction is calculated as $(1-t_{\mathrm{inference}}/t_{\mathrm{FE}})\times100\%$. }}
\endgroup
\end{table*}

\begin{figure}[htbp]
\centering
\includegraphics[width=\linewidth, trim=0mm 0mm 0mm 0mm, clip]{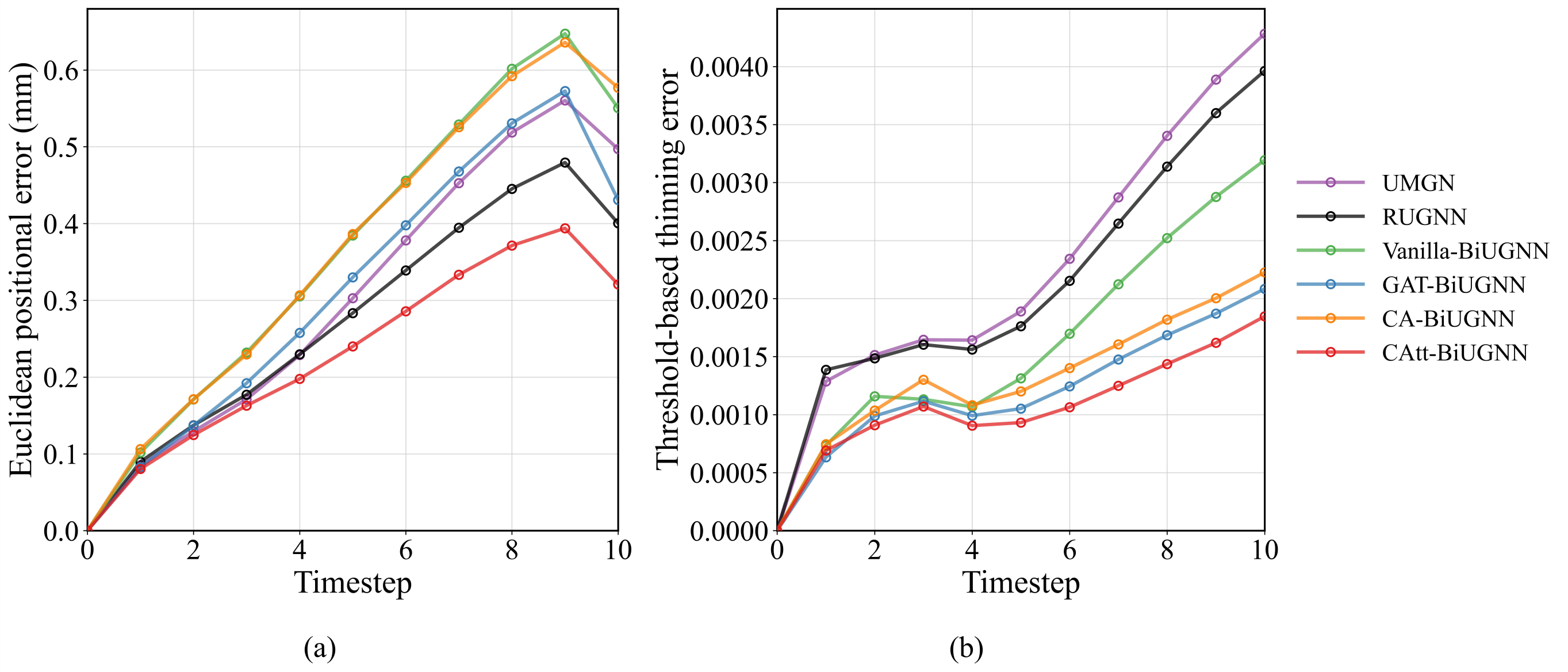}
\caption{Corner case: evolution of test-set errors over the 10 autoregressive
rollout transitions. Each point
shows the test-set mean over the 25 trajectories. (a) Euclidean positional
error in mm. (b) Threshold-based thinning error at $\tau=0.005$.}

\label{corner_mean_over_time}
\end{figure}

Figure~\ref{corner_mean_over_time} presents the evolution of the Euclidean positional error and threshold-based thinning error over the 10 rollout timesteps for the corner-shaped case study. The error trajectories are calculated from the seed-0 predictions of the six models shown. CAtt-BiUGNN achieves the lowest final-timestep error for both metrics and generally remains among the best-performing models across the rollout. Its error trajectory is also lower than those of the node-centred hierarchical baselines, UMGN and RUGNN, over most timesteps. Compared with the other bipartite variants, the advantage of CAtt-BiUGNN is more evident in the later rollout stages, particularly for the threshold-based thinning error. \rev{These trends are consistent with Table~\ref{tab:corner-ablations}, where CAtt-BiUGNN achieves the lowest error for both displacement measures, all three threshold-based thinning measures, and all three mean top-$p\%$ critical relative thinning measures among the architectural variants.}

\begin{figure}[htbp]
\centering
\includegraphics[width=\linewidth, trim=0mm 0mm 0mm 0mm, clip]{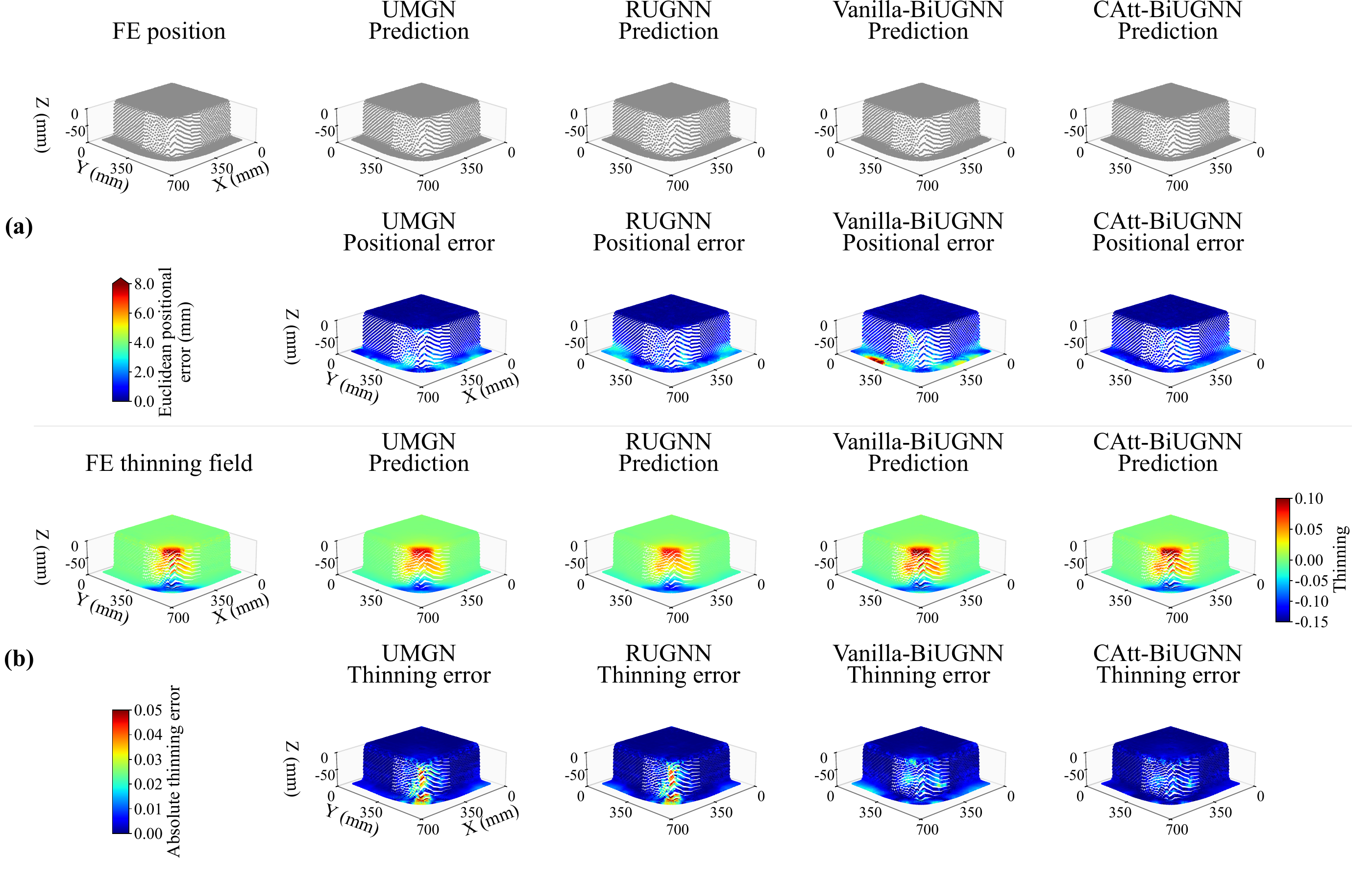}
\caption{Final-timestep visual comparison of selected models for the corner test case with the largest height. Panel (a) compares deformed nodal positions and Euclidean position-error maps. Panel (b) compares elemental thinning fields and absolute thinning-error maps. Position errors are reported in mm, and thinning errors are dimensionless.}
\label{corner_results_vis_all}
\end{figure}

Figure~\ref{corner_results_vis_all} shows the final rollout predictions for a corner-shaped test case with the largest height value in the test set. This case is used to visualise model behaviour under a relatively pronounced geometric configuration, where the corner region and contact-driven deformation are of particular interest. \rev{UMGN and RUGNN reproduce the overall formed geometry, but show larger local errors around geometric transitions. Vanilla-BiUGNN substantially reduces the thinning error, but the same level of improvement is not observed for the displacement field. Similar spatial concentrations are also visible in its displacement and thinning error maps. This suggests that direct node--element information exchange is useful, although the coupling may not yet be sufficiently selective to improve both fields simultaneously. } 

\rev{CAtt-BiUGNN shows the closest visual agreement with the FE reference for both fields. Compared with Vanilla-BiUGNN, the remaining displacement and thinning errors are both reduced. The edge-aware attention allows the contribution of each node--element interaction to vary with its local geometric relation. This may help the model retain useful coupling while reducing error propagation between the two prediction fields. }\rev{This visual comparison is consistent with Table~\ref{tab:corner-ablations}.}

\begin{figure}[htbp]
\centering
\includegraphics[width=\linewidth, trim=0mm 0mm 0mm 0mm, clip]{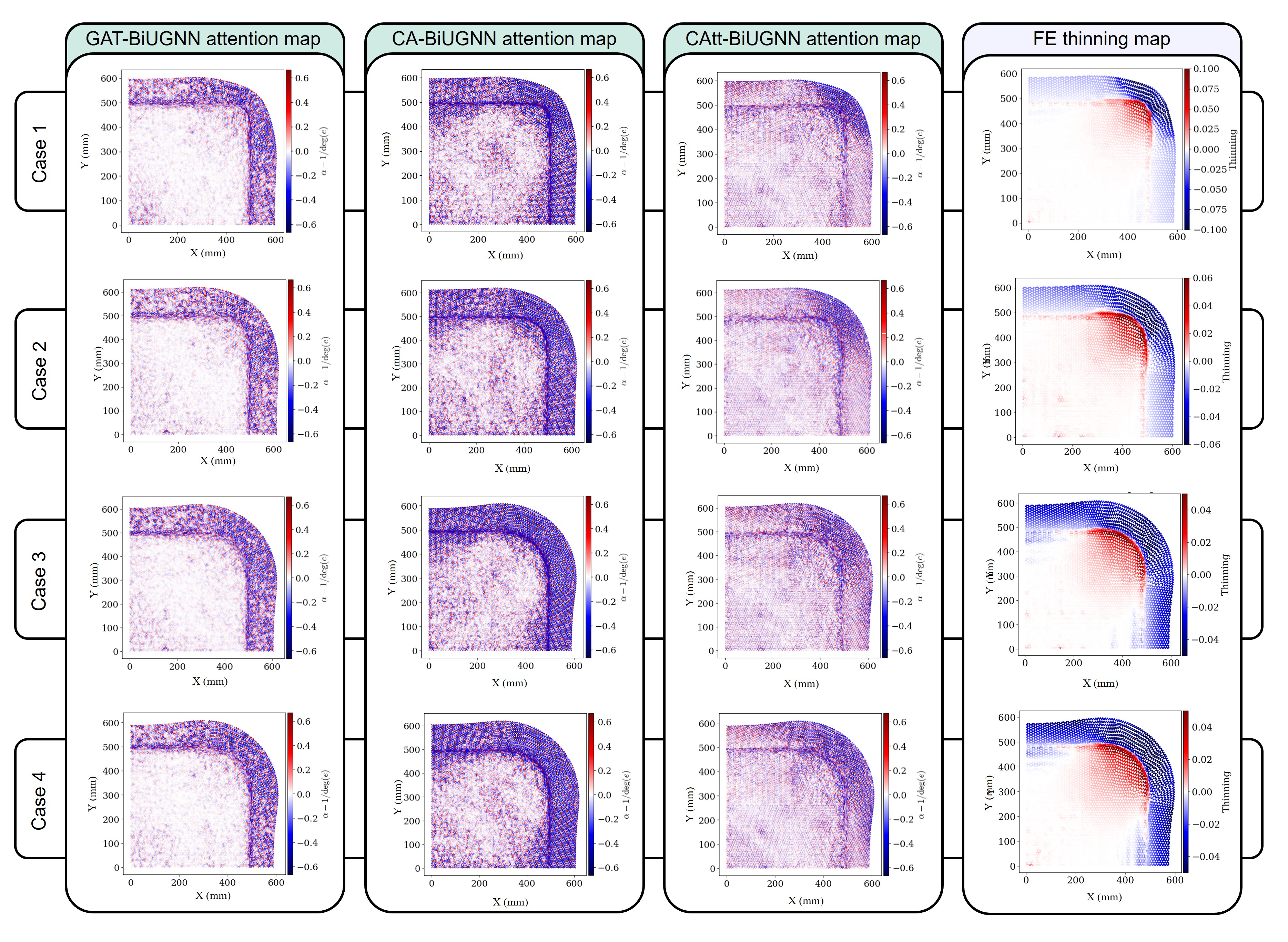}
\caption{Node-to-element attention-deviation maps for four corner-shaped test cases at a randomly selected rollout step, held fixed across all panels. Each row represents one test case. The first three columns show GAT-BiUGNN, CA-BiUGNN, and CAtt-BiUGNN; the last shows the corresponding FE thinning field as a spatial reference. Each directed node-to-element edge is coloured by its deviation from uniform weighting, with white indicating zero deviation.}
\label{visualise_attention}
\end{figure}

\rev{Figure~\ref{visualise_attention} shows the node-to-element attention deviations at the randomly selected rollout step used for all panels. Let \(\deg(i)\) denote the number of incoming node-to-element edges of element centre-node \(i\) at the plotted graph level. Under equal weighting of the incoming edges, each incoming edge has weight \(1/\deg(i)\). The deviation plotted on edge \(v\to i\) is
\begin{equation}
\Delta\alpha_{v\to i,L}^{\,n}
=
\alpha_{v\to i,L}^{\,n}
-
\frac{1}{\deg(i)}.
\label{eq:attention_deviation_visualisation}
\end{equation}
Positive and negative values indicate weights above and below equal weighting, respectively.}

\rev{The GAT-BiUGNN and CA-BiUGNN maps show relatively weak case-to-case variation. As reported in Table~\ref{tab:corner-ablations}, GAT-BiUGNN improves seven measures relative to Vanilla-BiUGNN, whereas CA-BiUGNN improves all six thinning-related measures but increases both displacement errors. Neither variant matches CAtt-BiUGNN, which gives the lowest error for all eight measures.} CAtt-BiUGNN produces more spatially structured attention patterns near geometric transitions and thinning-prone regions. \rev{These maps should be interpreted as qualitative evidence that edge-aware attention adapts to local node--element geometry, rather than as evidence that the network explicitly enforces finite-element equilibrium or physical consistency.} This interpretation helps explain why edge-aware message weighting improves the overall predictive balance between nodal displacement and elemental thinning.

\subsection{\rev{Ensemble-disagreement uncertainty analysis}}
\label{sec:ensemble_uncertainty}

\rev{An ensemble-disagreement analysis was performed using three independently trained instances of the proposed model for each case study. The analysis evaluates the final predictions over all 25 test trajectories and examines whether larger model disagreement is associated with larger prediction errors.}

\rev{For a predicted output component $y_{j,k}$ at spatial location $j$, component $k$, and ensemble member $r\in\{1,2,3\}$, the ensemble mean and sample standard deviation are}
\begin{equation}
\rev{
\bar{y}_{j,k}
=\frac{1}{3}\sum_{r=1}^{3}y_{j,k}^{(r)},
\qquad
s_{j,k}
=\sqrt{\frac{1}{2}\sum_{r=1}^{3}
\left(y_{j,k}^{(r)}-\bar{y}_{j,k}\right)^2}.
}
\label{eq:ensemble_mean_std}
\end{equation}
\rev{For nodal position, the scalar disagreement and absolute error are}
\begin{equation}
\rev{
U^{\mathrm{pos}}_v
=\sqrt{s_{v,x}^{2}+s_{v,y}^{2}+s_{v,z}^{2}},
\qquad
E^{\mathrm{pos}}_v
=\left\|\bar{\vecv{x}}_v-\vecv{x}^{*}_v\right\|_2.
}
\label{eq:position_uncertainty_error}
\end{equation}
\rev{For elemental thinning, $U^{\theta}_i$ is the sample standard deviation of the three final-step thinning predictions and $E^{\theta}_i=|\bar{\theta}_i-\theta_i^{*}|$. The mean disagreement and Spearman rank correlation $\rho(U,E)$ are calculated after pooling the retained spatial locations over all test trajectories. Position statistics use all nodes. Consistent with the main accuracy evaluation, thinning statistics retain elements satisfying $|\theta_i^{*}|>0.005$ for both the Dome and Corner cases.}

\begin{table}[htbp]
\caption{\rev{Final-step ensemble-disagreement uncertainty and its Spearman rank correlation with absolute prediction error over the 25-trajectory test sets. Position uncertainty is reported in mm; thinning uncertainty is dimensionless.}}
\label{tab:ensemble-uncertainty}
\centering
\begingroup
\small
\setlength{\tabcolsep}{8pt}
\renewcommand{\arraystretch}{1.15}
\begin{tabular}{@{}llrr@{}}
\toprule
\rev{\textbf{Case}} &
\rev{\textbf{Output}} &
\rev{\makecell[c]{\textbf{Mean}\\\textbf{uncertainty}}} &
\rev{\makecell[c]{\textbf{Spearman}\\$\boldsymbol{\rho(U,E)}$}} \\
\midrule
\rev{Dome}   & \rev{Position} & \rev{0.313828} & \rev{0.408989} \\
\rev{Dome}   & \rev{Thinning} & \rev{0.002515} & \rev{0.534934} \\
\rev{Corner} & \rev{Position} & \rev{0.213555} & \rev{0.727300} \\
\rev{Corner} & \rev{Thinning} & \rev{0.000960} & \rev{0.387456} \\
\bottomrule
\end{tabular}
\endgroup
\end{table}

\rev{Table~\ref{tab:ensemble-uncertainty} shows a positive association between ensemble disagreement and absolute prediction error for both outputs and both case studies. The association is strongest for Corner position ($\rho=0.727$), followed by Dome thinning ($\rho=0.535$), Dome position ($\rho=0.409$), and Corner thinning ($\rho=0.387$). Thus, spatial locations at which the ensemble members disagree more also tend to be locations with larger FE-reference errors. These ensemble-disagreement statistics provide an empirical indication of predictive uncertainty rather than calibrated confidence intervals.}

\subsection{\rev{Output-level physical consistency during autoregressive rollout }}
\label{Physical consistency and rollout error evolution}

\rev{The preceding results evaluate nodal geometry and elemental thinning
accuracy independently. To further assess whether these two predicted fields
remain physically consistent during autoregressive rollout, an element-wise
area--thickness consistency check is performed for the corner-shaped case
study. The same \(M=25\) test trajectories and \(N_p=10\) prediction
transitions are used.}

\rev{The predicted nodal positions are reconstructed from the displacement
increments using Eq.~\eqref{eq:position_update}. For the node-centred
baselines, elemental thinning is obtained by averaging the nodal thinning
predictions over the three nodes of each triangular element. Predictions that
are already defined at the element level are used directly.}

\rev{For a triangular element \(i\), with
\(\mathcal{N}(i)=\{v_1,v_2,v_3\}\), the current midsurface area is calculated
from the predicted nodal positions as}
\begin{equation}
\rev{\displaystyle
A_i^{\,m,n}
=
\frac{1}{2}
\left\|
\left(
\vecv{x}_{v_2}^{\,m,n}
-
\vecv{x}_{v_1}^{\,m,n}
\right)
\times
\left(
\vecv{x}_{v_3}^{\,m,n}
-
\vecv{x}_{v_1}^{\,m,n}
\right)
\right\|_2 .
}
\label{eq:predicted_element_area}
\end{equation}

\rev{According to Eq.~\eqref{eq:thinning_value},
\(1-\theta_i^{\,m,n}\) is the element thickness ratio. The local
area--thickness residual is therefore defined for the prediction and FE
reference as}
\begin{equation}
\rev{\displaystyle
r_i^{\,m,n}
=
\ln\left(
\frac{A_i^{\,m,n}}
     {A_i^{\,m,0}}
\right)
+
\ln\left(
1-\theta_i^{\,m,n}
\right),
\qquad
r_i^{*\,m,n}
=
\ln\left(
\frac{A_i^{*\,m,n}}
     {A_i^{*\,m,0}}
\right)
+
\ln\left(
1-\theta_i^{*\,m,n}
\right).
}
\label{eq:area_thickness_residual}
\end{equation}

\rev{This residual represents the logarithm of the local
midsurface-area--thickness volume ratio. It is expected to remain small for
an approximately incompressible sheet. The mean absolute FE residual is
\(5.17\times10^{-4}\) over the rollout and \(6.02\times10^{-4}\) at the
final state, supporting its use as an output-level consistency diagnostic.}

\rev{The model consistency gap is evaluated against the corresponding FE
residual at each predicted state:}
\begin{equation}
\rev{\displaystyle
G^{\,n}
=
\frac{1}{M N_e}
\sum_{m=1}^{M}
\sum_{i=1}^{N_e}
\left|
r_i^{\,m,n}
-
r_i^{*\,m,n}
\right|.
}
\label{eq:local_physical_consistency_gap}
\end{equation}

\rev{The rollout-average and final-state consistency gaps are then}
\begin{equation}
\rev{\displaystyle
G_{\upr{avg}}
=
\frac{1}{N_p}
\sum_{n=1}^{N_p}
G^{\,n},
\qquad
G_{\upr{final}}
=
G^{\,N_p}.
}
\label{eq:physical_consistency_summary}
\end{equation}

\begin{table}[htbp]
\caption{\rev{Element-wise area--thickness consistency for the corner-shaped
case study using seed-0 autoregressive rollouts. Lower values are better.}}
\label{tab:corner_physical_consistency}
\centering
\begingroup
\small
\setlength{\tabcolsep}{9pt}
\renewcommand{\arraystretch}{1.15}
\begin{tabular}{lcc}
\toprule
\rev{\textbf{Model}} &
\rev{\(\boldsymbol{G_{\upr{avg}}}\)} &
\rev{\(\boldsymbol{G_{\upr{final}}}\)} \\
\midrule
\rev{UMGN}            & \rev{0.00823} & \rev{0.01541} \\
\rev{RUGNN}           & \rev{0.00726} & \rev{0.01337} \\
\rev{Transolver}      & \rev{0.01138} & \rev{0.02919} \\
\rev{GNN-Transformer}       & \rev{0.00533} & \rev{0.01181} \\
\rev{Vanilla-BiUGNN}  & \rev{0.00883} & \rev{0.01653} \\
\rev{GAT-BiUGNN}      & \rev{0.00687} & \rev{0.01281} \\
\rev{CA-BiUGNN}       & \rev{0.00777} & \rev{0.01422} \\
\rev{CAtt-BiUGNN (proposed)}
                      & \rev{\textbf{0.00514}}
                      & \rev{\textbf{0.00949}} \\
\bottomrule
\end{tabular}
\endgroup
\end{table}

\begin{figure}[htbp]
\centering
\includegraphics[width=0.72\linewidth, trim=0mm 0mm 0mm 0mm, clip]{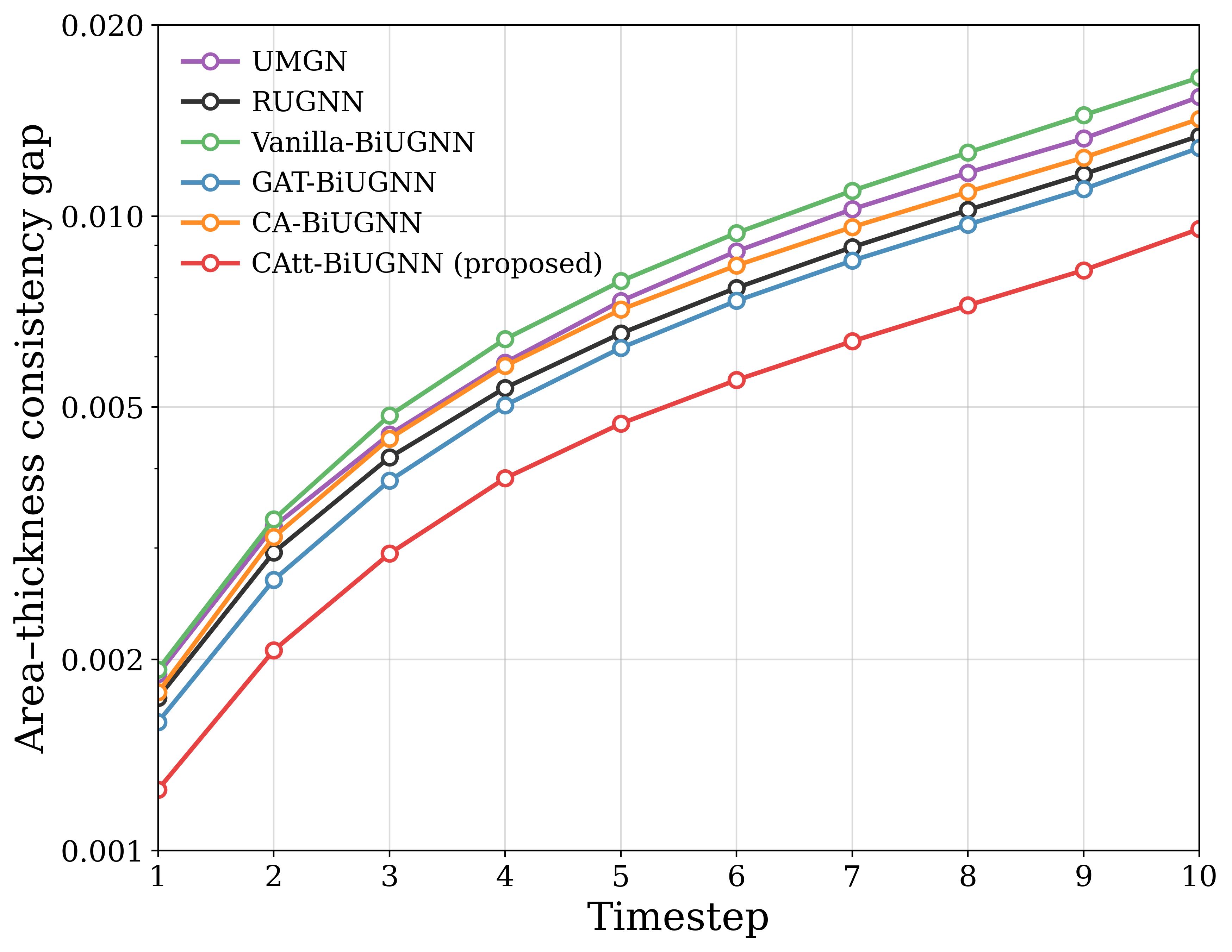}
\caption{\rev{Evolution of the mean local area--thickness consistency gap $G^{\,n}$ over the ten autoregressive prediction transitions for the corner-shaped case study using seed-0 predictions.}}
\label{fig:corner_physical_consistency_gap}
\end{figure}

\rev{Table~\ref{tab:corner_physical_consistency} shows that CAtt-BiUGNN
gives the lowest rollout-average and final-state consistency gaps among the
evaluated models. Figure~\ref{fig:corner_physical_consistency_gap} further
shows the evolution of \(G^{\,n}\) for the six representative models.
The consistency gap increases during rollout for all models shown, while CAtt-BiUGNN maintains the lowest gap over the later prediction stages. This trend is
consistent with the displacement and thinning error evolution in
Figure~\ref{corner_mean_over_time}.} \rev{The area--thickness consistency gap should be interpreted together with
the independent displacement and thinning errors, because errors in the two
fields can partially compensate within the same element. The present measure therefore provides an output-level physical-consistency diagnostic. These results cover the ten-step rollout over the complete forming process represented in the dataset.}

\section{Conclusion}
\label{sec:conclusion}
This study proposed CAtt-BiGNN, a cross-attention-based bipartite graph neural network for coupled nodal and elemental field prediction in large-deformation sheet material forming. \rev{The work was motivated by the observation that explicit dynamic FE simulations organise nodal kinematics and element-level deformation measures on distinct but coupled state domains, with bidirectional update dependencies between them.} The proposed bipartite graph representation retains mesh nodes and finite elements as distinct but coupled graph entities. This allows nodal displacement increments and element-level deformation states to be predicted on their native discretisation domains. In the sheet-forming instantiation considered in this work, the element-level state corresponds to elemental thinning. \rev{Treating nodal displacement increments and elemental thinning as coupled physical outputs provides complementary information for forming analysis: nodal displacement describes geometry evolution, whereas elemental thinning identifies local deformation and critical manufacturability regions. Feeding both predicted fields back during autoregressive rollout allows each field to inform the subsequent update of the other.} The model remains a learned simulator rather than an explicitly constrained FE solver, but the representation provides a finite-element-inspired inductive bias for learning nodal-elemental interactions from simulation data.

The proposed edge-aware cross-attention processor further improves the modelling of directional node-element coupling. \rev{The processor dynamically updates directed edge representations and weights the exchanged information according to the receiving state and the evolving local node-element relation. } For large forming graphs, the hierarchical CAtt-BiUGNN extension combines the same bipartite processor with \rev{separate nodal and elemental graph downsampling and upsampling}, enabling longer-range information propagation while preserving the nodal-elemental representation across graph levels.

\rev{The numerical results support the proposed modelling choices. CAtt-BiGNN gives the lowest displacement errors and threshold-based thinning errors among the evaluated models. It also gives the lowest top-2.5\% critical relative thinning error and the second-lowest top-5\% and top-10\% errors, ranking first in six of the eight reported measures. In the corner-shaped benchmark, which involves a larger graph, sharper geometric transitions, and more complex hot-forming conditions, CAtt-BiUGNN gives the lowest mean error for all eight measures in the main-model comparison. In the controlled seed-0 BiUGNN ablation, CAtt-BiUGNN gives the lowest error for all eight measures, including both displacement measures, all three threshold-based thinning measures, and all three mean top-$p\%$ critical relative thinning errors. Thus, CAtt-BiUGNN provides the best overall balance between geometry and thinning prediction. The output-level area–thickness analysis shows that CAtt-BiUGNN gives the smallest rollout-average and final-step consistency gaps among the eight Corner models evaluated with seed 0. This diagnostic indicates consistency between the predicted geometry and thinning fields over the evaluated rollout. It does not imply that equilibrium or constitutive relations are enforced, nor does it establish stability beyond the ten-step prediction horizon. The controlled ablations indicate that the bipartite representation alone is insufficient to explain the performance gain and that the proposed edge-aware node-element coupling contributes to the improvement in joint displacement and thinning prediction. The corner results further demonstrate that the dual-domain formulation can be retained within a hierarchical architecture for larger forming graphs.}

\rev{The proposed CAtt-BiGNN and CAtt-BiUGNN reduce online simulation runtime by 99.46\% and 99.73\%, respectively, relative to the representative FE runs for the Dome and Corner cases. The ensemble-disagreement analysis gives positive uncertainty--error correlations for both position and thinning in both cases, indicating that model disagreement can serve as an empirical indicator of spatially elevated error.}

Overall, the proposed framework provides a finite-element-inspired graph learned simulator for predicting manufacturability-oriented nodal and elemental fields during autoregressive rollout. The present evaluation focuses on geometry-driven generalisation within two sheet-forming benchmark families under fixed material and process settings. Future work will extend the graph state and dataset construction to include richer element-level variables, material descriptors, process parameters, contact conditions, and mesh-topology variations. \rev{The explicit separation of nodal and elemental state domains also provides a natural basis for incorporating physics-informed constraints in future work. In particular, nodal dynamic-equilibrium residuals and element-level constitutive relations could be incorporated as additional physics-informed constraints on their corresponding FE domains to strengthen physical consistency beyond the present output-level assessment. \rev{Further work will also investigate more parameter-efficient hierarchical coupling and rollout strategies to reduce offline training and inference costs while retaining the accuracy of coupled nodal--elemental prediction.}} Integrating the learned simulator into optimisation workflows may further support rapid manufacturability assessment and efficient design-space exploration in large-deformation forming analysis.

\section*{Acknowledgement}
The authors acknowledge funding support from UKRI (UKRI221: AI-Driven Design for Forming High-Performance Vehicle Parts), as well as PhD scholarships from Imperial College London. The authors would like to thank Keysight for their technical support with the PAM-STAMP software. For the purpose of open access, the authors have applied a Creative Commons Attribution (CC BY) license to any Author Accepted Manuscript version arising.

\section*{CRediT authorship contribution statement}

\textbf{Yingxue Zhao}: Conceptualization, Methodology, Software, Investigation, Data curation, Formal analysis, Validation, Visualization, Writing – original draft, Writing – review \& editing.
\textbf{Haoran Li}: Methodology, Formal analysis, Validation, Writing – review \& editing.
\textbf{Haosu Zhou}: Methodology, Formal analysis, Writing – review \& editing.
\textbf{Tobias Pfaff}: Methodology, Validation, Supervision, Writing – review \& editing.
\textbf{Nan Li}: Conceptualization, Methodology, Formal analysis, Validation, Supervision, Project administration, Funding acquisition, Writing – review \& editing.

\section*{Declaration of competing interest}

The authors declare that they have no known competing financial interests
or personal relationships that could have appeared to influence the work
reported in this paper.

\section*{Data availability}
The data supporting the findings of this study are available upon reasonable request.

\appendix
\section{Dataset-size assessment}
\label{app:dataset_size_assessment}

\rev{The effect of training dataset size was assessed for the Dome and Corner cases. In each panel, the red curve shows the prediction error using the left vertical axis. The blue and green curves show the training time and the dataset generation time, respectively, using the right vertical axis labelled ``Total time (h)''.}

\rev{For the Dome case, models were assessed using 12, 24, 36, and 50 training samples with random seed 0. The four panels show: (a) mean positional Euclidean error, (b) positional MAE, (c) threshold-based thinning error with $\tau=0.005$, and (d) top-$5\%$ critical relative thinning error. The reported training time corresponds to 1000 epochs, and the dataset generation time is based on 3 min per sample.}

\begin{figure}[H]
    \centering
    \includegraphics[width=\linewidth]{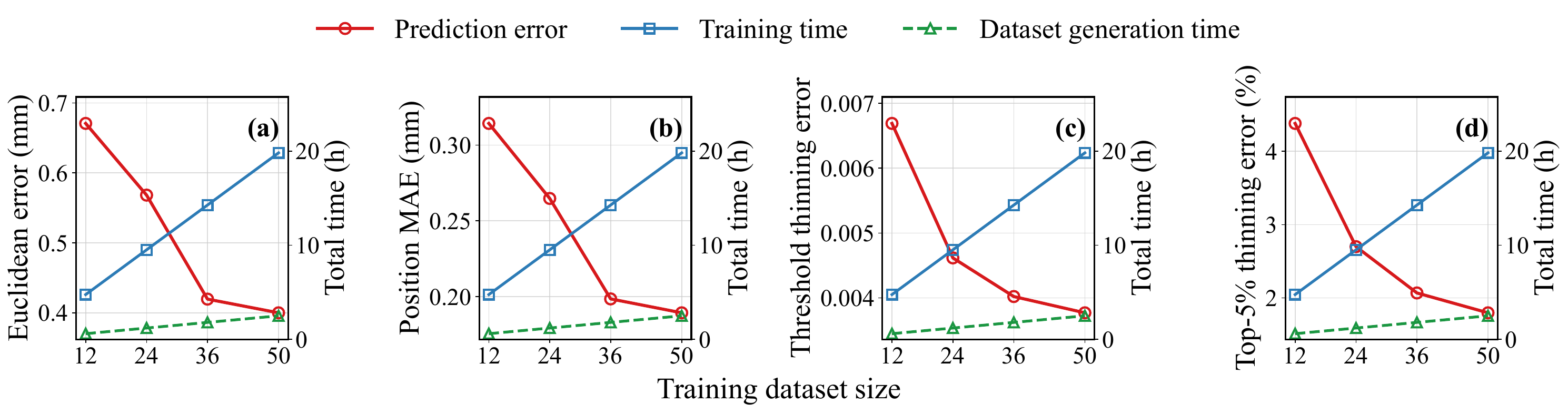}
    \caption{\rev{Dome dataset-size assessment. The red curves show the prediction errors, while the blue and green curves show the training time and the dataset generation time, respectively.}}
    \label{fig:dome_dataset_convergence}
\end{figure}

\rev{Figure~\ref{fig:dome_dataset_convergence} shows that the Dome prediction errors decrease substantially as the dataset size increases from 12 to 36 samples. The changes from 36 to 50 samples are smaller. Over the same increase, the training time rises from approximately 14.27 h to 19.82 h, while the dataset generation time rises from 1.80 h to 2.50 h. This indicates diminishing accuracy improvements beyond 36 samples relative to the additional time required. At 50 samples, the top-$5\%$ critical relative thinning error is below $5\%$.}

\rev{For the Corner case, models were assessed using 25, 50, 75, and 100 training samples with random seed 0. The four panels show: (a) mean positional Euclidean error, (b) positional MAE, (c) threshold-based thinning error with $\tau=0.005$, and (d) top-$1\%$ critical relative thinning error. The reported training time corresponds to 500 epochs, and the dataset generation time is based on 8 min per sample.}

\begin{figure}[H]
    \centering
    \includegraphics[width=\linewidth]{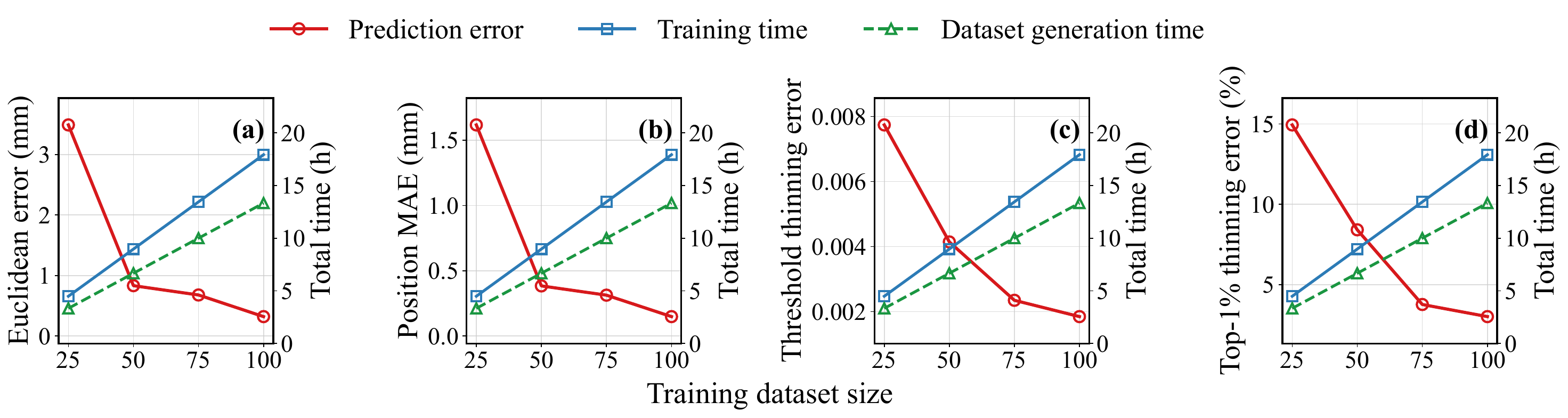}
    \caption{\rev{Corner dataset-size assessment. The red curves show the prediction errors, while the blue and green curves show the training time and the dataset generation time, respectively.}}
    \label{fig:corner_dataset_convergence}
\end{figure}

\rev{Figure~\ref{fig:corner_dataset_convergence} shows that all four
Corner prediction errors decrease as the training dataset size
increases. The largest reductions occur from 25 to 50 samples,
while the reductions from 50 to 75 and from 75 to 100 samples
are smaller. At 75 samples, the training time is approximately
13.45 h and the dataset generation time is 10.00 h. Increasing
the dataset to 100 samples further reduces the prediction errors,
but increases the training time to approximately 17.93 h and
the dataset generation time to 13.33 h. Considering these
trends and the additional computational costs of FE data
generation and model training, 100 training samples were
used for the Corner case.}

\section{Prediction results on dome-shaped case study}
\label{appA}

\begin{figure}[H]
  \centering
  \includegraphics[width=\linewidth,trim=0mm 0mm 0mm 0mm,clip]{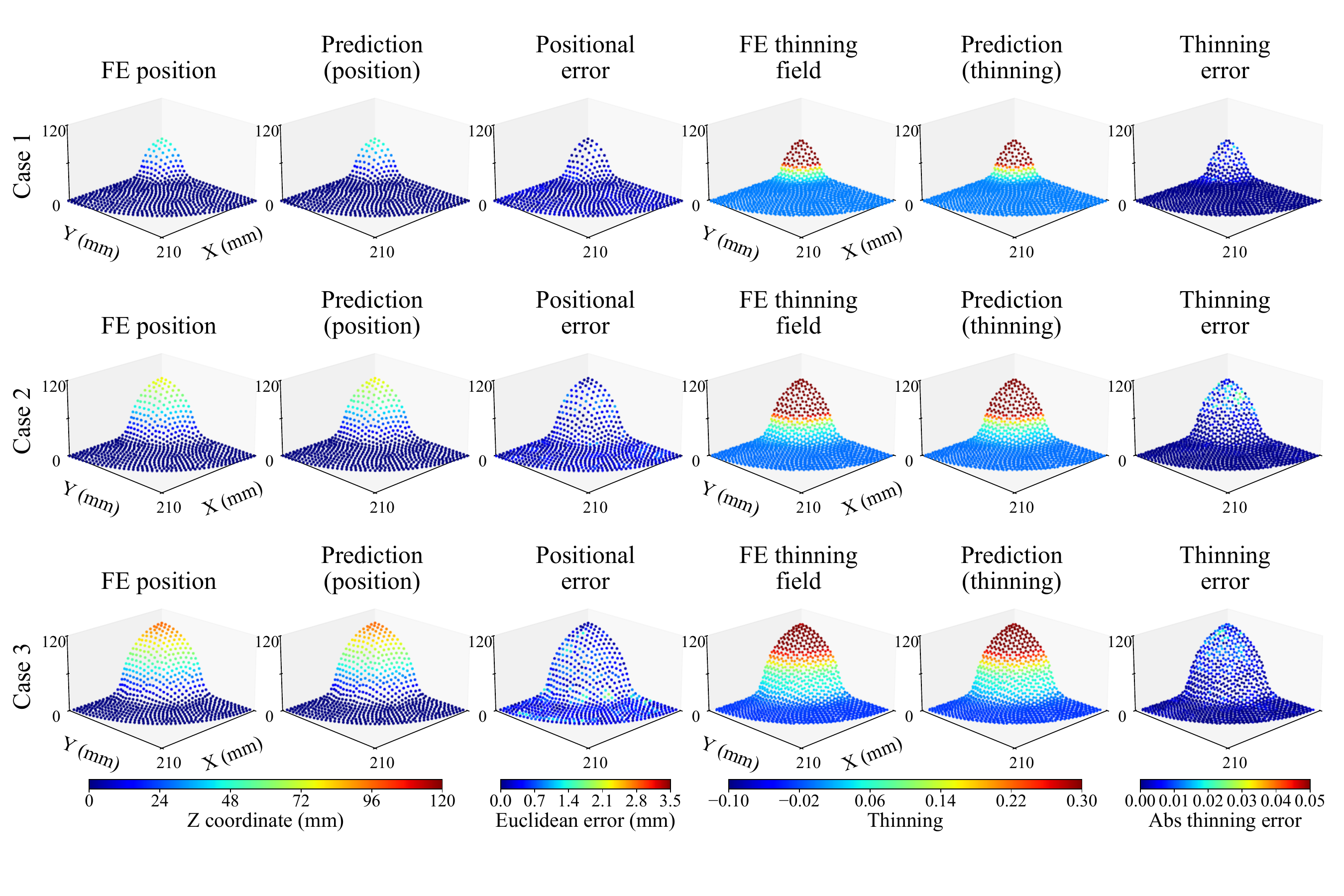}
  \caption{Final-timestep predictions of the CAtt-BiGNN for three dome test cases. Each row corresponds to one design case. Columns show FE reference position, predicted position, Euclidean position error, FE thinning field, predicted thinning field, and absolute thinning error. }
  \label{fig:dome_extreme_cases}
\end{figure}

Figure~\ref{fig:dome_extreme_cases} presents the prediction performance of the CAtt-BiGNN. Three representative dome-shaped cases, corresponding to the small, median, and large dome radii within the design space, are displayed. Across all three cases, the predicted geometries are in close agreement with the FE simulation results, with Euclidean positional errors at the final timestep remaining below approximately $2$~mm. The predicted thinning fields also exhibit a high level of consistency with the FE ground truth, capturing the spatial variation of the thinning field with strong fidelity. These results suggest that the model maintains reliable predictive accuracy across dome geometries with substantially different curvature conditions.

\section{Temporal rollout prediction results on dome-shaped case study}
\label{appB}

\begin{figure}[H]
  \centering
  \includegraphics[
    width=\linewidth,
    height=0.78\textheight,
    keepaspectratio
  ]{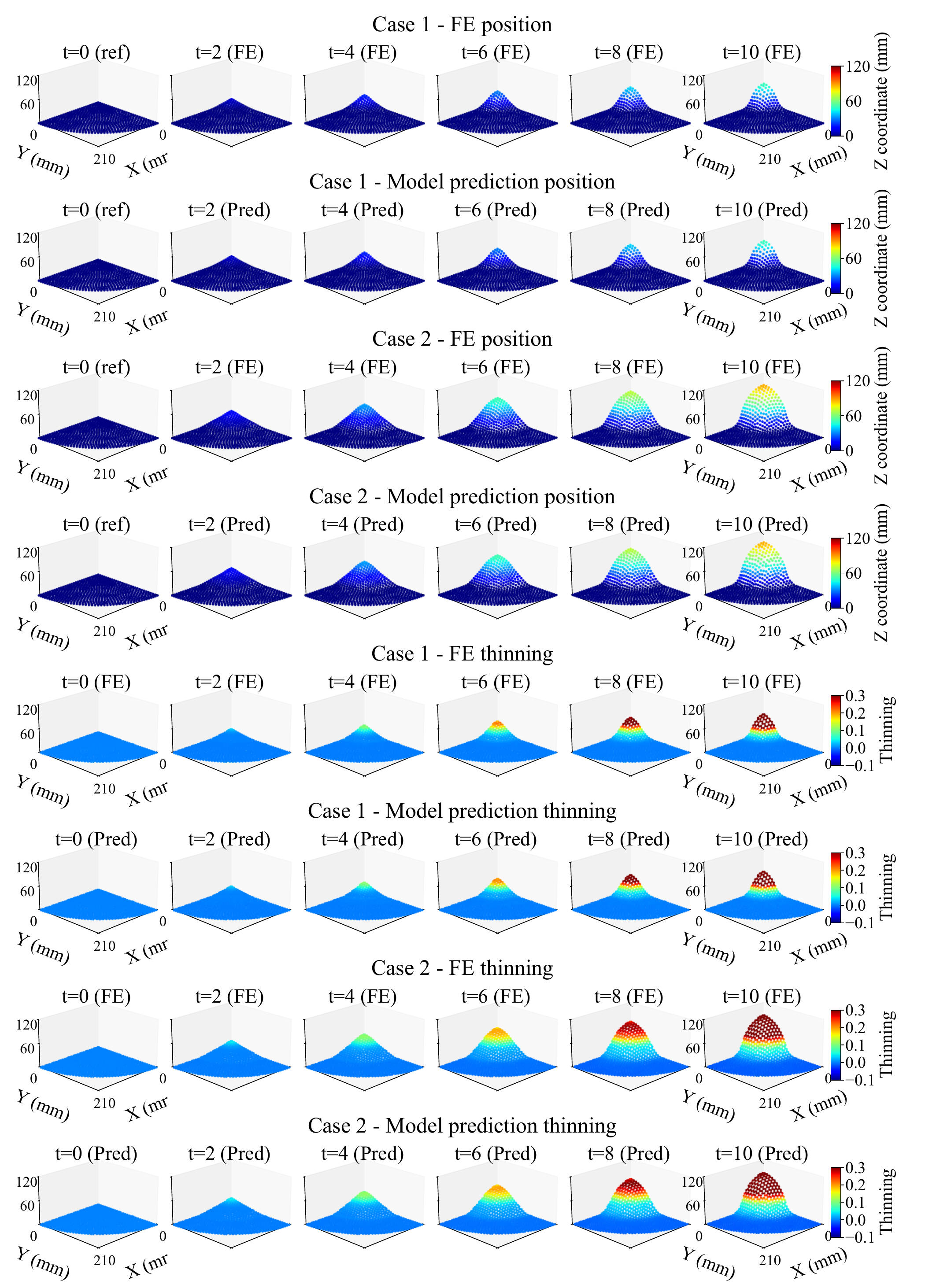}
  \caption{Temporal rollout predictions of the CAtt-BiGNN for two dome test cases. Columns correspond to rollout timesteps. Rows compare FE and predicted deformed positions for each case, followed by FE and predicted thinning fields. The figure illustrates temporal consistency during autoregressive rollout.}
  \label{fig:dome_rollout}
\end{figure}

Figure~\ref{fig:dome_rollout} illustrates the temporal rollout performance of the CAtt-BiGNN. Representative low- and high-radius dome cases are displayed. The results show that the model produces temporally coherent and stable predictions throughout the rollout sequence. The deformation and thinning fields evolve smoothly over time and remain closely aligned with the FE simulation results across all timesteps. The differences between the predicted and ground-truth thinning fields remain small, indicating that the model maintains stable predictions without visible rollout divergence in the displayed cases.

\section{Prediction results on corner-shaped case study}
\label{appC}

\begin{figure}[H]
  \centering
  \includegraphics[width=\linewidth,trim=0mm 30mm 0mm 10mm,clip]{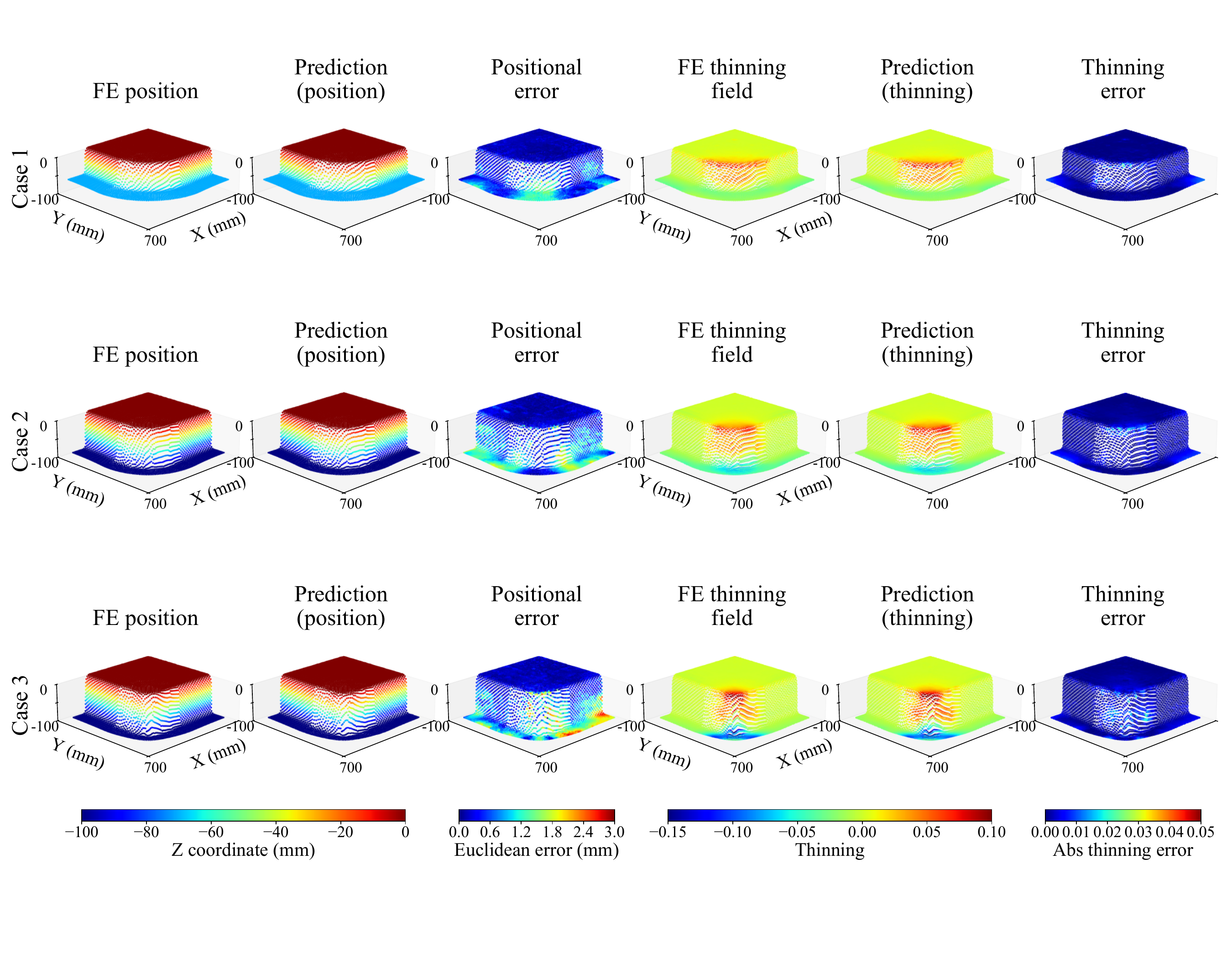}
  \caption{Final-timestep predictions of the CAtt-BiUGNN for three corner test cases. Each row corresponds to one design case. Columns show FE reference position, predicted position, Euclidean position error, FE thinning field, predicted thinning field, and absolute thinning error. }
  \label{fig:corner_extreme_cases}
\end{figure}

Figure~\ref{fig:corner_extreme_cases} presents the final-timestep prediction performance of the CAtt-BiUGNN variant for three representative corner test cases with the small, median, and large corner height.

Across the representative cases, the predicted geometries closely match the FE-simulated shapes, with maximum Euclidean positional errors below approximately $3$~mm. The predicted thinning fields reproduce the principal spatial patterns of the FE results, while the absolute thinning errors remain below $0.03$. These visual results indicate that the model maintains strong predictive accuracy across distinct deformation patterns.

\section{Temporal rollout prediction results on corner-shaped case study}
\label{appD}

\begin{figure}[H]
  \centering
  \includegraphics[
    width=\linewidth,
    height=0.78\textheight,
    keepaspectratio
  ]{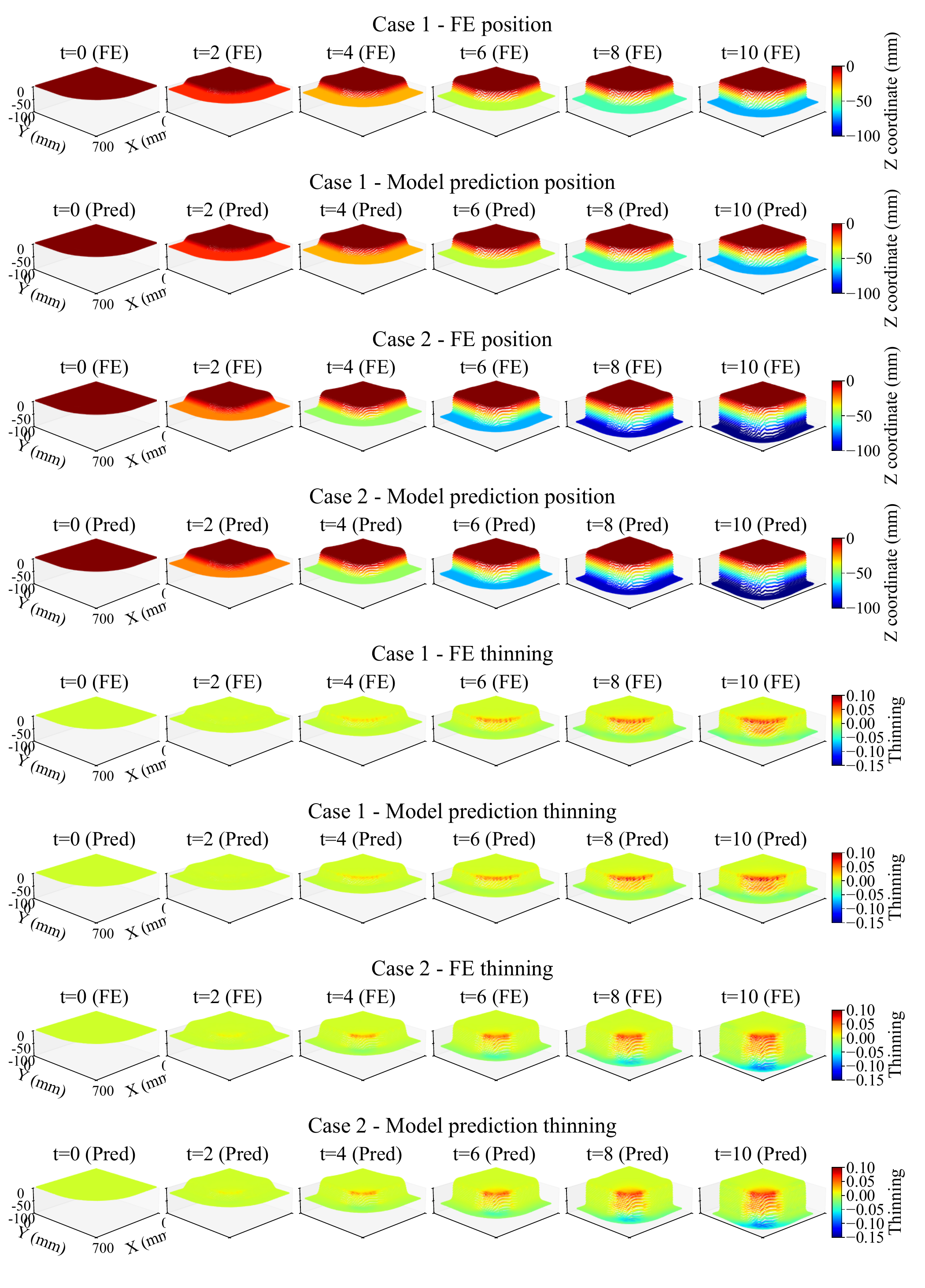}
  \caption{Temporal rollout predictions of the CAtt-BiUGNN for two corner test cases. Columns correspond to rollout timesteps. Rows compare FE and predicted deformed positions for each case, followed by FE and predicted thinning fields. The figure illustrates the stability of autoregressive predictions across the forming process.}
  \label{fig:corner_rollout}
\end{figure}

Figure~\ref{fig:corner_rollout} illustrates the temporal rollout performance of the CAtt-BiUGNN. 
Two representative corner test cases are used to assess the stability and accuracy of the model’s autoregressive predictions. 
The visualisations show the evolution of the predicted deformation and thinning fields from the initial to the final timestep. 

The predicted sequences exhibit smooth and realistic temporal transitions that align closely with the FE results. The differences between the predicted and ground-truth thinning fields remain small throughout the rollout, indicating that the model maintains stable predictions without visible rollout divergence in the displayed cases.
\section{Parameter-budget-controlled comparison}
\label{appF}

This appendix provides the parameter-budget-controlled comparison referred to
in Section~\ref{Corner shape case study}. The comparison is included to examine
whether the relative performance of the models changes when the trainable-parameter
counts of the baseline models are made comparable to those of the BiUGNN variants.

\rev{Table~\ref{tab:parameter-budget} compares prediction accuracy
and computational requirements under the parameter-budget-controlled setting. The threshold-based thinning error uses $\tau=0.005$. All prediction errors are single-run results obtained from models trained with random seed 0 and averaged over the 25 test cases. Lower values are better for all prediction errors, training time, GPU memory, and inference time; higher values are better for speed-up.
Increasing the hidden dimension of the node-centred baselines substantially
increases their model size and training cost, but does not reproduce the
prediction accuracy of CAtt-BiUGNN. The proposed model gives the lowest errors
for all four reported prediction measures under this comparison.}

\begin{table}[H]
\caption{\rev{Parameter-budget-controlled comparison of prediction accuracy and
computational efficiency for the corner-shaped case study.}}
\label{tab:parameter-budget}
\centering
\begingroup
\footnotesize
\setlength{\tabcolsep}{3.5pt}
\renewcommand{\arraystretch}{1.15}

\resizebox{\textwidth}{!}{%
\begin{tabular}{lccccccccc}
\toprule

\multicolumn{1}{c}{} &
\multicolumn{2}{c}{\textbf{Displacement evaluation}} &
\multicolumn{2}{c}{\textbf{Thinning evaluation}} &
\multicolumn{5}{c}{\textbf{Model size and computational efficiency}} \\

\cmidrule(lr){2-3}
\cmidrule(lr){4-5}
\cmidrule(l){6-10}

\textbf{Model} &
\makecell{\textbf{Mean positional}\\
          \textbf{Euclidean error}\\
          \textbf{(mm)}} &
\makecell{\textbf{Position}\\
          \textbf{MAE}\\
          \textbf{(mm)}} &
\makecell{\textbf{Threshold-based}\\
          \textbf{thinning error}} &
\makecell{\textbf{Top 1\% critical}\\
          \textbf{relative thinning}\\
          \textbf{error}} &
\makecell{\textbf{Trainable}\\
          \textbf{parameters}\\
          \textbf{(M)}} &
\makecell{\textbf{Training time}\\
          \textbf{(s/epoch)}} &
\makecell{\textbf{Peak training}\\
          \textbf{GPU memory}\\
          \textbf{(GiB)}} &
\makecell{\textbf{Inference time}\\
          \textbf{(s/trajectory)}} &
\makecell{\textbf{Speed-up}\\
          \textbf{over FE}\\
          \textbf{($\times$)}} \\

\midrule

UMGN-128
    & 0.497063
    & 0.229924
    & 0.004284
    & 14.198\%
    & 82.918
    & 49.25
    & 1.590
    & 1.025
    & 468.3 \\

UMGN-212
    & 0.362840
    & 0.169375
    & 0.003713
    & 11.783\%
    & 227.440
    & 74.75
    & 4.287
    & 1.025
    & 468.5 \\

RUGNN-128
    & 0.400468
    & 0.186470
    & 0.003963
    & 13.273\%
    & 85.269
    & 65.99
    & 1.753
    & 0.986
    & 486.9 \\

RUGNN-212
    & 0.479300
    & 0.221297
    & 0.004093
    & 12.107\%
    & 233.867
    & 88.93
    & 4.456
    & 1.141
    & 420.7 \\

\midrule

Vanilla-BiUGNN
    & 0.550565
    & 0.250835
    & 0.003194
    & 3.956\%
    & 237.290
    & 91.55
    & 4.477
    & 1.193
    & 402.5 \\

GAT-BiUGNN
    & 0.430733
    & 0.199060
    & 0.002084
    & 3.808\%
    & 237.630
    & 132.92
    & 4.485
    & 1.312
    & 365.8 \\

CA-BiUGNN
    & 0.576968
    & 0.263821
    & 0.002226
    & 3.371\%
    & 239.322
    & 125.56
    & 4.494
    & 1.307
    & 367.2 \\

CAtt-BiUGNN
    & \textbf{0.320854}
    & \textbf{0.148962}
    & \textbf{0.001846}
    & \textbf{3.019\%}
    & 239.490
    & 129.11
    & 4.497
    & 1.285
    & 373.5 \\

\bottomrule
\end{tabular}%
}

\endgroup
\end{table}

\rev{Most BiUGNN parameters lie in the edge-specific inter-level weight matrices of the separate nodal and elemental hierarchies. Because each inter-level edge has its own weights, the parameter count grows with mesh size. This increases the one-off offline
training cost. However, the inference times of the compared large-capacity
models remain of the same order, and all learned models retain a substantial
speed advantage over the FE simulation. The present implementation therefore
prioritises coupled displacement--thinning prediction accuracy, while reducing
the training and inference cost without degrading this accuracy remains an
important direction for further development.}

\FloatBarrier 



\nolinenumbers
 \biboptions{numbers,sort&compress}
 \bibliographystyle{elsarticle-num} 
 \bibliography{ref}






\end{document}